\documentclass{aa} 
\usepackage[varg]{txfonts}
\usepackage{graphicx,url}
\usepackage{natbib}

\usepackage{lscape}

\usepackage{latexsym}

\usepackage{multirow}

\begin{document}

\title{Supernova rates from the SUDARE VST-Omegacam search. I
\thanks{Based on observations made with ESO telescopes at the Paranal Observatory under programme ID 
088.D-4006, 088.D-4007, 089.D-0244, 089.D-0248, 090.D-0078, 090.D-0079, 
088.D-4013, 
089.D-0250, 090.D-0081 
}}

\author{E. Cappellaro\inst{\ref{OAPD}} \and M.T. Botticella\inst{\ref{OAC}} \and
  G. Pignata\inst{\ref{Chile},\ref{millenium}} \and  L. Grado\inst{\ref{OAC}} \and L. Greggio\inst{\ref{OAPD}} 
  \and L. Limatola\inst{\ref{OAC}} \and  M. Vaccari\inst{\ref{SA},\ref{oabo}} \and
A. Baruffolo\inst{\ref{OAPD}} \and S. Benetti\inst{\ref{OAPD}} \and F. Bufano\inst{\ref{oact}}
\and M. Capaccioli\inst{\ref{uniNA}} \and E. Cascone\inst{\ref{OAC}} \and G. Covone\inst{\ref{uniNA}} \and D. De Cicco\inst{\ref{uniNA}} \and S. Falocco\inst{\ref{uniNA}} \and M. Della Valle\inst{\ref{OAC}} \and  M. Jarvis\inst{\ref{oxford},\ref{SA}} \and L. Marchetti\inst{\ref{milton}} \and N. R. Napolitano\inst{\ref{OAC}} \and M., Paolillo\inst{\ref{uniNA},\ref{ASDC}} \and A. Pastorello\inst{\ref{OAPD}} \and  M. Radovich\inst{\ref{OAPD}} \and P. Schipani\inst{\ref{OAC}}
\and  S. Spiro\inst{\ref{OAPD}} \and L. Tomasella\inst{\ref{OAPD}} \and M. Turatto\inst{\ref{OAPD}}
}
\institute{
  INAF,  Osservatorio Astronomico di Padova, vicolo dell'Osservatorio 5, Padova, 35122 Italy
  \label{OAPD}
  \and 
  INAF, Osservatorio Astronomico di Capodimonte, Salita Moiariello 16, Napoli, 80131 Italy
	\label{OAC}
 \and   
   Departemento de Ciencias Fisicas, Universidad Andres Bello, Santiago, Chile \label{Chile}
\and  
   Millennium Institute of Astrophysics, Santiago, Chile\label{millenium}
\and
   Astrophysics Group, Physics Department, University of the Western Cape, Private Bag X17, 7535 Bellville, Cape Town, South Africa  \label{SA}   
   \and
   INAF - Istituto di Radioastronomia, via Gobetti 101, 40129 Bologna, Italy\label{oabo}
\and
INAF, Osservatorio Astronomico di Catania, via S.Sofia 78, 95123 Catania, Italy \label{oact}
 \and  
  Dipartimento di Fisica, Universit\'a Federico II, Napoli, Italy \label{uniNA}
 \and
  Astrophysics, University of Oxford, Denys Wilkinson Building, Keble Road, Oxford OX1 3RH, UK
   \label{oxford} 
\and
  Department of Physical Sciences, The Open University, Milton Keynes, MK7 6AA, UK \label{milton}
  \and
  ASI Science Data Center, via del Politecnico snc, 00133 Roma, Italy
  \label{ASDC}
  }
	
\date{Received: ????; Revised: ??????; Accepted: ????? }
\titlerunning{Sudare. I.}
\authorrunning{EC}

\abstract{} 
{We describe the observing strategy, data reduction tools and early results of a supernova (SN) search project, named SUDARE, conducted with the ESO VST telescope aimed at measuring the rate of the different types of SNe in the redshift range $0.2<z<0.8$. }
{The search was performed in two of the best-studied extragalactic fields, CDFS and COSMOS, for which a wealth of ancillary data are available in the literature or public archives. We  developed a pipeline for the data reduction and  rapid identification of transients. 
As a result of the frequent monitoring of the two selected fields we obtained light curve and colour information for the transients sources that were used for the selection and classification of SNe by means of a especially developed tool. For the accurate characterisations of the surveyed stellar population we exploit public data and our own observations to measure the galaxy photometric redshifts and rest frame colours. }
{We obtained a final sample of 117 SNe, most of which are SN~Ia (57\%) and the remaining core collapse events of which 44\% type II, 22\% type IIn and 34\% type Ib/c. In order to link the transients, we built a catalog of $\sim 1.3\times10^5$ galaxies in the redshift range $0<z\le1$ with a limiting magnitude ${\rm K}_{\rm AB}=23.5$ mag.
We measured the SN rate per unit volume for SN~Ia and core collapse SNe in different bin of redshifts. The values are consistent with other measurements from the literature. }
{The dispersion of the rate measurements for SNe-Ia is comparable with the scatter of the theoretical tracks for single (SD) and double degenerate (DD) binary systems models, therefore the data do not allow to disentangle among the two different progenitor scenarios. However, we may notice that among the three tested models, SD and two flavours of DD, either with a steep (DDC) or a wide (DDW) delay time distribution, the SD gives a better fit across the whole redshift range whereas the DDC better matches the steep rise up to redshift $\sim 1.2$. The DDW appears instead less favoured.
The core collapse SN rate is fully consistent, unlike recent claims,  with the prediction based on recent estimates of the star formation history, and standard progenitor mass range.
}


\maketitle

\section{Introduction}\label{intro}

The evolution of the SN rate with redshift provides the observational link between the cosmic star formation history (SFH), the initial mass function and the stellar evolutionary scenarios leading to the explosions. Until recent times, the available measurements were limited to the local Universe and to sparse, sometimes conflicting, high redshift measurements\citep[e.g.][ and reference therein]{dahlen:2012vn,maoz:2014kt}.
The new generation of panoramic detectors, now available in many observatories, has substantially improved the survey capabilities and as a consequence the number of SN searches and rate measurements. 
Most of the efforts were devoted to type Ia SNe whose progenitor scenario is still strongly debated but the interest for core collapse SNe (CC~SN) is also growing.  

The notion that measurement of the evolution of the SN~Ia rate with redshifts, in combination with measurement of the SFH, can be used to constrain the SN~Ia progenitor scenarios was  first illustrated by \citet{madau:1998kx} \citep[see also][]{Sadat:1998yq,Ruiz-Lapuente:1998fj}. 
Early measurements were puzzling, showing a very rapid raise of the SN~Ia rate up to redshift $\sim 1$ and then a decline at higher redshift \citep{dahlen:2004on,barris:2006vp, Dahlen:2008yq}. This implied a long delay time from star formation to explosion for the SN~Ia progenitors, that appears conflicting with the indications derived from the rate measurement in the galaxies of the local Universe \citep{mannucci:2005mb}.
Subsequent measurements did not confirm the early results, but the issue is still debated
\citep[e.g.][and reference therein]{kuznetsova:2008zn,rodney:2010xy,Smith:2012fk,perrett:2012uq,graur:2014fk}.

For what concerns CC SNe, the early measurements show that, as expected,  the rates tracks the cosmic star history \citep{botticella:2008fr}, though the scaling factor seems to be a factor 2 smaller than what was expected from the SFH. 
A possible explanation is that many dim CC SNe,  are missed by SN searches \citep{horiuchi:2011xv} and/or part of the missing SNe may be hidden in the dusty nuclear regions of star-burst galaxies  \citep[][and reference therein]{dahlen:2012vn}.  However, it is fair to say that the significance of the claimed discrepancy is still poor.

In addition, we note that in most cases, the cosmic SN rates were derived from  surveys  designed to identify un-reddened SN~Ia for the cosmological distance ladder. In these cases the specific observing strategy and/or candidate selection criteria may introduce biases in the event statistics that can be difficult to properly account.
As a consequence, the SN rates derived from these surveys  may be inaccurate.

In order to use the rate evolution to constrain the SN progenitor scenarios, the knowledge of the properties of the parent stellar population is of fundamental importance, so that the volume of Universe searched for SNe needs to be characterized in terms of the galaxy distribution as a function of redshift, mass and star formation history. 

Based on these considerations, we conceived a new SN search \citep[Supernova Diversity and Rate Evolution, SUDARE,][]{botticella:2013uq},  with primary goal to measure SN rates at medium redshift, that is $0.2<z<0.8$. To combine the requirements of good statistics ($>200$ events) and the availability of ancillary data for the surveyed fields, we planned for a four year project to monitor of two very well known extragalactic fields, the Chandra Deep Field South (CDFS) and the Cosmic Evolution Survey (COSMOS) fields. Thanks to the long term commitment of many different observing programs, extended multi-band photometry is available for these fields. These data allow an accurate characterization of the galaxy sample, which is crucial to infer general properties of the SN progenitors.

The present paper describes the SUDARE survey strategy, the procedures for the identification of transients, and the SN candidate selection and classification after the first two years of observations. Then, we discuss the definition of the galaxy sample and the procedure to derive photometric redshifts. Finally, we estimate the SN rates per unit volume at different redshifts and compare them to published estimates.  A detailed study of SN rates as a function of different galaxy parameters will be presented in a companion paper (Botticella et al., in preparation, hereafter PII). 

Along the paper, we will adopt  $H_{\circ} = 70  \: {\rm km \: s^{-1} \: Mpc^{-1}}$, $\Omega_{\rm M} = 0.3$ and $\Omega_{\rm \Lambda} = 0.7$. Magnitudes are in the AB system.

\section{The survey}

The SUDARE SN survey is performed using the VLT Survey Telescope \citep[VST,][]{capaccioli:2011fj} equipped with the OmegaCAM  camera \citep{kuijken:2011qy}, that started regular operation in October 2011 at ESO Paranal (Chile).  
The VST has a primary mirror of $2.6\,\mbox{m}$ and a f/5.5 modified Ritchey-Chretien optical layout designed to deliver a large, uniform focal plane. The camera is equipped with a  mosaic of $8\times4$ CCDs, each with 4k$\times$2k pixels, covering 1 square degree with a pixel scale of $0.214\,\arcsec{\mathrm{pix}}^{-1}$, allowing for an optimal sampling of the PSF even in good seeing conditions.  The thinned CD44-82 devices from E2V have the advantage of an excellent quantum efficiency in the blue bands but with the drawback that the i and z-bands suffer for significant fringing  contamination.

Most of the observing time at this facility is committed to ESO public surveys\footnote{\url{http://www.eso.org/public/teles-instr/surveytelescopes/vst/surveys/}} but a  fraction of the time is dedicated to Guaranteed Time Observations (GTO) made available to  the telescope and instrument teams in reward of their investments in the construction and installation of the instruments.

SUDARE is a  four-year program  and 
this paper is devoted to the analysis of first two observing seasons for VOICE-CDFS and one season for COSMOS. We are now completing the monitoring of both fields for the subsequent two seasons.

The time allocated to our project for monitoring  VOICE-CDFS was from the VST and OMEGAcam GTO. The observing strategy is to span $4\,\mbox{deg}^2$ in four pointings, with one pointing for each observing season (August to January). Here we present data for two of these pointings.

To extend the photometric coverage, we implemented a synergy with the  VOICE (VST Optical Imaging of the CDFS and ES1 Fields) project (Covone et al. in preparation). VOICE is a GTO program that has the aim to secure deep optical counterpart to existing multi-band photometry of selected fields. The multi-band catalog will be used to study the mass assembly and star formation history in galaxies by combining accurate photometric redshifts, stellar masses and weak lensing maps.

The monitoring of the COSMOS field relies instead on a proposal submitted for ESO VST open time (P.I. Pignata). For this field we maintained the same pointing coordinates from one season to the next. This allows us the detection of transients with very long time evolution, but, due to the limited area probed, the data may be prone to cosmic variance.  
 
The survey strategy consists of monitoring the selected fields every three days in the r-band, excluding only $\pm 5$ days around full moon. 
The exposure time is 30 min with the aim of reaching a magnitude limit of 25$^{\rm th}$ in average sky conditions.  Each observation is split into five 6 min exposures with a dithering pattern designed to fill the gaps between the detector chips (that range from 25 to 85 arcsec).
Because of the dithering, the effective area covered by combining the exposures of a given field is 1.15 deg$^2$, although with a reduced signal to noise ratio at the edge. 

With a more relaxed cadence (3-4 times per month), we also planned for g and i-band exposures. With these observations we can measure the colour of the transients that is essential for their photometric classification and to obtain an estimate of the extinction along the line of sight. Adapting to the rules for the observing blocks in service mode at ESO, the planned observing sequence is g-r, r, r-i with a three-day interval between each block. To ensure a good image quality we required a maximum seeing of $1.2\,\arcsec$ (FWHM) at the beginning of the exposure. This, along with the obvious requirement of clear sky, implied that the actual epochs of observations often deviate from the ideal plan mainly because of unsuitable sky conditions.  

Table~\ref{fieldobslog} lists for each field, the pointing coordinates along with the field size, the observing season, the number of available epochs in the different bands and the range and median value of the seeing measured for the r-band exposures. The full log of observations is given in Tab.~\ref{obslog} where, for each epoch, we list the seeing (in arcsec) and $m_{50}$, the magnitude corresponding with a transient detection efficiency of 50\% (cf. Sect. \ref{artstar}).

\begin{table*}
\caption{Field coordinates and compact log of observations}\label{fieldobslog}
\centering
\begin{tabular}{lcccccccc}
\hline        
field        &   \multicolumn{2}{c}{R.A.    (2000.0)       Dec. }   & field size        & observing &  \multicolumn{3}{c}{epochs}   & seeing \\
              &    hh:mm:ss & dd:mm:ss & $\rm deg^{2}$ &season    & r & g & i & range [median]\\                    
\hline
VOICE-CDFS1  &   03:33:34.506   &   -27:34:10.78  & 1.15 &Aug2012-Jan2013  & 29 & 7 & 11 & 0.51-1.44 [0.89]\\
VOICE-CDFS2  &   03:29:02.654   &   -27:34:00.70  & 1.15 &Oct2011-Jan2012  & 23 & 6 & 4  & 0.51-1.46 [0.82]\\
COSMOS       &   10:00:28.600   &   +02:12:21.00   & 1.15 & Dec2011-Apr2012 & 28 & 7 & 7  & 0.50-1.20 [0.84]\\
\hline
\end{tabular}
\end{table*}

\begin{table*} 
\caption{Log of observations. For each field we report the epoch of observation (civil date and MJD), the seeing (FWHM in arcsec and $m_{50}$ the magnitude corresponding with a transient detection efficiency of 50\%.)  }\label{obslog}
\begin{tabular}{rccc|rccc|rccc} 
\hline
\multicolumn{4}{c|}{VOICE-CDFS1}  & \multicolumn{4}{c|}{VOICE-CDFS2}  & \multicolumn{4}{c}{COSMOS} \\
\hline
Epoch & MJD&seeing &$m_{50}$ &Epoch & MJD&seeing &$m_{50}$ &Epoch & MJD&seeing &$m_{50}$\\
\hline
\multicolumn{4}{c}{r}  & \multicolumn{4}{c}{r}  & \multicolumn{4}{c}{r}  \\
2012-08-05 & 56144.38 &  1.31 &  23.0 & 2011-10-20 & 55854.15 &  1.17 &  23.2 & 2011-12-18 & 55913.30 &  0.65 &  23.2 \\
2012-08-13 & 56152.37 &  0.69 &  23.9 & 2011-10-25 & 55859.34 &  0.56 &  23.8 & 2011-12-22 & 55917.27 &  0.92 &  23.5 \\
2012-09-02 & 56172.23 &  1.02 &  21.7 & 2011-10-28 & 55862.16 &  0.92 &  23.5 & 2011-12-27 & 55922.24 &  1.03 &  23.3 \\
2012-09-05 & 56175.20 &  1.28 &  21.8 & 2011-10-30 & 55864.16 &  1.06 &  23.4 & 2011-12-31 & 55926.28 &  1.14 &  23.3 \\
2012-09-08 & 56178.31 &  1.00 &  23.0 & 2011-11-02 & 55867.10 &  0.78 &  23.5 & 2012-01-02 & 55928.32 &  0.63 &  23.7 \\
2012-09-14 & 56184.25 &  0.55 &  23.9 & 2011-11-04 & 55869.15 &  0.62 &  23.4 & 2012-01-06 & 55932.30 &  0.58 &  23.7 \\
2012-09-17 & 56187.29 &  1.06 &  23.2 & 2011-11-15 & 55880.06 &  0.61 &  23.6 & 2012-01-18 & 55944.20 &  0.63 &  23.7 \\
2012-09-20 & 56190.22 &  0.87 &  23.1 & 2011-11-18 & 55883.29 &  0.90 &  23.4 & 2012-01-20 & 55946.25 &  0.87 &  23.4 \\
2012-09-22 & 56192.22 &  0.89 &  23.1 & 2011-11-21 & 55886.23 &  0.68 &  23.5 & 2012-01-22 & 55948.25 &  0.77 &  23.6 \\
2012-09-24 & 56194.24 &  1.44 &  22.6 & 2011-11-23 & 55888.28 &  0.90 &  23.4 & 2012-01-24 & 55950.28 &  0.68 &  23.8 \\
2012-10-07 & 56207.34 &  0.93 &  23.1 & 2011-11-26 & 55891.28 &  0.64 &  23.8 & 2012-01-27 & 55953.19 &  0.92 &  23.5 \\
2012-10-08 & 56208.24 &  0.93 &  22.9 & 2011-11-28 & 55893.30 &  1.04 &  23.2 & 2012-01-29 & 55955.20 &  0.86 &  23.3 \\
2012-10-11 & 56211.32 &  0.92 &  23.1 & 2011-12-01 & 55896.22 &  0.82 &  23.6 & 2012-02-02 & 55959.31 &  0.88 &  23.3 \\
2012-10-14 & 56214.71 &  1.07 &  23.7 & 2011-12-03 & 55898.24 &  0.52 &  23.8 & 2012-02-16 & 55973.25 &  0.50 &  23.6 \\
2012-10-17 & 56217.21 &  0.92 &  23.7 & 2011-12-14 & 55909.21 &  0.88 &  23.3 & 2012-02-19 & 55976.11 &  0.97 &  23.3 \\
2012-10-21 & 56221.19 &  0.51 &  23.8 & 2011-12-17 & 55912.25 &  0.88 &  23.2 & 2012-02-21 & 55978.15 &  0.77 &  23.6 \\
2012-10-25 & 56225.11 &  0.86 &  22.9 & 2012-01-14 & 55940.15 &  0.77 &  23.3 & 2012-02-23 & 55980.18 &  0.74 &  23.6 \\
2012-11-04 & 56235.14 &  0.67 &  23.6 & 2012-01-18 & 55944.06 &  0.57 &  23.9 & 2012-02-26 & 55983.12 &  0.84 &  23.7 \\
2012-11-06 & 56237.26 &  0.83 &  23.5 & 2012-01-20 & 55946.134&  1.00 &  23.4 & 2012-02-29 & 55986.040 &  0.89 &  23.2 \\
2012-11-08 & 56239.27 &  0.88 &  23.6 & 2012-01-23 & 55949.11 &  0.59 &  23.5 & 2012-03-03 & 55989.196 &  0.93 &  23.2 \\
2012-11-10 & 56241.30 &  0.76 &  23.6 & 2012-01-25 & 55951.12 &  0.90 &  23.7 & 2012-03-06 & 55992.115 &  0.80 &  22.6 \\
2012-11-20 & 56251.14 &  0.78 &  24.0 & 2012-01-29 & 55955.06 &  0.67 &  23.6 & 2012-03-13 & 55999.033 &  0.68 &  23.5 \\
2012-12-03 & 56264.21 &  0.71 &  23.4 & 2012-02-02 & 55959.08 &  1.46 &  22.4 & 2012-03-15 & 56001.045 &  1.11 &  23.1 \\
2012-12-07 & 56268.05 &  0.81 &  23.8 & \multicolumn{4}{c}{g}  & 2012-03-17 & 56003.05 &  0.92 &  23.3 \\
2012-12-13 & 56274.06 &  0.55 &  23.8 & 2011-11-02 & 55867.12 &  0.91 &  23.4 & 2012-05-08 & 56056.000&  0.71 &  23.4 \\
2012-12-20 & 56281.13 &  0.96 &  23.3 & 2011-11-21 & 55886.25 &  0.89 &  23.5 & 2012-05-11 & 56058.10 &  0.85 &  23.3 \\
2013-01-03 & 56295.10 &  0.68 &  23.8 & 2011-12-01 & 55896.25 &  1.05 &  23.7 & 2012-05-17 & 56064.03 &  0.75 &  23.5 \\
2013-01-06 & 56298.13 &  0.91 &  23.7 & 2012-01-18 & 55944.09 &  0.56 &  23.6 & 2012-05-24 & 56071.07 &  1.20 &  22.8 \\
2013-01-10 & 56302.12 &  0.89 &  23.0 & 2012-01-25 & 55951.15 &  1.56 &  23.2 & \multicolumn{4}{c}{g}  \\
\multicolumn{4}{c}{g}  & 2012-12-08 & 56269.26 &  1.02 &  23.4 & 2011-12-27 & 55922.26 &  1.13 &  23.5 \\
2012-09-20 & 56190.24 &  1.17 &  23.6 & \multicolumn{4}{c}{i}  & 2012-01-22 & 55948.27 &  1.11 &  23.6 \\
2012-10-11 & 56211.35 &  1.13 &  23.4 & 2011-11-02 & 55867.15 &  0.59 &  23.1 & 2012-02-02 & 55959.33 &  0.88 &  23.8 \\
2012-10-21 & 56221.22 &  0.57 &  23.7 & 2011-11-21 & 55886.28 &  1.12 &  22.7 & 2012-02-16 & 55973.28 &  0.61 &  23.9 \\
2012-11-06 & 56237.28 &  0.91 &  23.6 & 2011-12-01 & 55896.27 &  0.92 &  22.8 & 2012-02-26 & 55983.14 &  1.04 &  23.8 \\
2012-12-07 & 56268.07 &  0.82 &  24.1 & 2012-01-18 & 55944.11 &  0.69 &  22.9 & 2012-03-17 & 56003.07 &  1.04 &  23.6 \\
2012-12-09 & 56270.28 &  1.08 &  23.5 &  & & &   & 2012-05-09 & 56056.02 &  0.78 &  23.8 \\
2013-01-06 & 56298.16 &  1.14 &  23.7 &  & & &   & \multicolumn{4}{c}{i}  \\
\multicolumn{4}{c}{i}  &  & & &   & 2011-12-27 & 55922.28 &  0.93 &  24.1 \\
2012-08-13 & 56152.39 &  0.54 &  23.0 &  & & &   & 2012-01-22 & 55948.30 &  0.95 &  23.2 \\
2012-09-02 & 56172.26 &  0.98 &  21.1 &  & & &   & 2012-02-02 & 55959.36 &  1.05 &  22.7 \\
2012-09-08 & 56178.34 &  1.45 &  22.0 &  & & &   & 2012-02-16 & 55973.30 &  0.60 &  23.3 \\
2012-09-17 & 56187.31 &  0.91 &  22.7 &  & & &   & 2012-02-26 & 55983.17 &  0.88 &  24.1 \\
2012-09-24 & 56194.27 &  1.42 &  22.2 &  & & &   & 2012-03-17 & 56003.10 &  0.53 &  24.2 \\
2012-10-15 & 56215.19 &  0.89 &  22.9 &  & & &   & 2012-05-09 & 56056.05 &  0.62 &  23.4 \\
2012-10-25 & 56225.14 &  0.85 &  22.2 &  & & &   &  & & &   \\
2012-11-08 & 56239.29 &  0.84 &  22.8 &  & & &   &  & & &   \\
2012-11-20 & 56251.16 &  0.69 &  23.2 &  & & &   &  & & &   \\
2012-12-20 & 56281.15 &  0.79 &  22.7 &  & & &   &  & & &   \\
2013-01-10 & 56302.15 &  0.97 &  22.0 &  & & &   &  & & &   \\
\hline
\end{tabular} 
\end{table*} 

The SN search was complemented by 3 runs of one night each at the ESO-VLT for the spectroscopic classification  of a dozen candidates. These observations, described in Sect.~\ref{spectra}, were intended as spot checks of the SN photometric classification  tool (Sect.~\ref{phclass}).

We remark that, as  by-product,  the SUDARE data archive was also used to explore the performances and completeness of AGN detection via variability \citep{De-Cicco:2015rm,falocco:2015fk}.

\subsection{Image calibration \label{datared}}

The raw data were retrieved from the ESO archive and transferred to the VST data reduction node in OAC-Naples. Here the first part of the data reduction is performed using the VSTTube pipeline \citep{grado:2012qy}. A description of the VSTtube data reduction process is reported in \cite{De-Cicco:2015rm}.  In short, the pipeline first performs flat fielding, gain harmonisation and illumination correction and  all images for a given field are registered to the same spatial grid and  photometric scale. 
Finally the dithered images for one epoch are median averaged producing one stacked image.
The pipeline also delivers  weight pixel masks tracking for each pixel the number of dithered exposures contributing to the combined image after accounting for CCD gaps, bad pixels and cosmic rays rejection.

The pipeline was also used to produce deep stacked images by combining all the exposures in a given filter   with the best image quality, those with $\mbox{seeing}\le0.8 \arcsec$. These stacked images, reaching a limiting magnitude $\sim 1 \mbox{mag}$ fainter than good single epoch exposures (the 3-$\sigma$ magnitude limits are 26.2, 25.6, 24.9 mag for r,g and i-band respectively) were used to extract galaxy photometry to complement the public multi-band  catalogs.
(Sect.~\ref{galaxysample}).

\subsection{Transient detection}

For the detection of transient sources and the selection of SN candidates, the mosaic images were processed with an ad-hoc pipeline. This is a collection of {\em python} scripts that makes use of {\em pyraf} and {\em pyfits}\footnote{\url{http://www.stsci.edu/institute/software_hardware/}} tasks and incorporate other publicly available software for specific tasks, in particular {\em SExtractor} \citep{bertin:1996fj} for source detection and characterisation, {\em hotpants}\footnote{A package provided by A. Becker (\url{http://www.astro.washington.edu/users/becker/v2.0/hotpants.html})\label{hotpants}} for PSF match and image difference, {\em daophot} \citep{stetson:1987yq} for accurate point spread function (PSF) fit photometry, {\em stilts}\footnote{\url{http://www.star.bris.ac.uk/~mbt/stilts/}} for catalog handling and {\em mysql} for the transient database. The flowchart of the SUDARE pipeline is the following:

\begin{enumerate}

\item We produced a mask for saturated stars that is combined with the weight map produced by {\em VSTTube}, to build a bad pixel mask for each mosaic image. Those flagged as bad pixels were excluded from further analysis.

\item For each image we computed the difference from a selected template image. 
This required the derivation of the convolution kernel matching the PSF of the two images.
The method is described in \cite{Alard:2000fk} though we used the {\em hotpants} implementation.
We note that for the first observing season we do not have earlier templates and we therefore used as templates, images acquired on purpose a few months after completion of the transient survey campaign.

\item The transient candidates were identified, using {\em SExtractor}, as positive sources in the difference image. Depending on the image quality and detection threshold, the candidate list starts with several thousands objects for each epoch. Most detections are artefacts due to poorly  masked CCD defects, poorly removed cosmic rays, residual from the subtraction of bright sources, reflection ghosts from bright sources, etc. 

\item The transient candidates were ranked based on a custom algorithm that uses a number of measured {\em sextractor} metrics of the detected sources.  The most informative parameters and the ranking scores were selected and calibrated through extensive artificial star experiments. In these experiments, a number of fake stars were placed in the search image that is then processed through the detection and ranking pipeline. The success rate of artificial star recovery was compared to the number of residual spurious sources. We found that the most informative parameters are the source $FWHM$, $flux\_ratio$, $isoarea$ and magnitudes, measured at different apertures. With a proper selection of these parameters, we can drastically reduce the number of spurious events while limiting the number of good candidates improperly rejected. The performance of the ranking algorithm depends on the image quality. On average, we found that we can eliminate $\sim95\%$ of the spurious transients at the cost of losing $\sim5\%$ of  good candidates. Correction for the lost SNe is incorporated in the detection efficiency  (c.f. Sec.~\ref{artstar}) since we used the same algorithm to select real and fake SNe. After this selection, typically a hundred candidates per field and epoch are left to the next step of human inspection and validation.

\item
To associate each transient to its possible host galaxy, we cross correlated the transient list with the galaxy catalogs derived from the deep r-band stacked image (cf. Sec.~\ref{galaxysample}). A galaxy is adopted as the host for a given transient when the latter appears engulfed in the galaxy boundaries. The boundaries are those of the ellipse defined by the
{\em SExtractor} parameters CXX, CYY and CXY through the equation:

$$ {\rm CXX}(x - x_c)^2 +{\rm CYY}(y - y_c)^2 +{\rm CXY}(x - x_c)(y - y_c) = R^2$$

where $x_c$, $y_c$ is the galaxy center and following the {\em SExtractor}'s manual, we assume that the isophotal limit corresponds to $R=3$.

In four cases the transient/host galaxy pairing  was ambiguous because of the overlap of the ellipses of different galaxies. In these cases we also considered the consistency of the host galaxy redshift with the indication of SN photometric classification (c.f. Sect.~\ref{phclass}).

Only for two transients  no counterparts was detected in the deep stacked image (cf. Tab.~\ref{snlist}). 

\item  The available information for the best ranked candidates are posted on WEB pages where the user can inspect the images and candidate metrics for the search and template epochs and selects the good candidates assigning each of them a preliminary classification according to different classes (SN, AGN, variable star, moving object). The selected candidates are then archived in a {\em mysql} database.

\item For all selected candidates, we derive accurate light curves measuring the source magnitude at all available epochs. We measure both aperture and PSF-fit photometry on the original search images and on the difference images. We verified that the PSF-fit gives more reliable measurements that plain  aperture integration mainly because PSF photometry is less sensitive to the background noise.
\end{enumerate}

The transient search process was performed in the r-band for each epoch. Because of the dense temporal sampling, in general good SN candidates will have multiple detections in the database. In principle, we can easily implement a candidate selection based on multiple occurrences of a given source in the database that would further reduce spurious candidates. However, at the present stage of the project we adopted a conservative approach accepting the burden of the visual examination of many candidates  to maximise the completeness fraction.   

\section{Detection efficiency \label{artstar}}

To derive the SN rates we need to obtain an accurate estimate of the completeness of our search. This is done by extensive artificial stars experiments exploring a range of  magnitudes and positions on the images. 

For every search image, we first obtain the PSF from the analysis of isolated field stars. Then,  a number of fake stars of a given magnitude, generated scaling  the  PSF  and adding the  proper Poisson  noise, were injected on the search image.
To mimic the range of properties of real sources, three different criteria were adopted for positioning the fake stars, with roughly the same number of star for each class. The three classes are:

\begin{itemize}
\item events associated with galaxies. From the source catalog on the field (cf. Sect.~\ref{galaxysample}) we picked up a random sample of galaxies and one fake star was placed in each of them. The position inside the galaxy was chosen randomly following the distribution of $r$-band flux intensity.

\item events coincident with persistent, {\bf point-like sources}. Fake stars were added at the same position of existing sources in the field. This mimics SNe in the nucleus of {\bf compact} host galaxies, variable AGNs and variable stars.

\item events with no counterpart in the template image. These were placed at random positions across the field of view irrespective of existing sources or to the possible coincidence with CCD defects or gaps.
\end{itemize} 

The images with fake stars were processed with the search pipeline and the number of detected events surviving after the ranking procedure were counted. The fraction of detected over injected events gives the detection efficiency for the given magnitude. The experiment is repeated sampling the magnitude range of interest ($18<r<26$) to derive the detection efficiency as a function of magnitude.

We inject 500 fake stars per experiment per image and repeat the experiment 5 times for a given magnitude. In fact, adding a large number of fake stars could bias the computation of the convolution kernel, hence the image subtraction process, making the experiments less reliable.

\begin{figure}
\includegraphics[width=\hsize]{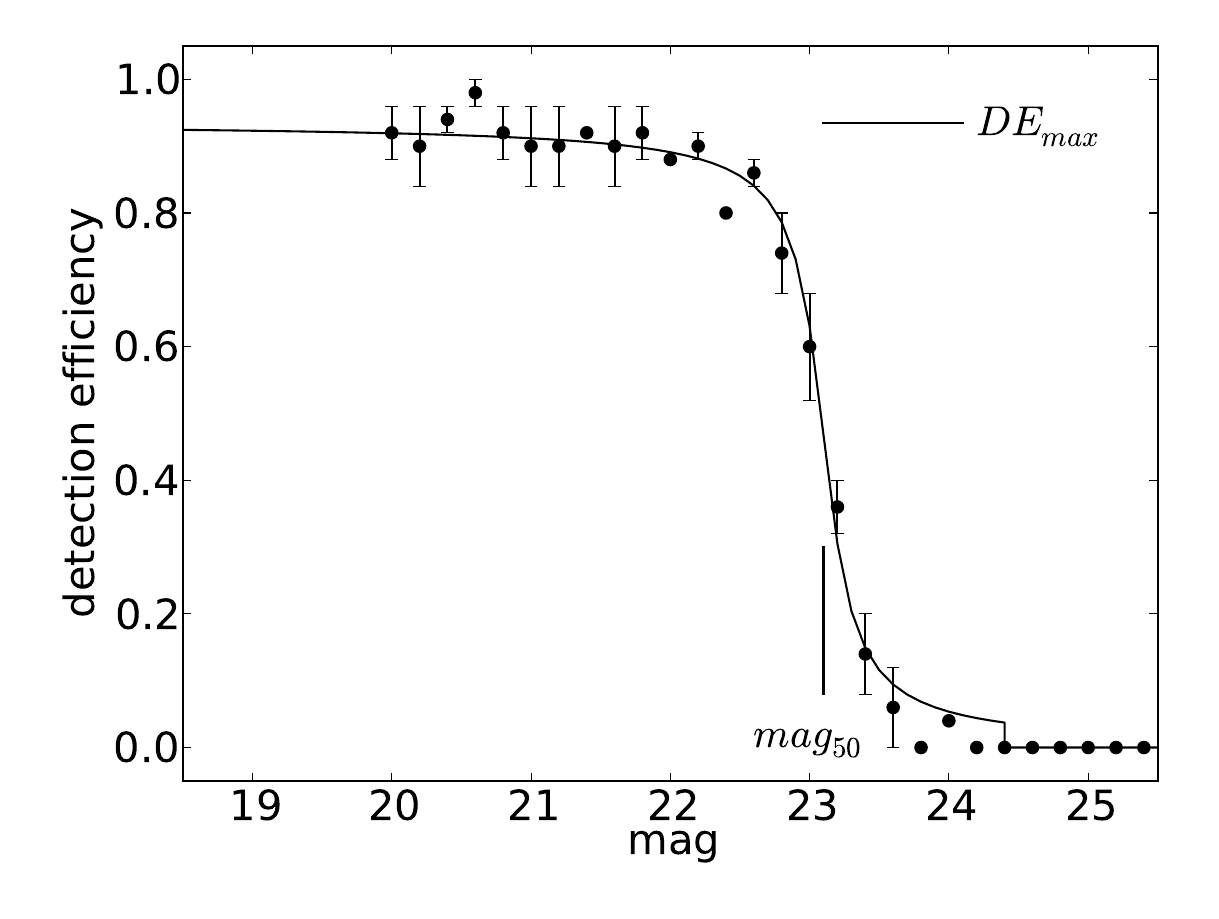}
\caption{Transient detection efficiency as a function of magnitude for the r-band observation of COSMOS on 2012/03/15. The dots are the averages of three artificial star experiments with the errorbar being the dispersion whereas the line is the adopted efficiency curve after the fit with the Eq.~\ref{eff} ($DE_{max}=93\%$, $mag_{50}=23.0$ mag and $\beta=6.1$).}\label{efficiency}
\end{figure}

An example of the derived detection efficiencies as a function of magnitude for one epoch and field is shown in Fig.~\ref{efficiency},  where the errorbars show the dispersions out of three experiments. As can be seen in Fig.~\ref{efficiency},  the detection efficiency as a function of magnitude can be represented by the following analytical function:

\begin{equation}\label{eff}
DE = DE_{max}\, \left( \frac{\arctan{\beta\,(mag_{50}-mag)}}{\pi} + \frac{1}{2} \right)
\end{equation}

where $DE_{max}$ is the maximum value of the detection efficiency, $mag_{50}$ is the magnitude corresponding with the 50\% drop of the detection efficiency    and $\beta$ measures the decline rate of the $DE$. The best fit parameters are determined through least square minimisation of the residuals (e.g. Fig.~\ref{efficiency}). 
Note that the maximum detection efficiency is $\sim95\%$, even at the bright magnitude end. This is because not all the pixels of the image are useful and indeed the bad pixel mask (cf. step 1 of the {\em SUDARE} pipeline) flags $5-10\%$ of the image area. 

We verified that the detection efficiencies measured independently for each of the three classes of fake stars described above are similar with a dispersion of $mag_{50}$ values of $\sim 0.3$ mag and hereafter we will use the average values. Also, we found that the values of  $DE_{max}$  and $\beta$ are very similar for each epoch and field with a mean value of  $ 95\% \pm 3$ and $ 4\pm 2$, respectively.

The artificial star experiment described above was repeated for all epochs and fields of the search, and the resulting detection efficiencies are used in the rate calculation.

\section{Transient classification \label{phclass}}

The result of the transient search was a list of $\sim 350$ SN candidates. A fraction of these 
transients are coincident with persistent sources and therefore can be SNe in the nucleus of the host galaxies but also variable stars or AGNs.  In fact, after the analysis of \cite{falocco:2015fk} and \cite{De-Cicco:2015rm}, three  candidates  with slow evolving light curves  were found to coincides with X-ray sources. These were classified as AGN and removed from the SN candidate sample. We also removed from the SN candidate list all transients with spectroscopic or photometric redshift $z>1$ ($\sim25\%$).

We used the measured light curve and colour evolution to constrain the nature of the transients and to classify SNe in their different types. Conservatively, only candidates with at least 5 photometric measurements at different epochs (even in different filters) were considered. Because of the frequent monitoring of our survey, this criterium excludes only four candidates with a negligible effect on the SN counts.

The photometric classification is more reliable if the redshift of the host galaxy is available. 
When the transient is not associated with a host galaxy or  when the host redshift is not available, the redshift is left as a free parameter in the transient light curve fitting (see next section).

\subsection{Photometric classification of SNe}\label{classtool}

For the photometric classification we used a tool developed for the SUDARE project. The tool compares the SN candidate multi-color light curves with those of SN templates, and identifies the best-matching template, redshift, extinction and luminosity class. The tools was developed following the strategy  of the SN classification tool {\em PSNID} \citep{Sako:2011gf}. We developed our own tool because we want to explore different priors for the fitting parameters and different classification scheme. 

We collected a sample of templates for different SN types for which both multicolour light curves and sequence of spectra are available (Tab.~\ref{templateslist}). The spectra are needed to estimate the $K$-correction. The templates were retrieved from a database of SN light curves and spectra that we have collected in the study of SNe at ESO and ASIAGO Observatory \citep[the template spectra can be downloaded also  from {\em WISEREP}\footnote{\url{http://wiserep.weizmann.ac.il}}, the SN spectra database;][]{yaron:2012qf}.

The templates were selected to represent the well established SN types, namely Ia, Ib, Ic, IIb, II Plateau and Linear, IIn, with the addition of representative peculiar events (see individual references for details).
In particular, we included SN 2008es as representative of the recently discovered class of very luminous SNe  \citep[SLSN,][]{quimby:2007fk,quimby:2013kx,gal-yam:2012uq} that, although intrinsically very rare, may be detected in high redshift searches because of their large volume sampling.

\begin{figure}
\includegraphics[width=\hsize]{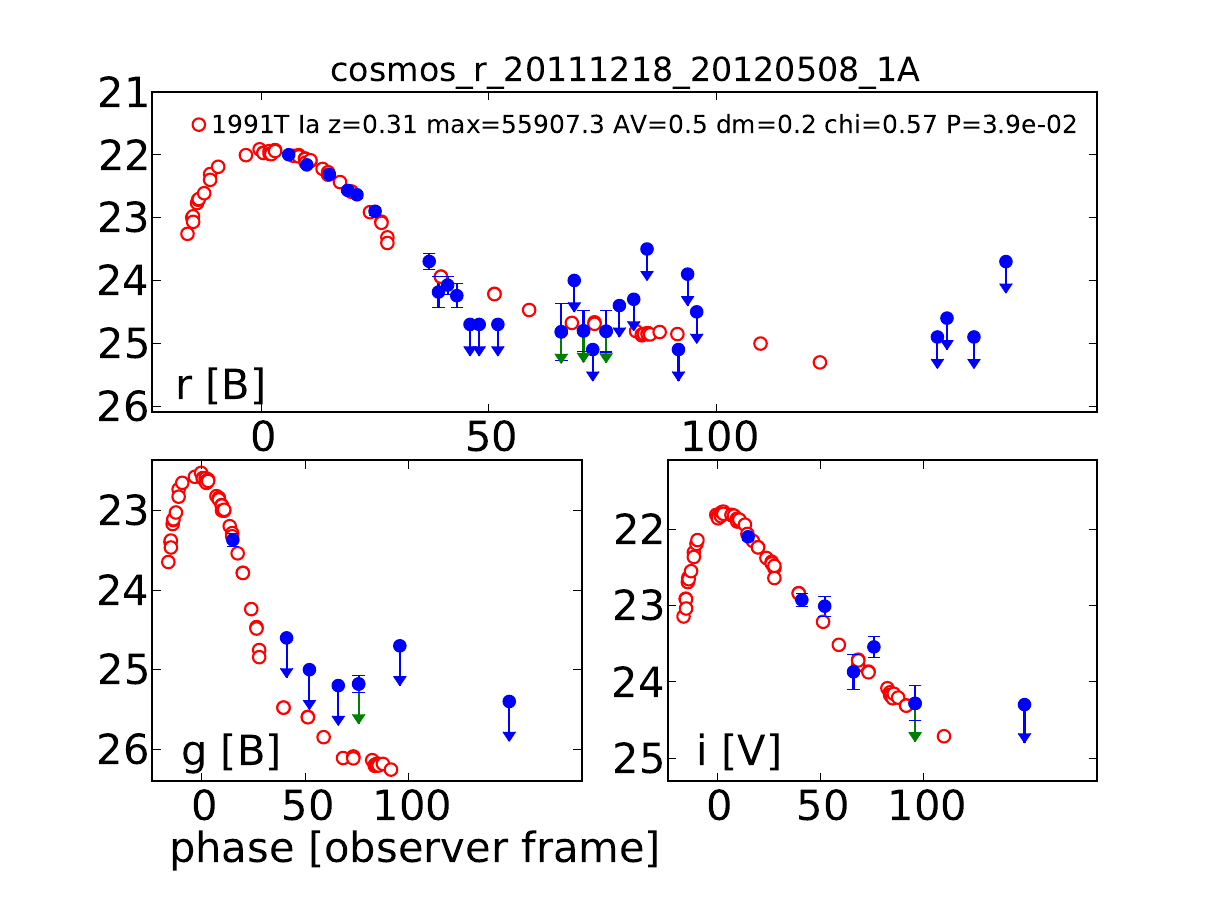}
\caption{Example of the output of the SN typing procedure. The top panel shows  the observed $r$-band light curve that in this case  is compared with the template $K$-corrected $B$-band light curve. The bottom panels show the observed light curve and template fit for the the $g$-band (left) and $i$-band (right). Blue dots are the SN candidate observed magnitudes (arrows indicate upper limits) while red open circles are the template photometry. The legend identifies the best fitting template and parameters. }\label{sntyping}
\end{figure}

The steps of the photometric typing are:
\begin{itemize}

\item for each template, we derived $K$-correction tables as a function of phase from maximum and redshift (in the range $0<z<1$). $K$-corrections are obtained as the difference of the synthetic photometry measured on the rest frame spectra and on the same spectra properly redshifted. The redshift range for which we can derive accurate $K$-correction is limited by the lack of UV coverage from most templates. This is a problem in particular for the g-band where we are forced to accept uncertain extrapolations.

\item  the K-corrected light curves of template SNe were used to predict the observer frame light curves in the $gri$-bands exploring the $0<z<1$ redshifts range  and the $-0.3<E_{B-V}<1$ mag extinction range (the negative lower limit for the $E_{B-V}$ range allows for uncertainties in the template extinction correction and for  variance in the intrinsic SN colour). With the goal to minimise the uncertainties on the $K$-correction, the template input band was taken to best match  the observer frame band for the given redshift, e.g. we use the template  $V, B, U$ bands to predict the observer frame $r$-band light curve of SNe at redshift $z \sim$ 0.1, 0.4, 0.7, respectively.

\item we estimate the goodness of the fit of the template to the observed light curve computing the sum of the square of flux residuals weighted by the photometric errors ($\chi^2$) for each simulated light curve of the grid. Besides the redshift and extinction ranges we explore
a range of epochs of maximum, $T_{\rm max}$ (the initial guess is the epoch of the observed $r$-band brightest point) and of intrinsic luminosities, $\Delta(\mu)$ (allowing for a $\pm 0.3$ mag flux scaling of the template).
The residuals for all bands are summed together and therefore each band contributes to the overall $\chi^2$ with a weight proportional to the number of measurements.

\item for the selection of the best fitting template we use Bayesian model selection \cite[e.g.][]{Poznanski:2007rt,Kuznetsova:2007fr,Rodney:2009mz}. In particular following \citet{Sako:2011gf} we compute for each SN type the Bayesian evidence:

$$E_{\rm type} = \sum_{\rm template} \int_{\rm pars.\,range} P(z) \, e^{-\chi^2/2} 
dz\,dA_V\,dT_{\rm max}\, d\Delta(\mu)$$ 

where the fitting parameters are the redshift $z$, with $P(z)$ its probability distribution, the extinction $A_V$, the time of maximum $T_{\rm max}$ and the flux scaling factor $\Delta(\mu)$. 

The spectroscopic redshift is used as a prior if available and in this case for $P(z)$ we adopt a normal distribution centered at the spectroscopic redshift and with $\sigma=0.005$. Otherwise, if a photometric redshift estimate is available, we use as redshift prior the $P(z)$ provided by the photometric redshift code (cf. Sect. \ref{zconf}). In the worst case, either when the host galaxy is not detected, or when the photometric redshift is poorly constrained, we adopt a flat prior in the range $0<z<1$.

In all cases we adopted flat prior for the extinction distribution and for the flux scaling. 
 
More critical is choice of templates.
As emphasised by \citet{Rodney:2009mz}, the Bayesian approach relies on an appropriate template list that should be as complete as possible but, at the same time, avoid duplicates. When the template list includes rare, peculiar events, especially if they mimic  the properties of a more frequent SN type, it is appropriate to use frequency priors. 
Alternatively, for specific applications one may exclude ambiguous cases or rare, peculiar SN types \cite[c.f.][]{Sako:2011gf} from the template list. 

Our template list, given in Tab.~\ref{templateslist}, is intended to represent the full range of the most frequent SN type with a number of templates for each class that is broadly consistent with their frequency in a volume limited SN sample \citep{li:2011zr}. After that we adopted flat priors for the relative rate of each template within a given class and for the relative rates of the different SN types.

We computed the Bayesian probability for each of the main SN types as:

$$ P_{\rm type} = \frac{E_{\rm type}}{E_{\rm Ia}+E_{\rm Ib/c}+E_{\rm II}+(E_{\rm SLSN})} $$\label{bayeseq}

Notice that for the purpose to assign probability, we merged regular type II and type IIn templates. However, in the subsequent analysis we indicated when the best fitting template (the one with the highest probability) is a type IIn. 
Also, after verifying that none of our candidate has a significant probability of matching a SLSN, the corresponding template was dropped from the fitting list and the $E_{\rm SLSN}$ term in Eq.~\ref{bayeseq} cancelled. This was done to allow a direct comparison with 
{\em SNANA} (see next Section).

\item for the most probable SN type, we record the best fitting template along with the fit parameters corresponding to the $\chi^2$ minimum.
Of the about 250 transients, 117 were classified as SNe. Most of the remaining have erratic light curves consistent with that of AGNs.    

An example of the output of the SN typing procedure is shown in Fig.~\ref{sntyping}.
 
\end{itemize}

Table \ref{snlist} lists our SNe; for each event we report the coordinates, the host galaxy redshift if available (col.~4), for photometric redshift, the 95\% lower and upper limits of the P(z) (col.~5), the method of redshift measurement (col.~6) the most probable SN type (col.~7) and corresponding Bayesian probability (col~8), the best fitting template (col.~9),  redshift (col.~10), extinction (col.~11),  flux scaling (col~12) and epoch of maximum (col.~13). 
We also list the $\chi^2_n$ (col.~14), the number of photometric measurements with, in parenthesis, the number of measurement with $SNR>2$ (col.~15) and the integrated right tail probability of the $\chi^2$ distribution  ($P_{\chi^2}$, col~16).

In a number of cases the $P_{\chi^2}$ probability is fairly low (15 SNe have $P \chi^2 <10^{-4}$). In some cases this is due to one/two deviant measurements, in other cases there is evidence of some variance in the light curve not fully represented by the adopted template selection. We have to consider the possibility that these events are not SNe. 

Also for some candidates with a small $\chi^2$, the number of actual detections (photometric measurements with signal to noise ratio $SNR>2$) is also small that it is not possible to definitely assess the SN nature of the transient (for 12 candidates the number of detection is $N_{\rm det}<=7$). 

To these probable SNe (indicated with PSN in the last column of Tab.~\ref{snlist}) we will attribute a weight 0.5 in the rate calculation. The impact of the arbitrary thresholds for $P\chi^2$ and the $N_{\rm det}$ and the adopted PSN weight will be estimated in Sect.~\ref{Syst}.

\begin{table*}
\caption{List of template SNe used for the SUDARE's SN photometric classification tool}\label{templateslist}
\begin{tabular}{lcccl}
\hline
SN & type & vel  & $m-M$& reference \\
   &      &[${\rm km}\,{\rm s}^{-1}$]\\
\hline
 1990N  & Ia  &    998  &31.73 & \cite{lira:1998ys}; \cite{mazzali:1993vn}\\
 1992A &  Ia  &   1845  &31.14 & \cite{kirshner:1993zr}; \cite{altavilla:2004ly}\\
 1994D &  Ia  &    450  &30.92 & \cite{richmond:1995qf}; \cite{patat:1996ve}\\
 2002bo&  Ia  &   1289  &31.77 & \cite{benetti:2004nx}\\
 1999ee & Ia bright  &   3407  &33.42 & \cite{stritzinger:2002dq};\cite{hamuy:2002cr}\\
 1991T &  Ia bright  &   1732  &30.74& \cite{lira:1998ys}; \cite{altavilla:2004ly}; \cite{ruiz-lapuente:1992oq}\\
 1991bg&  Ia faint &    913  &31.44 & \cite{filippenko:1992kl}; \cite{leibundgut:1993tg}; \cite{turatto:1996hc}\\
 2000cx&  Ia pec & 2421  &32.39 & \cite{candia:2003ij}; \cite{matheson:2008bs}\\
 2002cx&  Ia pec & 7183  &35.09 & \cite{li:2003fv}\\
 1987A  & II   &   320  &18.48 & \cite{catchpole:1989dz}; \cite{hamuy:1990fu}; ESO/Asiago Archive\\          
 1992H  & IIL  &  1021  &30.97 & \cite{clocchiatti:1996kl}\\
 2009bw & IIP  &  1155  &31.45 & \cite{inserra:2012qo}\\
 1999em & IIP  &   710  &29.47 & \cite{hamuy:2001qa}; \cite{leonard:2002mi}; \cite{elmhamdi:2003pi}\\
 2004et & IIP  &    40  &28.85 &\cite{maguire:2010lh}\\
 1999br & II faint & 1021 & 30.97 & \cite{pastorello:2004kx}\\
 1999gi & II faint   &   592  &29.80 &\cite{leonard:2002ff}\\
 2005cs & II faint  &   600  &29.26 & \cite{pastorello:2006fu,pastorello:2009ye}\\
 1993J  & IIb  &   -35  &27.80 & \cite{filippenko:1992kl}; \cite{richmond:1994il}; \cite{barbon:1995zt}\\
 2008ax & IIb  &   579  &29.92 & \cite{pastorello:2008jl}; \cite{taubenberger:2011gb}\\
 1997cy & II pec & 17700 &37.03 & \cite{turatto:2000mb}\\
 1998S  & IIn  &   895  &31.18 & \cite{fassia:2000cq}\\
 2010jl & IIn  &  3207  &34.92 &\cite{pozzo:2004kh}\\
 2005gj & IIn (Ia)  & 17988  &37.15 &\cite{aldering:2006rq}\\
 2008es & SLSN-II & 0.202* &39.70  &\cite{gezari:2009fc} \\ 
 2009jf & Ib   &  2379  &32.65 &\cite{valenti:2011ss}\\
 2008D  & Ib (XRF)   &  1955  &32.29 &\cite{mazzali:2008qc}\\
 1994I  & Ic   &   461  &29.60 &\cite{wheeler:1994mw}; \cite{clocchiatti:1996ys}\\
 1998bw & Ic (GRB)   &  2550  &32.76 & \cite{galama:1998if}; \cite{patat:2001ud}\\
 2004aw & Ic   &  4742  &34.17 &\cite{taubenberger:2006kc}\\
 2007gr & Ic   &   492  & 29.84 &\cite{valenti:2008hs}; \cite{hunter:2009lo}\\
\hline
\end{tabular}

* redshift instead of velocity 
\end{table*}

\setlength{\tabcolsep}{3pt}
\begin{longtab}    
\begin{landscape} 
\begin{longtable}{lccccccccccrccccc}  
\caption{List of supernovae} \label{snlist}  \\
\hline\hline
designation    &     RA   & DEC  & z(host) & err & source & type & P$_{\rm type}$ & template &z(SN) & $A_V$ & $\Delta(\mu)$ & $T_{\rm max}$ [MJD]  & $\chi^2_n$ & npt  & P$\chi^2$ \\
\hline
\endfirsthead
\caption{continued.}\\
\hline\hline
designation    &     RA   & DEC & z(host) & err & source &type & P$_{\rm type}$ & template & z(SN) & $A_V$ & $\Delta(\mu)$ & $T_{\rm max}$ [MJD]   &$\chi^2_n$ & Npt & P$\chi^2$\\
\hline
\endhead
\hline 
\endfoot
cdfs1\_r\_20120805\_20130110\_1A    &  3:34:56.9 & -27:53:33.0 & 0.35  & [.27,.43] & phot & Ia   & 1.00 & 1991T  & 0.39 & -0.3 &  0.0 & 56161 & 1.3 & 35 (25) & 1E-1  \\
cdfs1\_r\_20120805\_20130110\_69A   &  3:35:45.9 & -27:22:56.4 & 0.09  & [.03,.17] & phot & Ia   & 0.64 & 1992A  & 0.14 & -0.3 &  0.3 & 56064 & 1.8 & 24 (17) & 1E-2  \\
cdfs1\_r\_20120805\_20130110\_9A    &  3:31:36.7 & -28:01:28.3 & 0.15  &           & spec & Ib/c & 1.00 & 2009jf & 0.14 &  1.3 & -0.3 & 56153 & 3.4 & 33 (27) & 3E-10 & PSN\\
cdfs1\_r\_20120813\_20130103\_175A  &  3:35:16.4 & -27:29:49.2 & 0.35  &           & spec & Ia   & 0.99 & 2002bo & 0.35 &  0.3 & -0.3 & 56166 & 1.0 & 32 (15) & 5E-1 & $a$\\
cdfs1\_r\_20120813\_20130103\_205A  &  3:35:21.0 & -27:14:34.8 &   -   &           &  -   & Ib/c & 1.00 & 2009jf & 0.19 & -0.3 & -0.3 & 56152 & 1.1 & 33 (17) & 3E-1   \\
cdfs1\_r\_20120914\_20121213\_4A    &  3:34:59.0 & -27:51:55.4 & 0.40  &           & spec & Ia   & 0.78 & 1990N  & 0.39 &  0.6 &  0.2 & 56186 & 0.7 & 35 (11) & 9E-1 & $b$\\
cdfs1\_r\_20121007\_20130110\_26A   &  3:35:28.3 & -27:36:21.6 & 0.28  &           & spec & Ib/c & 0.60 & 1998bw & 0.28 & -0.3 &  0.3 & 56212 & 1.9 & 33 (24) & 1E-3 & $c$\\
cdfs1\_r\_20121007\_20130110\_2A    &  3:32:32.9 & -27:51:21.1 & 0.67  &           & spec & Ia   & 1.00 & 1991T  & 0.66 & -0.3 &  0.3 & 56322 & 0.9 & 34 (17) & 6E-1 & $d$\\
cdfs1\_r\_20121007\_20130110\_7A    &  3:31:38.5 & -27:13:51.9 & 0.24  &           & spec & II   & 1.00 & 2009bw & 0.24 & -0.3 &  0.2 & 56206 & 1.6 & 34 (17) & 1E-2 & $e$\\
cdfs1\_r\_20121011\_20130110\_215A  &  3:32:43.2 & -27:10:33.3 & 0.64  &           & spec & IIn  & 1.00 & 1998S  & 0.65 &  0.0 &  0.3 & 56213 & 1.5 & 34 (13) & 3E-2 & $f$\\
cdfs1\_r\_20121021\_20120813\_13A   &  3:35:40.7 & -27:26:19.2 & 0.45  &           & spec & Ia   & 1.00 & 1990N  & 0.45 &  0.4 & -0.2 & 56233 & 1.4 & 31 (23) & 7E-2 & $g$**\\
cdfs1\_r\_20121021\_20120813\_15A   &  3:34:38.2 & -27:20:20.6 & 0.53  &           & spec & Ia   & 1.00 & 1991T  & 0.52 &  0.3 & -0.2 & 56230 & 1.5 & 31 (18) & 4E-2 & $h$ \\
cdfs1\_r\_20121021\_20120813\_2A    &  3:33:08.7 & -28:05:04.2 & 0.49  &           & spec & Ia   & 1.00 & 2002bo & 0.48 &  0.1 & -0.1 & 57229 & 1.8 & 24 (16) & 9E-3 & $i$ \\
cdfs1\_r\_20121104\_20120813\_137A  &  3:34:16.3 & -27:40:47.8 & -     &           &  -   & II   & 1.00 & 2004et & 0.26 &  0.0 &  0.5 & 56254 & 0.5 & 25 (13) & 1.0      \\
cdfs1\_r\_20121104\_20120813\_45A   &  3:31:37.1 & -27:21:18.2 & 0.57  &           & spec & Ia   & 0.99 & 1990N  & 0.56 &  0.3 & -0.3 & 56243 & 1.3 & 24 (24) & 1E-1 & $j$\\
cdfs1\_r\_20121104\_20120813\_4A    &  3:31:18.1 & -27:32:20.7 & 0.45  & [.20,.65] & phot & Ia   & 0.96 & 1990N  & 0.39 &  0.4 & -0.3 & 56235 & 0.5 & 23 (19) & 1.0   \\
cdfs1\_r\_20121104\_20120813\_58A   &  3:33:26.8 & -27:08:50.8 & 0.15  & [.04,.24] & phot & IIn  & 1.00 & 1998S  & 0.17 &  2.0 &  0.3 & 56248 & 2.4 & 20 (20) & 4E-4  \\
cdfs1\_r\_20121104\_20120813\_6A    &  3:34:37.5 & -27:27:58.7 & -     &           & -    & II   & 0.64 & 1993J  & 0.11 &  1.4 &  0.3 & 56220 & 0.7 & 24 (14) & 9E-1  \\
cdfs1\_r\_20121106\_20120813\_66A   &  3:33:16.8 & -27:55:36.9 & 0.77  &           & spec & Ia   & 1.00 & 1999T  & 0.76 & -0.3 &  0.2 & 56244 & 1.6 & 26 (17) & 3E-2  \\
cdfs1\_r\_20121120\_20120813\_101A  &  3:31:30.5 & -27:51:44.2 & 0.68  &           & spec & Ia   & 1.00 & 1990N  & 0.67 & -0.1 & -0.1 & 56255 & 0.6 & 33 (14) & 1.0 \\
cdfs1\_r\_20121120\_20120813\_2A    &  3:32:08.8 & -27:55:44.8 & 0.37  &           & spec & Ia   & 0.99 & 2002bo & 0.36 &  0.7 & -0.2 & 56255 & 0.6 & 22 (16) & 9E-1  \\
cdfs1\_r\_20121120\_20120813\_38A   &  3:34:17.7 & -27:39:34.6 & 0.15  &           & spec & Ib/c & 0.89 & 1994I  & 0.14 &  0.5 & -0.3 & 56250 & 1.0 & 24 (9)  & 5E-1  \\
cdfs1\_r\_20121120\_20120813\_66A   &  3:33:49.8 & -27:09:16.7 &  -    &           & -    & Ib/c & 0.90 & 2008D  & 0.19 &  0.7 & -0.1 & 56259 & 0.6 & 19 (10) & 9E-1  \\
cdfs1\_r\_20121207\_20120813\_28A   &  3:32:15.8 & -27:58:22.7 & 0.74  &           & spec & Ia   & 1.00 & 1990N  & 0.73 & -0.3 &  0.0 & 56271 & 0.6 & 17 (7)  & 9E-1 & PSN  \\
cdfs1\_r\_20121207\_20120813\_49A   &  3:31:50.8 & -27:35:35.5 & 0.38  & [.24,.51] & phot & Ia   & 0.80 & 1994D  & 0.44 &  0.6 &  0.3 & 56280 & 0.8 & 14 (12) & 7E-1\\
cdfs1\_r\_20121207\_20120813\_61A   &  3:32:17.5 & -27:25:41.3 & 0.51  & [.32,.65] & phot & Ia   & 0.96 & 1990N  & 0.38 &  0.1 &  0.1 & 56276 & 0.8 & 14 (14) & 7E-1  \\
cdfs1\_r\_20121207\_20120813\_64A   &  3:35:15.4 & -27:23:50.6 & -     &           & -    & Ib/c & 0.89 & 2007gr & 0.19 &  0.6 & -0.3 & 56264 & 0.9 & 14 (14) & 6E-1  \\
cdfs1\_r\_20121213\_20120914\_47A   &  3:33:58.2 & -27:28:12.3 & 0.69  &[.47,.86]  & phot & Ia   & 1.00 & 1991T  & 0.79 &  0.1 & -0.3 & 56273 & 1.2 & 14 (8)  & 4E-1  \\
cdfs1\_r\_20130103\_20120813\_78A   &  3:32:02.0 & -27:04:59.0 & -     &           & -    & Ib/c & 0.83 & 1998bw & 0.16 &  0.1 &  0.3 & 56292 & 0.5 & 12 (7)  & 9E-1 & PSN\\
cdfs1\_r\_20130106\_20120813\_150A  &  3:33:36.4 & -27:50:49.5 & -     &           & -    & II   & 1.00 & 2004et & 0.06 &  0.9 & -0.3 & 56267 & 1.1 & 14 (14) & 4E-1 \\
cdfs1\_r\_20121120\_20120813\_87A   &  3:31:59.7 & -28:00:03.1 & 0.53  &           & spec & IIn  & 1.00 & 1998S  & 0.52 &  1.5 &  0.3 & 56253 & 0.6 & 22 (8)  & 9E-1  \\
cdfs1\_r\_20120805\_20130110\_3A    &  3:32:08.6 & -27:46:48.2 & 0.31  &           & spec & Ia   & 0.86 & 1990N  & 0.31 &  0.6 & -0.3 & 56151 & 0.7 & 32 (12) & 9E-1  \\
cdfs1\_r\_20130103\_20120813\_15A   &  3:34:49.6 & -28:00:49.7 & 0.38  & [.25,.49] & phot & Ia   & 0.96 & 1991T  & 0.37 &  0.0 & -0.3 & 56310 & 0.6 & 11 (5)  & 8E-1 &PSN \\
cdfs1\_r\_20130103\_20120813\_1A    &  3:32:15.4 & -28:01:23.2 & 0.53  &           & spec & Ib/c & 1.00 & 1998bw & 0.52 & -0.3 & -0.3 & 56298 & 1.7 & 12 (8)  & 6E-2 &PSN\\
cdfs1\_r\_20130103\_20120813\_59A   &  3:33:58.4 & -27:22:10.3 & 0.61  & [.48,.65] & phot & Ib/c & 0.99 & 1998bw & 0.49 & -0.3 & -0.3 & 56300 & 3.1 &  9 (6)  & 1E-3  \\
cdfs1\_r\_20130103\_20120813\_8A    &  3:34:47.9 & -27:21:42.8 & 0.51  & [.39,.64] & phot & Ia   & 1.00 & 2002bo & 0.39 & -0.3 &  0.3 & 56300 & 1.0 &  9 (8)  & 4E-1 \\
cdfs1\_r\_20120813\_20130103\_162A  &  3:35:24.9 & -27:36:55.6 & 0.43  & [.30,.68] & phot & Ia   & 0.90 & 1990N  & 0.47 &  0.3 &  0.0 & 56130 & 1.0 & 33 (11) & 5E-1  \\
cdfs1\_r\_20121007\_20130110\_43A   &  3:33:08.2 & -27:14:52.6 & 0.65  & [.53,.76] & phot & IIn  & 1.00 & 1998S  & 0.64 &  0.0 &  0.3 & 56196 & 3.9 & 37 (18) & 1E-14 &PSN \\
cdfs1\_r\_20121203\_20120813\_72A   &  3:33:51.3 & -27:35:59.0 & 0.18  &           & spec & IIn  & 1.00 & 1998S  & 0.18 &  2.0 &  0.3 & 56277 & 4.9 & 12 (12) & 4E-8 &PSN \\
cdfs2\_r\_20111020\_20120125\_111A  &  3:28:30.0 & -27:59:55.7 & 0.22  & [.15,.60] & phot & IIn  & 1.00 & 1998S  & 0.29 &  0.3 &  0.3 & 55811 & 1.2 & 23 (18) & 2E-1 \\
cdfs2\_r\_20111020\_20120125\_34A   &  3:27:24.7 & -28:00:10.7 & -     &           & -    & Ia   & 0.91 & 1991T  & 0.39 &  0.6 &  0.3 & 55849 & 0.7 & 23 (10) & 9E-1   \\
cdfs2\_r\_20111020\_20120125\_44A   &  3:29:43.8 & -27:49:07.2 & 0.15  & [.04,.26] & phot & Ia   & 1.00 & 1999ee & 0.11 &  0.1 &  0.1 & 55839 & 2.0 & 27 (21) & 2E-3 \\
cdfs2\_r\_20111020\_20120125\_48A   &  3:26:55.0 & -27:45:36.7 & 0.50  & [.40,.66] & phot & Ia   & 0.99 & 1990N  & 0.47 &  0.0 &  0.2 & 54837 & 3.8 & 30 (17) & 1E-11 &PSN \\
cdfs2\_r\_20111020\_20120125\_50A   &  3:29:56.7 & -27:41:22.7 & 0.62  & [.58,.68] & phot & Ia   & 1.00 & 1991T  & 0.60 & -0.3 &  0.3 & 55847 & 0.8 & 31 (7)  & 8E-1  &PSN\\
cdfs2\_r\_20111020\_20120125\_8A    &  3:29:55.2 & -27:58:04.4 & -     &           & -    & Ia   & 0.61 & 1991T  & 0.39 &  0.5 &  0.3 & 55856 & 0.5 & 23 (12) & 1.0 \\
cdfs2\_r\_20111025\_20120123\_141A  &  3:27:31.4 & -27:52:17.2 & 0.75  & [.16,1.2] & phot & Ia   & 0.57 & 1990N  & 0.47 &  0.6 &  0.3 & 55856 & 0.9 & 23 (7)  & 6E-1  \\
cdfs2\_r\_20111025\_20120123\_25A   &  3:30:29.4 & -27:56:37.6 & 0.60  & [.49,.70] & phot & Ia   & 1.00 & 1991T  & 0.58 &  0.6 &  0.3 & 55871 & 0.7 & 30 (16) & 9E-1 &PSN \\
cdfs2\_r\_20111025\_20120123\_29A   &  3:29:23.0 & -27:52:59.5 & 0.46  & [.39,.54] & phot & II   & 1.00 & 1992H  & 0.39 & -0.3 &  0.3 & 55829 & 1.3 & 23 (17) & 2E-1 \\
cdfs2\_r\_20111025\_20120123\_32A   &  3:27:49.4 & -27:50:15.0 & -     &           & -    & Ia   & 0.99 & 1991T  & 0.47 &  0.0 &  0.3 & 55867 & 2.8 & 30 (18) & 5E-7 &PSN \\
cdfs2\_r\_20111025\_20120123\_90A   &  3:30:54.6 & -27:03:40.7 & 0.29  & [.23,.34] & phot & II   & 1.00 & 2009bw & 0.26 &  0.3 & -0.3 & 55863 & 3.0 & 31 (21) & 4E-8 &PSN \\
cdfs2\_r\_20111028\_20120125\_181A  &  3:30:26.0 & -27:31:22.3 & 0.33  & [.26,.40] & phot & II   & 1.00 & 2009bw & 0.31 & -0.3 &  0.1 & 55867 & 1.1 & 30 (17) & 3E-1  \\
cdfs2\_r\_20111028\_20120125\_225A  &  3:29:07.5 & -27:04:57.4 & 0.10  & [.01,.21] & phot & Ib/c & 0.61 & 2008D  & 0.12 &  1.8 &  0.3 & 55860 & 1.0 & 27 (10) & 5E-1 \\
cdfs2\_r\_20111102\_20120129\_156A  &  3:26:54.1 & -27:29:24.8 & 0.16  & [.04,.26] & phot & Ia   & 0.70 & 1991T  & 0.15 & -0.3 &  0.3 & 55741 & 0.7 & 27 (21) & 9E-1 \\
cdfs2\_r\_20111115\_20120123\_3A    &  3:30:06.4 & -27:19:48.4 & 0.49  & [.33,.63] & phot & Ia   & 0.90 & 1990N  & 0.39 &  0.9 & -0.3 & 55880 & 0.6 & 31 (12) & 1.0 \\
cdfs2\_r\_20111115\_20120123\_42A   &  3:30:27.3 & -27:28:13.0 & 0.41  & [.22,.64] & phot & Ia   & 0.95 & 1990N  & 0.39 &  0.6 &  0.0 & 55890 & 1.0 & 27 (14) & 5E-1  \\
cdfs2\_r\_20111118\_20120125\_136A  &  3:30:39.1 & -27:26:47.5 & 0.28  & [.16,.51] & phot & Ib/c & 0.88 & 2000jf & 0.18 &  0.8 &  0.3 & 55882 & 0.8 & 31 (11) & 8E-1  \\
cdfs2\_r\_20111118\_20120125\_140A  &  3:31:04.8 & -27:20:32.4 & 0.61  & [.48,.77] & phot & Ia   & 1.00 & 1991T  & 0.60 &  0.4 & -0.3 & 55896 & 1.4 & 31 (14) & 7E-2  \\
cdfs2\_r\_20111118\_20120125\_21A   &  3:30:51.2 & -27:44:11.3 & 0.60  & [.39,.78] & phot & Ia   & 0.53 & 1990N  & 0.57 &  0.7 & -0.3 & 55881 & 0.8 & 30 (10) & 8E-1  \\
cdfs2\_r\_20111118\_20120125\_5A    &  3:31:06.7 & -27:34:14.6 & 0.64  & [.51,.76] & phot & Ia   & 1.00 & 19901T & 0.62 &  0.1 &  0.2 & 55885 & 1.4 & 31 (15) & 7E-2  \\
cdfs2\_r\_20111123\_20120125\_38A   &  3:31:14.2 & -27:34:12.3 & 0.53  &           & spec & Ib/c & 0.52 & 1998bw & 0.53 & -0.2 & -0.3 & 55901 & 3.9 & 27 (15) & 3E-11 &PSN\\
cdfs2\_r\_20111126\_20120129\_50A   &  3:29:19.9 & -27:07:09.3 & -     &           & -    & Ia   & 0.78 & 1994D  & 0.40 & -0.3 &  0.1 & 55901 & 0.5 & 27 (11) & 1.0 \\
cdfs2\_r\_20111201\_20111025\_12A   &  3:28:53.2 & -27:07:55.0 & -     &           & -    & Ib/c & 0.89 & 2008D  & 0.22 &  0.3 & -0.3 & 55900 & 0.9 & 24 (16) & 6E-1  \\
cdfs2\_r\_20111203\_20111025\_145A  &  3:29:30.2 & -27:07:15.8 & 0.76  & [.18,1.4] & phot & II   & 0.75 & 2009bw & 0.66 &  0.0 &  0.0 & 55891 & 0.5 & 27 (11) & 1.0 \\
cdfs2\_r\_20111203\_20111025\_33A   &  3:30:51.1 & -27:11:07.9 & -     &           & no host& Ia   & 0.82 & 1990N  & 0.39 &  0.7 &  0.1 & 55903 & 0.8 & 24 (13) & 7E-1  \\
cdfs2\_r\_20120114\_20111025\_1A    &  3:29:02.7 & -27:43:48.1 & -     &           & -    & Ia   & 0.65 & 2000cx & 0.16 &  0.9 &  0.3 & 55937 & 0.3 &  9  (9) & 1.0  \\
cdfs2\_r\_20120114\_20111025\_20A   &  3:28:45.1 & -27:34:37.5 & 0.29  & [.19,.41] & phot & Ia   & 0.92 & 1990N  & 0.33 &  0.3 &  0.1 & 55946 & 0.8 &  9  (9) & 6E-1  \\
cdfs2\_r\_20120114\_20111025\_2A    &  3:27:36.0 & -27:26:18.6 & 0.23  &           & spec & Ia   & 1.00 & 2002bo & 0.23 &  1.2 & -0.3 & 55924 & 0.9 &  9  (9) & 5E-1& *\\
cdfs2\_r\_20120114\_20111025\_5A    &  3:31:11.2 & -27:56:17.5 & 0.27  & [.19,.36] & phot &IIn  & 1.00 & 2005gj & 0.26 &  1.0 &  0.3 & 55934 & 0.3 &  9  (9) & 1.0 \\
cdfs2\_r\_20120114\_20111025\_7A    &  3:31:12.8 & -27:54:46.5 & 0.57  & [.44,.69] & phot & Ia   & 1.00 & 1991T  & 0.56 & -0.3 &  0.3 & 55941 & 4.9 &  9  (9) & 1E-6 &PSN \\
cdfs2\_r\_20120118\_20111025\_10A   &  3:29:43.5 & -27:06:29.9 & 0.68  & [.57,.76] & phot & Ib/c & 0.45 & 1998bw & 0.67 & -0.3 &  0.3 & 55946 & 1.4 &  9  (9) & 2E-1 \\
cdfs2\_r\_20120118\_20111025\_66A   &  3:30:56.5 & -27:10:47.9 & 0.30  &           & spec & Ia   & 0.59 & 1992A  & 0.29 &  0.3 &  0.3 & 55947 & 1.5 &  9  (9) & 1E-1 &* \\
cdfs2\_r\_20120118\_20111025\_49A   &  3:27:58.5 & -27:25:32.2 & 0.37  & [.13,.61] & phot & II   & 0.98 & 2009bw & 0.19 & -0.3 &  0.2 & 55948 & 0.5 &  9  (7) & 9E-1 &PSN\\
cdfs2\_r\_20111025\_20120123\_148A  &  3:29:03.2 & -27:49:07.8 & 0.60  &[.14,.83]  & phot & Ia   & 0.97 & 1991T  & 0.72 & -0.3 & -0.3 & 55835 & 1.0 & 23  (7) & 5E-1 &PSN \\
cdfs2\_r\_20111025\_20120123\_37A   &  3:28:50.9 & -27:42:58.2 & 0.15  & [.03,.23] & phot & Ia   & 0.60 & 1992A  & 0.10 & -0.3 &  0.3 & 55762 & 1.5 & 23  (18)& 6E-2 \\
cdfs2\_r\_20111028\_20120125\_124A  &  3:30:47.2 & -27:59:17.4 & 0.69  &           & spec & IIn  & 0.86 & 2005gj & 0.69 &  0.6 &  0.2 & 55831 & 6.5 & 31  (16)& 3E-27&PSN \\
cdfs2\_r\_20111028\_20120125\_8A    &  3:29:14.5 & -27:18:04.2 & 0.18  & [.11,.27] & phot & II   & 1.00 & 2009bw & 0.18 &  0.0 &  0.3 & 55813 & 3.8 & 26  (20)& 2E-10&PSN \\
cdfs2\_r\_20120118\_20111025\_196A  &  3:28:47.1 & -27:15:54.1 & 0.09  & [.05,2.8] & phot & II   & 0.98 & 1999em & 0.23 & -0.3 & -0.3 & 55935 & 0.7 &  9   (9)& 7E-1  \\
cdfs2\_r\_20120123\_20111025\_83A   &  3:31:14.2 & -27:06:57.6 & -     &           & -    &  Ia   & 0.88 & 1990N  & 0.57 & -0.1 &  0.3 & 55957 & 1.5 &  9   (9)& 1E-1  \\
cdfs2\_r\_20111020\_20120125\_76A   &  3:30:51.5 & -27:12:54.8 & 0.49  & [.22,.63] & phot & Ia   & 1.00 & 1991T  & 0.26 &  0.0 &  0.0 & 55865 & 4.5 & 26 (15) & 2E-13 &PSN \\
cdfs2\_r\_20111020\_20120125\_78A   &  3:28:16.9 & -27:10:17.0 & 0.21  &           & spec & II   & 1.00 & 1992H  & 0.21 &  0.6 &  0.3 & 55838 & 3.4 & 27 (24) & 5E-9 &PSN \\
cosmos\_r\_20111218\_20120508\_1A   &  9:58:34.3 &  1:51:36.4 & 0.33   & [.07,.53]  & phot & Ia   & 1.00 & 1991T  & 0.31 &  0.5 &  0.2 & 55907 & 0.5 & 32 (20) & 1.0  \\
cosmos\_r\_20111218\_20120508\_24A  & 10:00:45.1 &  2:06:23.4 & 0.27   & [.12,.47]  & phot & II   & 1.00 & 1992H  & 0.17 &  0.5 &  0.3 & 55916 & 1.1 & 32 (32) & 3E-1  \\
cosmos\_r\_20111218\_20120508\_2A   &  9:59:21.1 &  2:05:33.0 & 0.40   & [.28,.51]  & phot & Ia   & 1.00 & 2002bo & 0.39 &  0.1 & -0.3 & 55909 & 0.7 & 31 (24) & 9E-1  \\
cosmos\_r\_20111218\_20120508\_3A   & 10:00:20.6 &  2:12:34.7 & 0.24   &            & spec & II   & 1.00 & 2004et & 0.23 &  0.2 & -0.1 & 55889 & 1.0 & 36 (28) & 5E-1  \\
cosmos\_r\_20111218\_20120508\_41A  &  9:59:32.0 &  2:22:47.7 & 0.35   & [.30,.40]  & phot & Ia   & 1.00 & 1990N  & 0.34 &  0.0 &  0.3 & 55924 & 0.5 & 35 (18) & 1.0  \\
cosmos\_r\_20111218\_20120508\_46A  & 10:01:03.6 &  2:26:01.5 & 0.19   & [.05,.32]  & phot & Ia   & 1.00 & 1990N  & 0.22 &  0.6 & -0.3 & 55900 & 0.8 & 32  (32)& 8E-1  \\
cosmos\_r\_20111218\_20120508\_56A  & 10:01:06.3 &  2:31:34.7 & 1.22   & [.66,1.3]  & phot & IIn  & 1.00 & 2005gj & 0.69 &  0.0 &  0.3 & 55921 & 1.6 & 33  (33) & 2E-2 \\
cosmos\_r\_20111222\_20120511\_10A  &  9:58:49.5 &  1:42:20.8 & 0.31   & [.27,.35]  & phot & Ia   & 1.00 & 1990N  & 0.29 &  0.1 &  0.3 & 55923 & 0.8 & 36  (15) & 8E-1  \\
cosmos\_r\_20111222\_20120511\_178A &  9:59:02.7 &  2:21:31.1 & 0.35   & [.31,.39]  & phot & Ib/c & 0.55 & 1998bw & 0.34 &  1.0 & -0.3 & 55920 & 0.7 & 33  (15) & 9E-1  \\
cosmos\_r\_20111227\_20120511\_82A  & 10:02:14.5 &  1:42:40.2 & -      &            & -    & II   & 0.90 & 2009bw & 0.29 &  0.0 &  0.3 & 55918 & 0.8 & 36 (13) & 8E-1  \\
cosmos\_r\_20120102\_20120517\_9A   & 10:02:15.3 &  2:32:10.1 & 0.47   & [.43,.52]  & phot & Ib/c & 1.00 & 1998bw & 0.47 & -0.2 & -0.1 & 55930 & 1.3 & 36 (18) & 1E-1  \\
cosmos\_r\_20120106\_20120517\_130A &  9:58:40.5 &  1:49:36.2 &-       &            & no host & II   & 1.00 & 2009bw & 0.21 & -0.3 &  0.3 & 55934 & 1.0 & 35 (23) & 5E-1  \\
cosmos\_r\_20120120\_20120511\_160A &  9:59:00.9 &  2:15:05.1 & 0.56   &            & spec & Ia   & 0.94 & 1990N  & 0.56 &  0.4 & -0.2 & 55950 & 1.3 & 31 (14) & 1E-1 \\
cosmos\_r\_20120122\_20120517\_18A  &  9:59:33.5 &  1:51:45.9 &  0.66  & [.59,.73] & phot & Ia   & 1.00 & 1990N  & 0.65 & -0.3 &  0.2 & 55960 & 0.6 & 31 (14) & 1.0\\
cosmos\_r\_20120124\_20120517\_175A &  9:59:05.5 &  2:33:20.8 &  0.71  & [.65,.75] & phot & IIn  & 1.00 & 2005gj & 0.70 &  0.8 &  0.3 & 55932 & 1.1 & 34 (18) & 3E-1 \\
cosmos\_r\_20120127\_20120511\_118A &  9:58:31.5 &  1:48:04.1 & 0.38   &           & spec & II   & 1.00 & 2004et & 0.37 & -0.3 &  0.1 & 55938 & 1.1 & 36 (18) & 3E-1  \\
cosmos\_r\_20120127\_20120511\_87A  & 10:00:20.8 &  2:43:57.3 & 0.75   & [.71,.79] & phot & Ia   & 1.00 & 1991T  & 0.73 & -0.1 &  0.3 & 55962 & 0.6 & 34 (10) & 1.0  \\
cosmos\_r\_20120216\_20111218\_2A   &  9:59:07.1 &  1:42:56.5 & 0.43   & [.37,.48] & phot & Ia   & 1.00 & 1990N  & 0.42 &  0.4 &  0.3 & 55968 & 2.0 & 27 (22) & 2E-3  \\
cosmos\_r\_20120216\_20111218\_4A   &  9:58:40.6 &  2:04:26.6 & 0.34   &           & spec & II   & 1.00 & 2004et & 0.33 &  0.3 &  0.3 & 55960 & 1.5 & 28 (22) & 4E-2 \\
cosmos\_r\_20120221\_20111218\_33A  & 10:01:11.0 &  2:02:26.4 & 0.72   &           & spec & Ia   & 1.00 & 1991T  & 0.70 & -0.1 &  0.3 & 55990 & 0.9 & 26 (14) & 6E-1  \\
cosmos\_r\_20120221\_20111218\_62A  & 10:01:26.1 &  2:32:05.4 & 0.48   & [.44,.53] & phot & Ia   & 1.00 & 2002bo & 0.47 & -0.1 & -0.1 & 55994 & 1.6 & 23 (15) & 3E-2 \\
cosmos\_r\_20120221\_20111218\_8A   & 10:00:52.2 &  2:39:34.0 & 0.66   & [.63,.70] & phot & Ia   & 1.00 & 2002bo & 0.65 & -0.3 &  0.1 & 55950 & 6.0 & 28 (26) & 7E-22 &PSN \\
cosmos\_r\_20120223\_20111218\_3A   &  9:59:42.0 &  1:44:07.7 & 0.22   & [.17,.27] & phot & II   & 1.00 & 2009bw & 0.21 &  0.1 &  0.3 & 55993 & 1.6 & 25 (18) & 3E-2  \\
cosmos\_r\_20120223\_20111218\_40A  & 10:02:29.5 &  2:13:35.1 & 0.35   & [.29,.41] & phot & II   & 0.93 & 2004et & 0.32 & -0.3 & -0.3 & 55985 & 1.6 & 25 (14) & 3E-2  \\
cosmos\_r\_20120226\_20111218\_59A  &  9:59:51.0 &  2:19:02.0 & 0.31   & [.27,.35] & phot & Ia   & 1.00 & 2002bo & 0.31 &  0.1 &  0.2 & 55997 & 0.6 & 23 (16) & 9E-1 \\
cosmos\_r\_20120226\_20111218\_67A  & 10:00:01.5 &  2:25:51.9 & 0.22   &           & spec & Ia   & 1.00 & 1990N  & 0.21 & -0.1 &  0.3 & 56001 & 1.5 & 23 (18) & 6E-2  \\ 
cosmos\_r\_20120313\_20111218\_12A  & 10:00:41.7 &  1:49:53.7 & 0.36   & [.31,.42] & phot & Ia   & 1.00 & 2002bo & 0.36 & -0.2 & -0.2 & 56006 & 0.9 & 23 (11) & 6E-1  \\
cosmos\_r\_20120313\_20111218\_92A  & 10:00:44.8 &  1:42:33.4 & 0.16   & [.13,.20] & phot & Ia   & 1.00 & 1990N  & 0.17 &  0.8 &  0.3 & 56011 & 0.8 & 23 (10) & 7E-1  \\
cosmos\_r\_20120303\_20111222\_162A & 10:01:59.2 &  1:50:22.4  & 0.50  &           & spec & IIn  & 0.84 & 1998S  & 0.49 &  1.2 &  0.1 & 55992 & 1.7 & 26 (10) & 1E-2 \\
cosmos\_r\_20120313\_20111218\_107A & 10:02:20.8 &  1:52:36.6  & 0.61  & [.56,.66] & phot & Ia   & 0.99 & 1990N  & 0.59 &  0.1 &  0.2 & 56003 & 0.8 & 23 (5)  & 7E-1 &PSN \\
cosmos\_r\_20120313\_20111218\_127A & 10:00:58.7 &  2:04:33.3  & 0.67  &           & spec & Ia   & 0.99 & 1991T  & 0.65 & -0.2 & -0.1 & 56020 & 0.8 & 23 (13) & 7E-1 \\
cosmos\_r\_20111218\_20120508\_107A & 10:01:41.0 &  1:56:42.8  & 0.02  & [.01,.06] & phot & II   & 1.00 & 1999em & 0.04 &  0.3 &  0.3 & 55887 & 2.9 & 31 (31) & 8E-8 &PSN\\
cosmos\_r\_20111218\_20120508\_121A &  9:58:50.7 &  2:08:42.7  & 0.09  & [.05,.11] & phot & II   & 1.00 & 2005cs & 0.11 &  0.0 &  0.3 & 55908 & 2.6 & 28 (28) & 1E-7 &PSN \\
cosmos\_r\_20120508\_20111218\_5A   & 10:02:06.9 &  2:18:46.5  & 0.62  & [.57,.69] & phot & Ia   & 0.91 & 1991T  & 0.60 &  0.5 &  0.3 & 56058 & 0.8 &  6 (5)  & 6e-1 &PSN \\
cosmos\_r\_20120517\_20111218\_165A &  9:59:12.7 &  2:06:58.4  & 0.49  & [.46,.52] & phot & Ia   & 1.00 & 1990N  & 0.47 &  0.0 &  0.3 & 56067 & 0.3 & 6 (6)   & 9E-1  &PSN \\
cosmos\_r\_20120524\_20111227\_164A & 10:00:53.8 &  2:32:53.1  & 0.22  & [.18,.25] & phot & Ib/c & 0.94 & 1998bw & 0.22 &  0.3 & -0.3 & 56070 & 0.8 & 6 (6)   & 6E-1 &PSN\\
cosmos\_r\_20111218\_20120508\_218A & 10:00:33.8 &  2:43:55.1  & 0.71  & [.66,.75] & phot & Ia   & 1.00 & 1999ee & 0.70 & -0.3 & -0.3 & 55922 & 2.2 & 31 (15) & 1E-4 &    \\

\hline 
\end{longtable}

\noindent $P$ = Probable SNe (see text)
 $a$ - SN 2012ez; $b$ - SN 2012fa IAUC 3236 \\
$c$ - SN 2012fn; $d$ - SN 2012fp; $e$ - SN 2012fq; $f$ - SN 2012fo IAUC 3274\\
$g$ - SN 2012gu; $h$ - SN 2012gs; $i$ -SN 2012gv; $j$ -SN 2012gt  IAUC 3311\\
* Host galaxy redshift obtained in this program \\
** SN type revised:  based only on spectrum was classified as Ic)
\end{landscape}    
\end{longtab}    

To evaluate the uncertainties of our classification tool in the next two sections we compare our derived SN types with $a)$ photometric classifications obtained using the public software package {\em SNANA} \citep{kessler:2009fk} and $b)$  with the spectroscopic observations of a small sample of "live" transients, observed while still in a bright state.

\subsection{Comparison with the photometric classifications by PSNID in SNANA}       

With the aim to check our procedure and evaluate the related uncertainties,
we performed the photometric classification of our SN candidates using the public code PSNID in the {\em SNANA}\footnote{\url{http://das.sdss2.org/ge/sample/sdsssn/SNANA-PUBLIC. We used {\em SNANA} version 10\_39k.}implementation \citep{Sako:2011gf,kessler:2009fk}. Overall, the  approach of {\em PSNID} is similar to that adopted here; besides the implementation of the computation algorithm, the main difference is in the  template list.  In particular, for SN~Ia we adopted the fitting set-up of \cite{Sako:2011gf}, while for CC SNe we used the extended list of 24 templates available in the {\em SNANA} distribution\footnote{We used the Nugent SED templates updated by D.Scolnic as illustrated in 
\url{http://kicp-workshops.uchicago.edu/SNphotID_2012/depot/talk-scolnic-daniel.pdf} .}} 

Also for the fit with {\em PSNID} we set the host galaxy redshift as a prior with the same range of uncertainty as in our procedure. In this case however, also for photometric redshift we assume a normal distribution for $P(z)$ with the $\sigma$ provided by the photometric redshift code.

\begin{figure}
\includegraphics[width=\hsize]{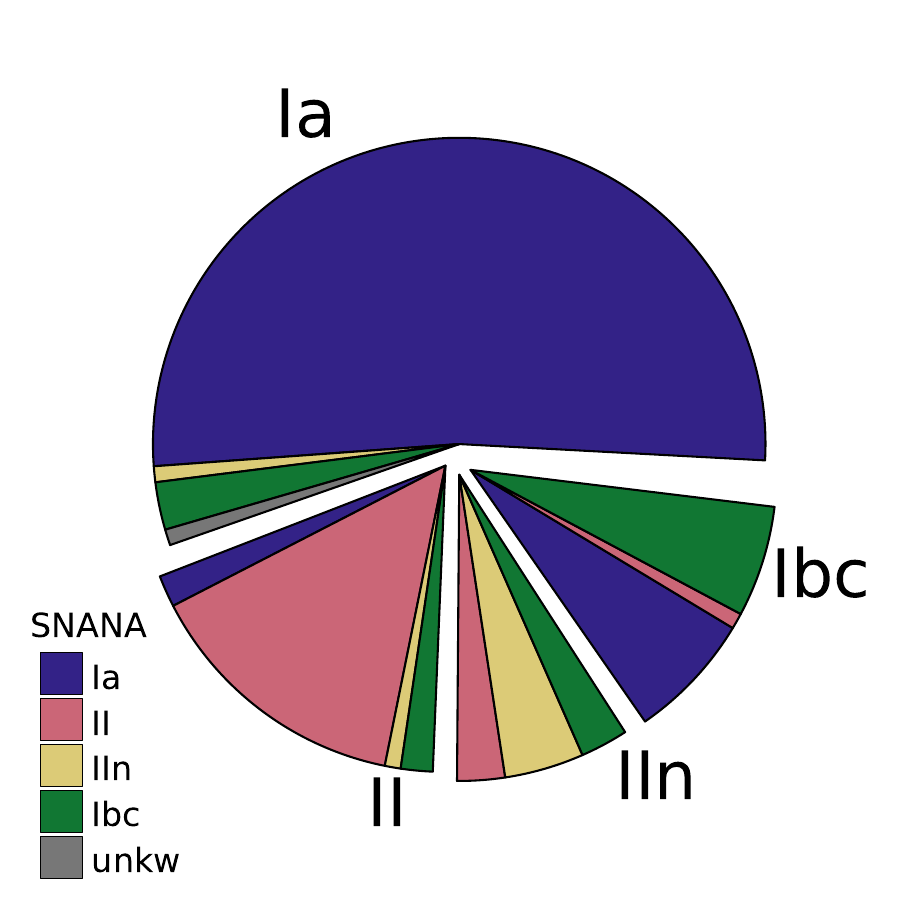}
\caption{Comparison of the SN classifications obtained with the different tools. The exploded wedges are the SN type fractions obtained with our SUDARE tool and the coloured sectors are the {\em SNANA} classification.}\label{classif}
\end{figure}

A comparison of the classifications obtained with the two tools is illustrated in Fig.~\ref{classif}. The pie chart shows the SN classifications in the four main types using the {\em SUDARE} tool and  the sectors with different colours within a given wedge show the {\em PSNID} classifications. For a two events, marked in grey, the fit with {\em PSNID} fails. 

The figure shows that the identification of SN~Ia is quite consistent ($92$\% of the type Ia classified by our tool are confirmed by {\em PSNID}) and there is a good agreement also for the normal type II ($77\%$ of the classifications are matched). The agreement is poor for type Ib/c  and for type IIn (only $~40\%$ are matched in both cases).
The latter result is not surprising considering the wide range of luminosity evolution: for these classes of SNe the choice of input templates is crucial.  

However, we remark that despite the discrepancy in the classification of individual events of specific sub-types, there is an excellent agreement of the event counts in each class, but for type IIn, as shown in Tab.~\ref{snana}.  This implies that, as far as the SN rates are concerned, using either classification tools makes a little difference, with the exception  the exception of type IIn where the difference is  $\sim 40\%$.

In Tab~\ref{snana} we also report the SN count for the different classes using the Bayesian probability. It  appears that with respect to the count of the most probable type, the number of SN~Ia is slightly reduced while the number of Ib/c increases. This is not unexpected given the similarity of the light curves of type Ia and Ib/c (in many cases an event can have a significant probability of being either a type Ia or a type Ib/c) and the fact that type Ia are intrinsically more frequent than Ib/c. The effect however is small, $<5\%$ in both cases.

\begin{table}
\caption{Comparison of photometric classification with the different tools. In parenthesis we report the events labelled as probable SNe. For column 2 and 3 we count the SN with respect to the most probable SN type. Column 4 lists instead sum the Bayesian probability for each SN type (in parenthesis is the number of PSN). }\label{snana}
\centering
\begin{tabular}{lcrcrrr}
\hline
          & SUDARE  &  &   SNANA  && \multicolumn{2}{c}{Bayesian} \\
\hline    
 Ia       &   67  &  &  72 & & 64.7 &(12.7)  \\
 II       &   22  &  &  21 & & 23.2 &(6.5)  \\
 IIn      &   11  &  &   7 & & 10.7 &(2.9)  \\
 Ib/c     &   17  &  &  15 & & 18.6 &(4.8)  \\
 All      &  117  &  & 115 & & 117.0  &(27.0)   \\
\hline
\hline
\end{tabular}
\end{table}

\subsection{Comparison with spectroscopic classification\label{spectra}}

For a small sample of the SN candidates we obtained immediate spectroscopic classification.  

Observations were scheduled at the ESO VLT telescope equipped with FORS2 at three epochs for a total allocation of  2 nights. The telescope time allocation, that was fixed several months in advance of the actual observations, dictated the choice of the candidates: 
we selected transients that were "live" (above the detection threshold) at the time of observations and among these we gave a higher priority to the brightest candidates with the aim to secure a higher S/N for the spectra.

For the instrument set-up we used two different grisms, namely  GRIS\_300V and GRIS\_300I covering the wavelength range 400-900 nm and  600-1000 nm respectively with similar resolution of about 1 nm. The choice of the grism for a particular target was based on the estimated redshift of the host galaxy with the GRISM\_300I used for redshift $z>0.4$. 

We were able to take the spectrum of 17 candidates. Spectra were reduced using standard recipes in {\em IRAF}. In three cases the S/N was too low for a conclusive transient classification and we were only able to obtain the host galaxy redshifts.
Four of the candidates turned out to be variable AGNs, in particular Seyfert galaxies at redshifts between $0.25<z<0.5$.
We stress that, to maximise the chance of obtaining useful spectra, we tried to observe the candidate shortly after discovery. This means that at the time of observations we had not yet a full light curve and hence a reliable photometric classification. Eventually, all the four AGN
 exhibit a erratic luminosity evolution that, if known at the time of spectroscopic observations, would have allowed us to reject them as SN candidates.

Ten transients were  confirmed as  SNe and their spectral type were assigned  through cross-correlation with libraries of SN template spectra using  {\em GEneric cLAssification TOol} \citep[GELATO,][]{Harutyunyan:2008fk} and the {\em Supernova Identification} code \citep[SNID,][]{Blondin:2007uq}. The spectroscopically classified SNe, identified with a label in Tab.~\ref{snlist}, turned out as 6 type Ia, 2 type Ic, 1 type II and a type IIn. In all cases the SN type was coincident with the independent photometric classification with one exception (SN~2012gs) classified Ic from spectroscopy and Ia from photometry. 

\begin{figure}
\includegraphics[width=\hsize]{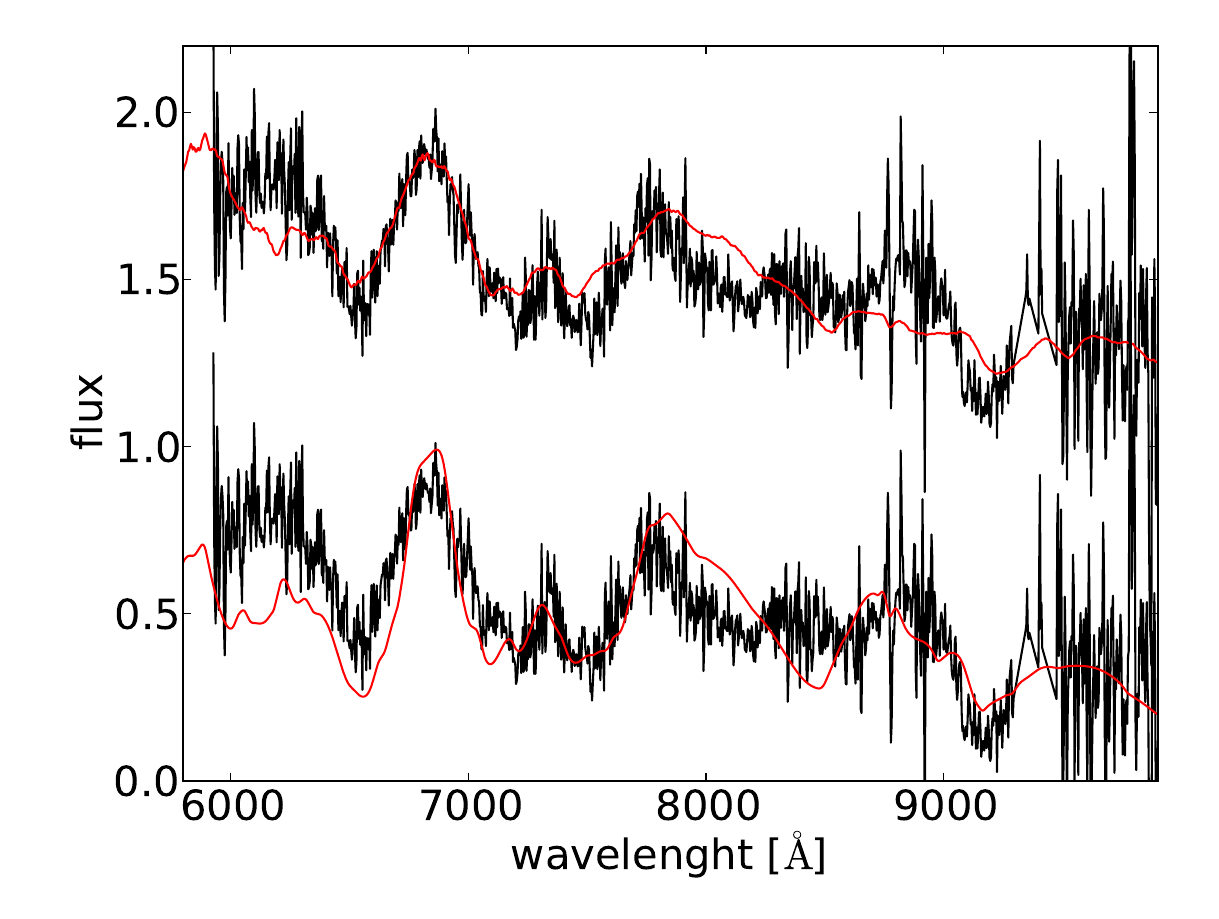}
\caption{The observed spectrum of SN~2012gs obtained with FORS2 on MJD 56252.0 (black line) is compared with that of the SN~Ic 2007gr at phase $-9$d (top)  and of the type Ia SN~1991T at phase $+14$d (bottom). In both cases it is adopted for SN~2012gs a redshift $z=0.5$ as measured from the narrow emission lines of the host galaxy. }\label{sn2012gs_spec}
\end{figure}

\begin{figure}
\includegraphics[width=\hsize]{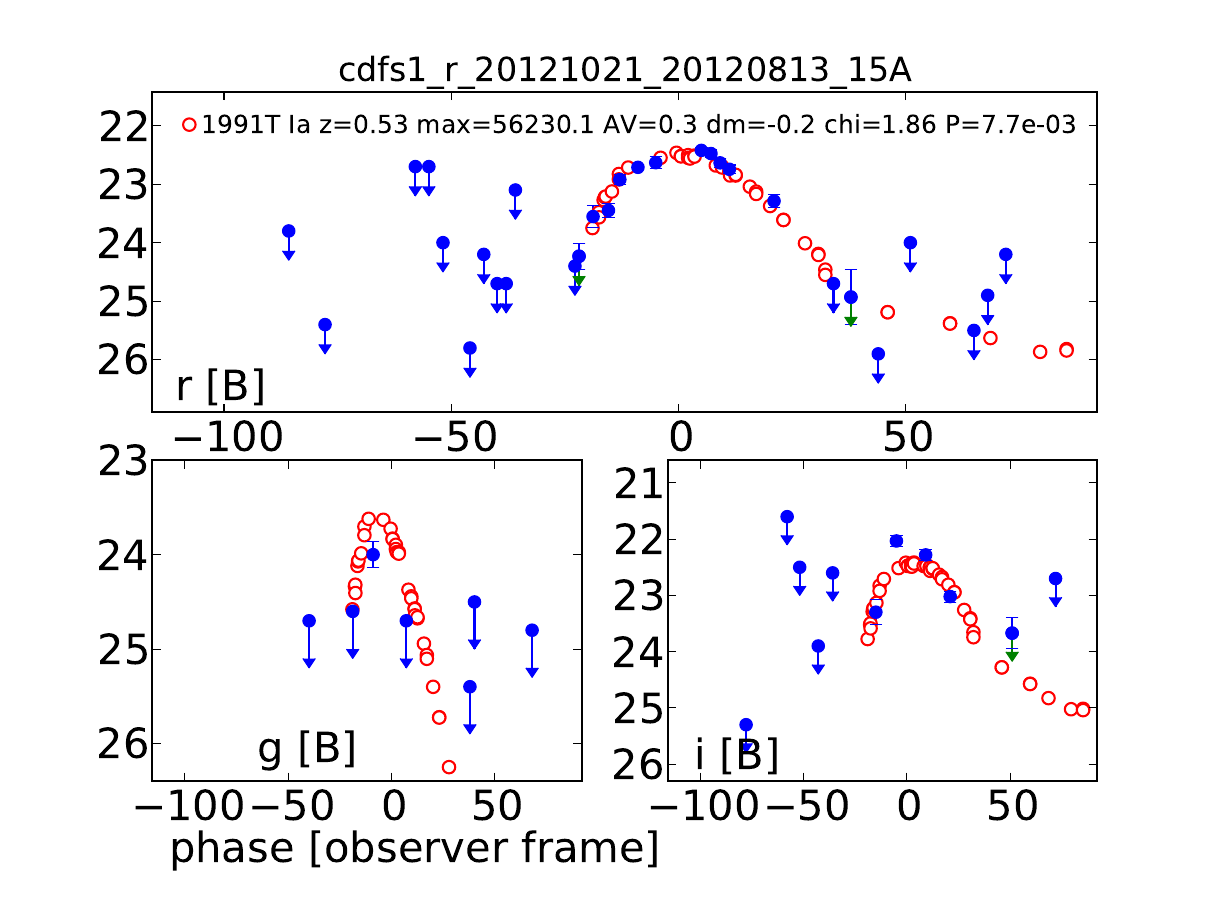}
\caption{SN 2012gs light curve fit obtained using our tool. The best match is obtained with SN~1991T and the maximum is estimated to occur on  MJD 56235.1. }\label{sn2012gs_lc}
\end{figure}

As shown in Fig.~\ref{sn2012gs_spec},  the spectrum of SN~2012gs can be fitted both by a template of type Ic SN well before maximum or by a type Ia SN two weeks after maximum, in both cases the redshift was $z \sim 0.5$. On the other hand, when we consider the light curve (Fig.~\ref{sn2012gs_lc}) it results that the spectrum was  obtained two weeks after maximum and therefore the first alternative can now be rejected. Therefore, revising the original spectroscopic classification, SN~2012gs is classified as type Ia. 

\subsection{Classification uncertainties\label{sne}}

The  comparison of the photometric and spectroscopic classifications, even if for a very small sample, confirms that photometric typing is a reliable in particular when the redshift of the host galaxy is known.  For our photometric tools we did not yet performed a detailed testing that instead has been performed for {\em PSNID}. In particular, \cite{Sako:2011gf} 
show that {\em PSIND} can identify SN~Ia with a purity of $90\%$. This appears consistent with the results obtained from the comparison of {\em PSNID} and {\em SUDARE} tools.
More difficult is to quantify the performances of photometric classification for CC~SN 
both because of the lack of suitable spectroscopic samples \citep{Sako:2011gf} and for the limitation of simulated samples \citep{Kessler:2010kq}.
From the comparison of the CC~SN classification of the {\em PSNID} and {\em SUDARE} tools we found differences in the individual classifications of $25\%$ for type II events and $40\%$ for type  Ib/c events. These should be considered  as lower limit of the uncertainty because the two codes adopt similar approaches the main difference being the choice of templates.
On the other hand, the discrepancy on the overall SN counts of a given type  is much lower, typically a few percents, though for type IIn it is about $40\%$. Based on these considerations and waiting for a more detailed testing we adopt the following uncertainties for SN classification: 10\% for Ia, 25\% for II, 40\% for Ib/c and IIn.

\section{The SN sample}

\begin{figure}
\includegraphics[width=\hsize]{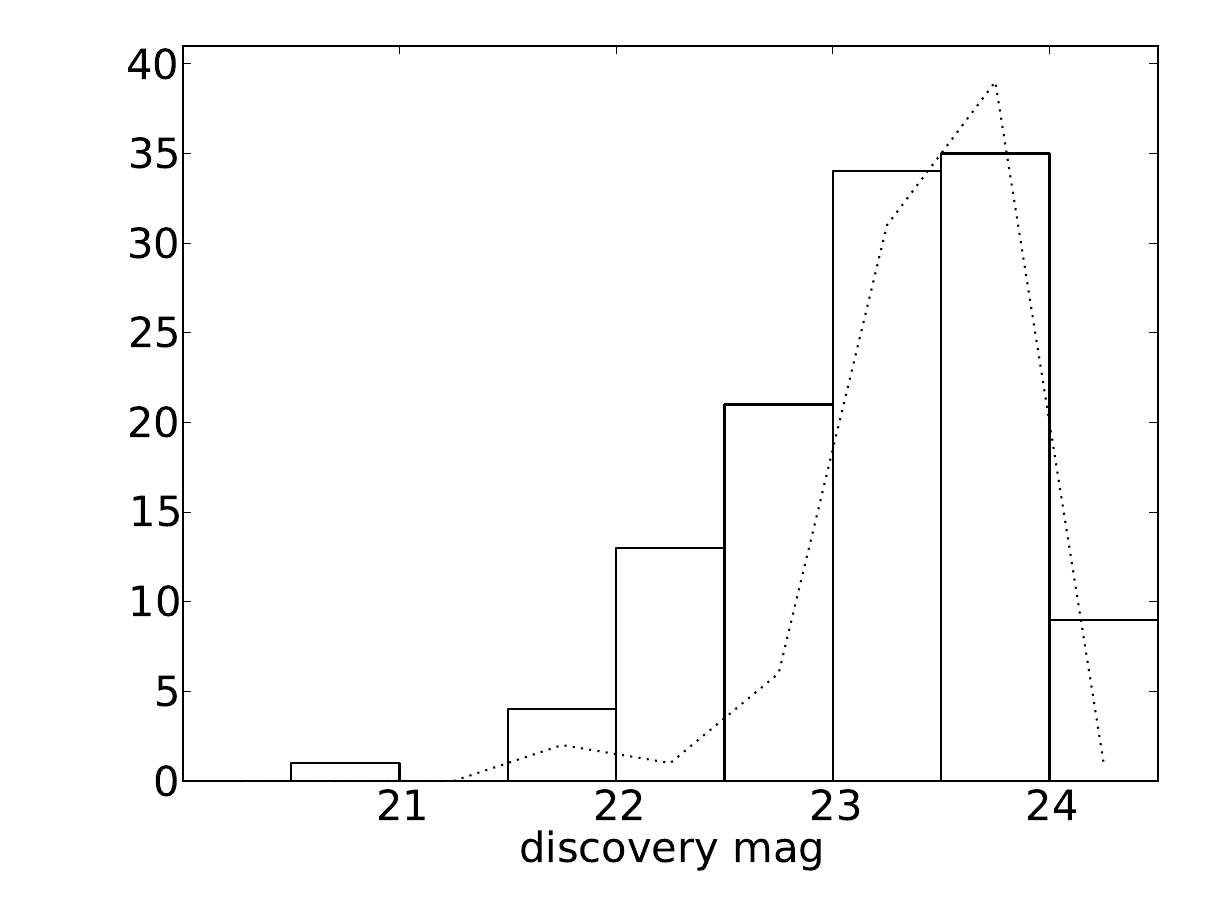}
\caption{Histogram of the r-band SN magnitudes at discovery. The dotted line shows the distribution of $m_{50}$ (the magnitude where the detection efficiency  is 0.5) for the r-band observations (as reported in Tab 2).} \label{magdist} 
\end{figure}

\begin{figure}
\includegraphics[width=\hsize]{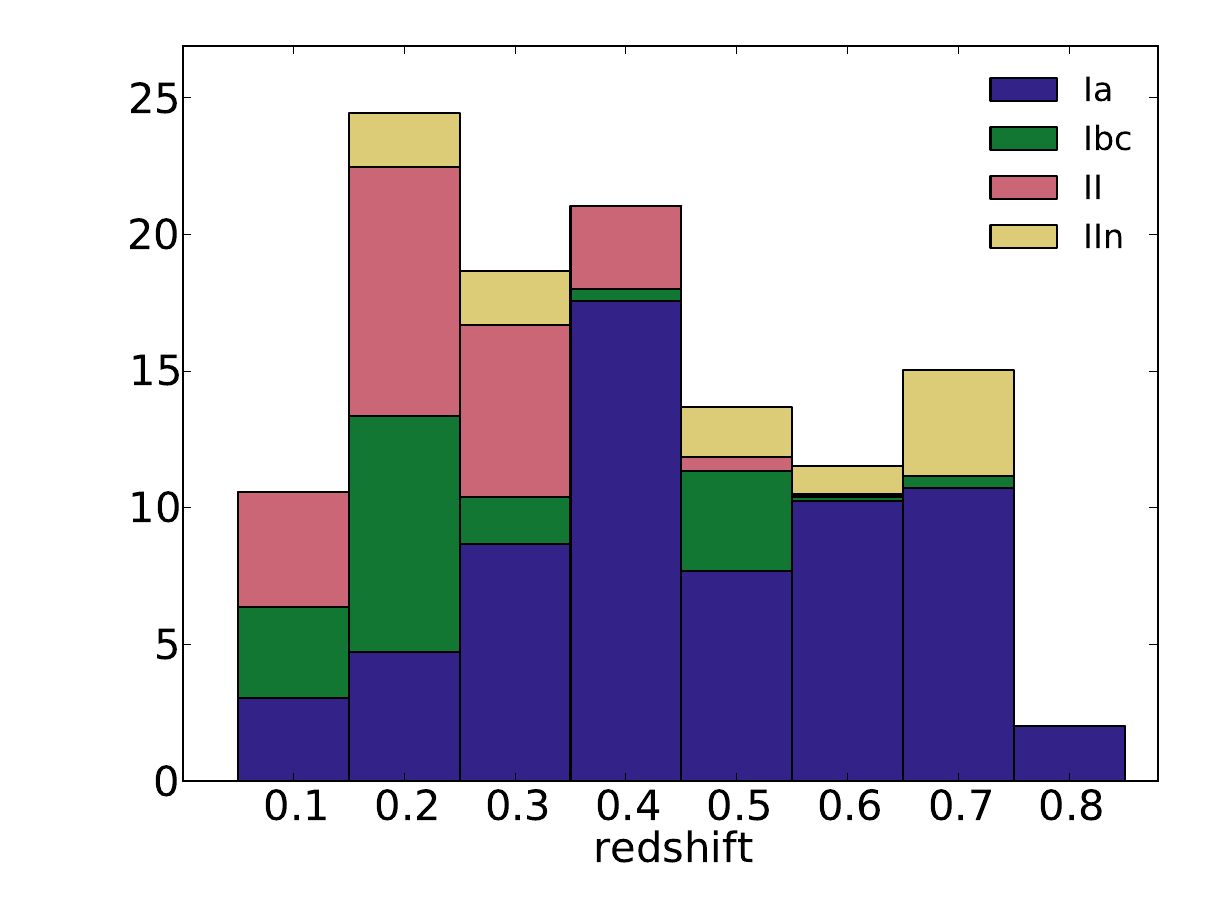}
\caption{Redshift distribution of the discovered SNe for the different types} \label{snredshift}
\end{figure}

As a result of the selection and classification process, we obtained a sample of 117 SNe, 27 of which are marked as probable SNe (PSNe). The distribution of the SN apparent magnitude at discovery, plotted in Fig.~\ref{magdist}, shows a peak at mag $r=23.5-24$ that is consistent with what expected given the detection efficiencies (see Sect.\ref{artstar}).

We found that 57\% of the SNe are of type Ia, 19\% of type II, 9\% of type IIn and  of 
15\% type Ib/c. 
We  notice that the percentage in different subtypes is quite close to the fraction of SN types in magnitude limited samples. For instance  the updated Asiago SN Catalog\footnote{\url{http://sngroup.oapd.inaf.it/asnc.html}}  includes 56\% Ia, 27\% II, 4\% IIn and 10\% Ib/c (counting only SNe discovered since 2000) with some difference from our sample only for the most uncertain events classified as type IIn.  
The result is encouraging when considering that we did not make any assumption on the fraction of  the different SN types in our typing procedure.
This also implies that the relative rates of the different SN types are similar in the local Universe and at $z\sim 0.5$.

At the same time the SN type distribution in our sample is very different from that derived in a volume limited sample like that derived for the LOSS survey \citep{li:2011zr}  which gave the following SN type fraction:  Ia 24\%, Ib/c 19\%, II 52 \% and IIn 5\%.  The much higher fraction of SN~Ia in our sample is a natural consequence of the high luminosity of SN~Ia compared with other types that makes it possible to discover SN~Ia in a much larger volume. This also explains  the SN redshift distribution shown in Fig.~\ref{snredshift}.  While SN~Ia are found  to $z\sim0.8$, the redshift limit for the discovered SN~II is only $z\sim0.4$. The relatively rare but bright type IIn are, on average, discovered at higher redshifts.

\section{The galaxy sample\label{galaxysample}}

To relate the occurrence of SN events to their parent stellar population,  we need to characterise the galaxy population in the survey fields and in the redshift range explored by the SN search.  To this aim, the extensive  multi-wavelength coverage of both COSMOS and CDFS gives a unique opportunity. In particular  the analysis of  deep multi-band surveys of the COSMOS field has already been published \citep{Muzzin2013a}, and we could retrieve the required information, such as photometric redshifts, galaxy masses and star formation rates  directly from public catalogs. For the CDFS fields we  performed instead our own analysis but closely following the method  described by \cite{Muzzin2013a}. In the following, we describe the detection and characterisation of the galaxies in our search field. 
   
\subsection{COSMOS field\label{cosmos}}

A  photometric catalog of the sources in the COSMOS field has been produced by \cite{Muzzin2013a}, and  available trough the UltraVISTA survey Web site\footnote{http://www.strw.leidenuniv.nl/galaxyevolution/ULTRAVISTA/ \label{muzzin}}.  The catalog covers an area of 1.62 deg$^2$  and encompasses the entire 1.15 deg$^2$  area monitored by SUDARE.
The catalog includes photometry in 30 bands obtained from: $i)$ optical imaging from Subaru/SuprimeCam ($grizBV$ plus 12 medium/narrow   bands IA427 -- IA827) and CFHT/MegaCam (u$^{*}$) \citep{Taniguchi2007,Capak2007}, $ii)$ NIR data from VISTA/ VIRCAM  \citep[$YJHK$ bands,][]{McCracken2012}, $iii)$ $UV$ imaging from {\it GALEX } \citep[FUV and NUV channels,][]{Martin2005}, $iv)$ MIR/FIR data from {\it Spitzer's} IRAC+MIPS cameras (3.6, 4.5, 5.8, 8.0, 24 and 70, 160 $\mu$m channels from \citealt{sanders:2007ul} and \citealt{Frayer2009}).

The optical and NIR imaging  for COSMOS have comparable  though not identical PSF widths (FWHMs are in the range  $ 0.5\arcsec -1.2\arcsec$). For an accurate measurement of galaxy colours, 
\cite{Muzzin2013a} performed the PSF homogenisation  by degrading the image quality of all bands to the image quality of the band with worst seeing (with seeing of 1\arcsec -- 1.2\arcsec). 
Source detection and photometric measurements  were performed using the {\sc SExtractor}  package in dual image mode with the non-degraded  K image  adopted as the reference for source detection.  
The  \texttt{flux\_auto}  in all bands was  measured with an aperture of 2.5 times the Kron radius  that includes  $ > 96 \%$ of the total flux of the galaxy \citep{Kron1980}. Hereafter, the K-band magnitude was corrected to the total flux by measuring the growth curve of bright stars out to a radius of 8$^{\prime\prime}$ (depending on the magnitude this correction ranges between 2\%-4\%). 

The space-based imaging from  {\it GALEX}, IRAC and MIPS have more complicated PSF shapes and larger FWHM, therefore photometry for these bands was performed separately \citep[see Sec. 3.5 and 3.6 in][]{Muzzin2013a}. 

The photometry in all bands is corrected for Galactic dust attenuation using dust maps from \cite{schlegel:1998uq} and using the Galactic Extinction Curve of \citet{Cardelli:1989fk}. The corrections were of the order of 15\% in the {\it GALEX } bands, 5\% in the optical and < 1\% in the NIR and MIR.
 
Star vs. galaxy separation was performed in the  $J  - K$ versus $u - J $ color space where there is a clear segregation between the two components \citep[Fig.~3 of][]{Muzzin2013a}. Sources were classified as  galaxies if they match the following criteria:

\begin{eqnarray}\label{colorselect}
J - K > 0.18 \times (u - J) - 0.75 & \mbox{for}\,\, u - J < 3.0 \nonumber \\
J - K > 0.08 \times (u - J) - 0.45 & \mbox{otherwise}
\end{eqnarray}

The photometric redshifts for the galaxy sample were obtained with the {\sc EAZY}\footnote{ http://www.astro.yale.edu/eazy/} code \citep{Brammer2008}.  {\sc EAZY} fits the galaxy SEDs with a linear combinations of  templates and includes optional flux- and redshift-based priors.  In addition, {\sc EAZY} introduces a rest frame template error function to account for wavelength dependent template mismatch. 
This function gives different weights to different wavelength regions, and ensures that the formal redshift uncertainties are realistic. 

The template set adopted by \cite{Muzzin2013a}  includes: $i)$ six templates derived from the PEGASE models \citep{Fioc:1999fk}, $ii)$ a red template from the models of \cite{Maraston:2005uq}, $iii)$ a 1 Gyr old single-burst \citep{Bruzual:2003kx} model to improve the fits for  galaxies with post starburst-like SEDs and $iv)$ a slightly dust-reddened young population to improve the fits for a  population of UV-bright galaxies.   \cite{Muzzin2013a} chose to use the v1.0 template error function, and the K magnitude prior, and allowed photometric redshift solutions in the range $0<z<6$.

Photometric redshifts are extremely sensitive to errors in photometric zeropoints.
 A common procedure to address this problem is to refine the zeropoints  using a subsample of galaxies with spectroscopic redshifts \citep[e.g.,][]{ilbert:2006vn,brammer:2011kx}.
 \cite{Muzzin2013a} used an iterative code developed for the NMBS survey \citep[see][]{Whitaker:2011uq} and found zeropoint offsets of the order of $\sim$ 0.05  mag for the optical bands  and of 0.1-0.2 mag for the NIR bands.

To remain above the 90\% completeness limit and guarantee the consistency with the CDFS catalog (see next section) we  selected  from the full COSMOS catalogue all galaxies with  K band magnitude  $\le 23.5$. We further restrict the catalog to the sky area coverage of our field of view (1.15 deg$^2$) and redshift range  $0<z<1$ of interest for the SN search, obtaining a final count of 67417 galaxies.

\subsection{VOICE-CDFS}

Areas of different sizes around the original CDFS field have been variously observed at different depths from the X-ray through the UV, Optical, IR to the Radio. The 0.5 deg$^2$ Extended CDFS (ECDFS) multi-wavelength dataset has been carefully reduced and band-merged over the years 
\citep[e.g.][and reference therein]{cardamone:2010zr,Hsu:2014vn}.
Conversely, most public multi-wavelength data over the VOICE-CDFS 4 deg$^2$ area have been collected  recently, and are not available as an homogeneous database for our study. For our purposes we thus collected, merged and analysed most existing data ourselves.

Available data over the VOICE-CDFS area include the following:

\begin{itemize}
\item GALEX UV deep imaging \citep{martin:2005eu}.
The GALEX photometry is from the GALEX GR6Plus7 data release\footnote{http://galex.stsci.edu/GR6/}.
\item SUDARE/VOICE $u,g,r,i$ deep imaging (this work, Vaccari et al. in prep.)
\item VISTA Deep Extragalactic Observations \citep[VIDEO,][]{Jarvis2013} Z,Y,J,H,K deep imaging
\item SERVS Spitzer Warm 3.6 and 4.5 micron deep imaging \citep{mauduit:2012qe}
\item SWIRE Spitzer IRAC and MIPS 7-band (3.6, 4.5, 5.8, 8.0, 24, 70, 160 micron) imaging \citep{lonsdale:2003rw}
\end{itemize}

While  data products are available as public catalog for most of the multi-wavelength surveys listed above, SERVS and SWIRE data were re-extracted and band-merged with all other datasets as part of the Spitzer Data Fusion project \citep[][http://www.mattiavaccari.net/df/]{vaccari:2010fk}.

With the VIDEO survey still in progress,
at the moment the sky areas covered by SUDARE and VIDEO do not fully overlap: restricting our analysis to the overlapping region, for galaxy detection  we lack  a small portion of our VOICE-CDFS1 and VOICE-CDFS2
(0.14 and 0.05 deg$^2$, respectively) . However, for the estimate of SN rates in the  cosmic volume surveyed by SUDARE we use the full area covered by the two fields.

\begin{figure}
\begin{center}
\includegraphics[width=9.cm]{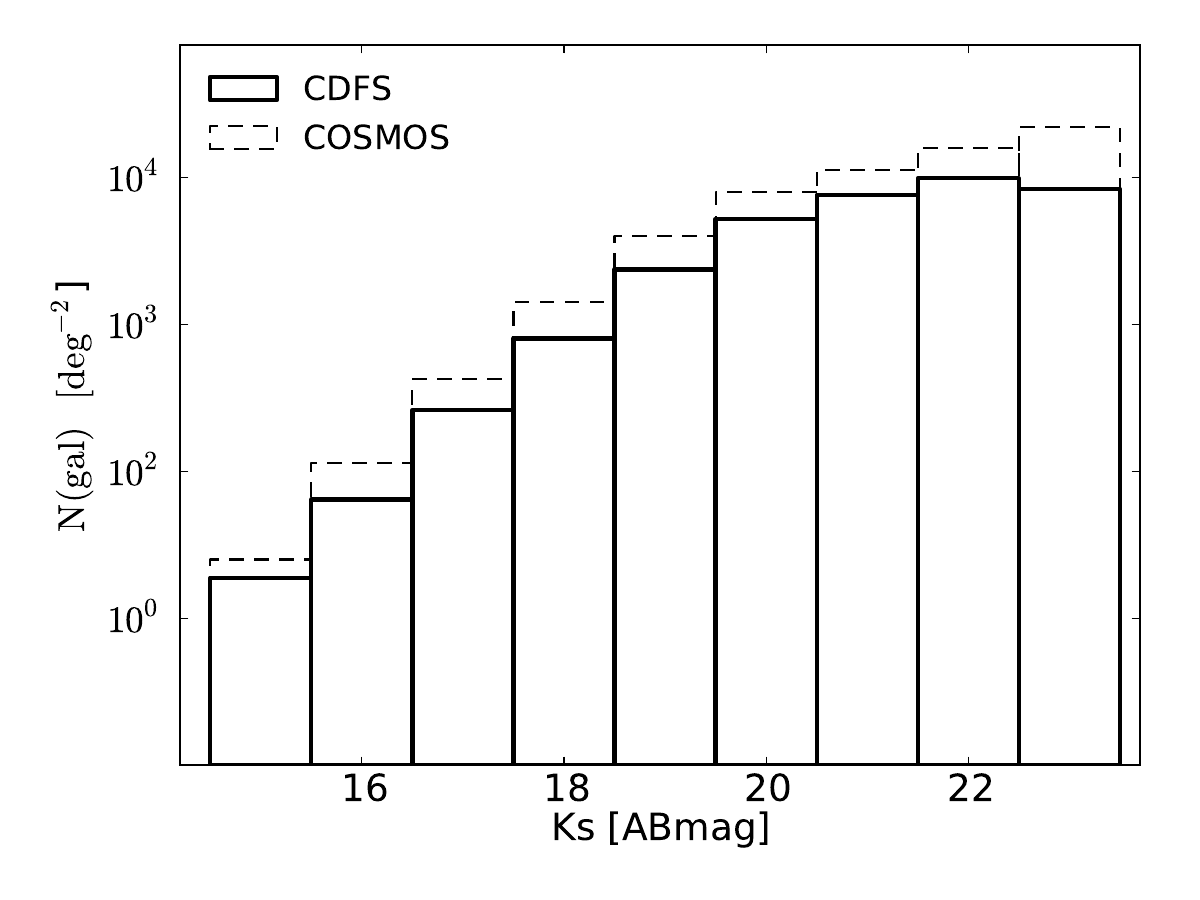} 
\end{center}
\caption{Distribution of $K$ magnitude for  galaxies in CDFS (solid line) and COSMOS (dashed line) catalogues. }
\label{magK}
\end{figure}

\begin{figure}
\begin{center}
\includegraphics[width=9.cm]{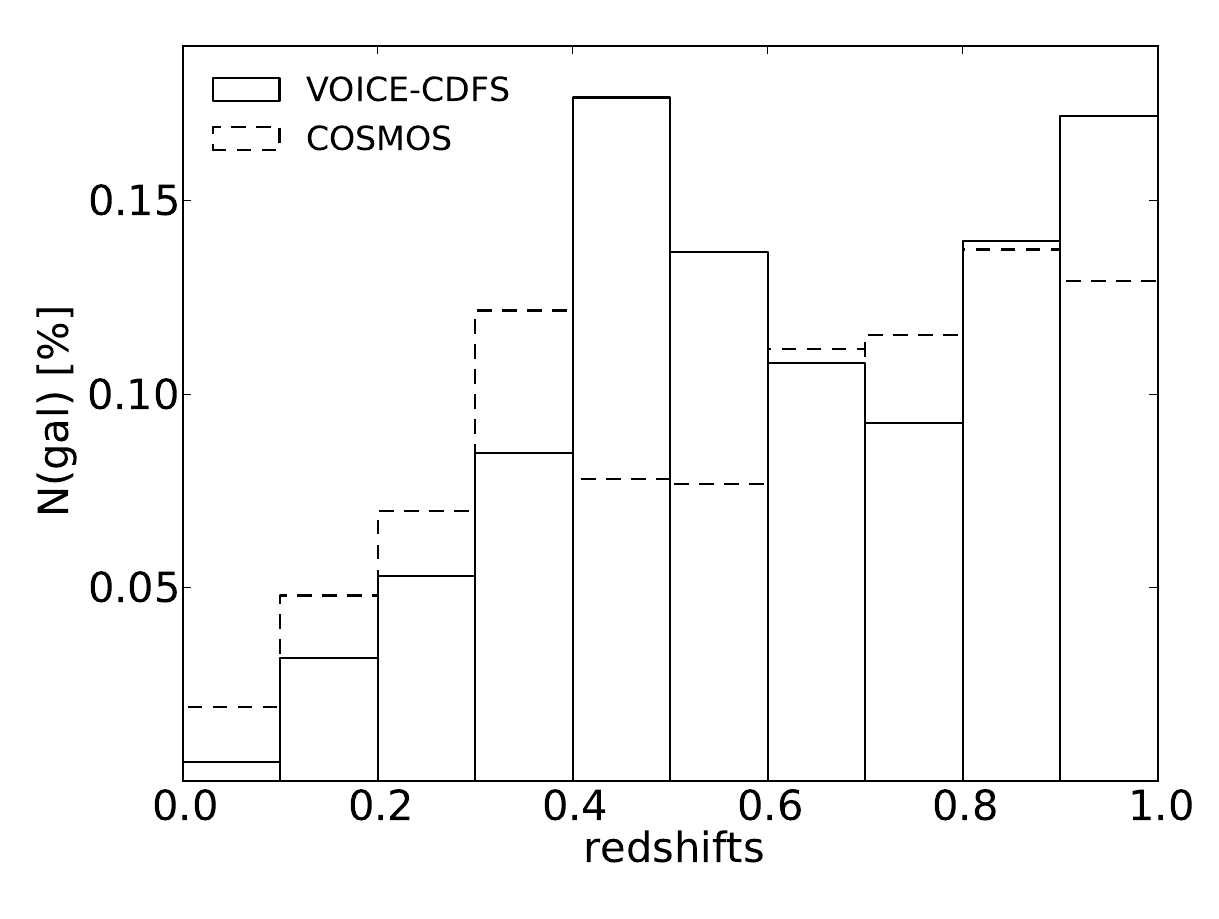} 
\end{center}
\caption{Redshift distribution ($z_{\rm peak}$) for  galaxies in CDFS (solid line) and COSMOS (dashed line) catalogues. }
\label{redshift}
\end{figure}

Deep image stacks have been obtained from SUDARE/VOICE data  as described in Sect.~\ref{datared}.  The VIDEO exposures were processed at the Cambridge Astronomical Survey Unit (CASU)  using the pipeline developed specifically to the reduction of VIRCAM data, as part of the VISTA Data Flow System (VDFS)\footnote{http://casu.ast.cam.ac.uk/surveys-projects/vista/technical/data-processing}  \citep{irwin:2004fk}. The stacks produced by CASU were weighted mean combined using {\sc SWarp}.

CDFS stacks in optical and NIR filters show a very small variation in seeing (ranging from 0.8\arcsec to 0.9\arcsec) and we do not need to perform PSF homogenisation for measuring colours, but only to resample both VST and VISTA images  to the same pixel scale of 0.21\arcsec pixel$^{-1}$ (for this we used {\sc SWarp}). 

Source detection and photometry for VOICE-CDFS  were performed with {\sc SExtractor} in dual image mode with the K band image  used as reference for  source detection.
   
The photometry was corrected for Galactic extinction corresponding to a flux correction of 3\% in the optical and   $<1$\% in the NIR. Then, we separated galaxies from stars using Eq.~\ref{colorselect}.

For all galaxies in the catalog we obtained photometric redshift using the  {\sc EAZY} code  adopting the same parameters and templates 
described in Sect.~\ref{cosmos}. The main difference between the two fields is the number of filters available for the analysis, 12 filters for CDFS and 30 for COSMOS.  To reduce catastrophic failures, we do not compute photometric redshifts for the  sources that have detection in  less than 6 filters ($<5$\%).

Similar to \cite{Muzzin2013a}, the magnitude zero-points were verified using the iterative procedure developed by Brammer (p.c.). The procedure is based on the comparison of  photometric to spectroscopic redshifts: systematic deviations are translated into zero-point offsets corrections using  K  as the "anchor" filter , the photometric scale is adjusted and  {\sc EAZY} re-run. We did not calculate offsets for {\it GALEX } and  {\it Spitzer} bands.  We found that $g,r,i$ bands require small offsets ($\le$ 0.05 mag)  while the $u$ band requires an offset of 0.14 mag, and NIR bands  of about 0.1 mag.

Also for VOICE-CDFS we selected all galaxies with  K band magnitude  $<23.5$ and redshift $0<z\le1$ that results in a final catalog of  92324 galaxies for VOICE-CDFS that, allowing also for a small overlap of the two pointings, covers 2.05 deg$^2$ . 

The distribution of K band magnitudes and  photometric redshifts  for the COSMOS and VOICE-CDFS  galaxy samples are shown in Fig.~\ref{magK} and in Fig.~\ref{redshift} respectively.

\begin{figure}
$
\begin{array}{c@{\hspace{.1in}}c@{\hspace{.1in}}c}
\includegraphics[width=9cm]{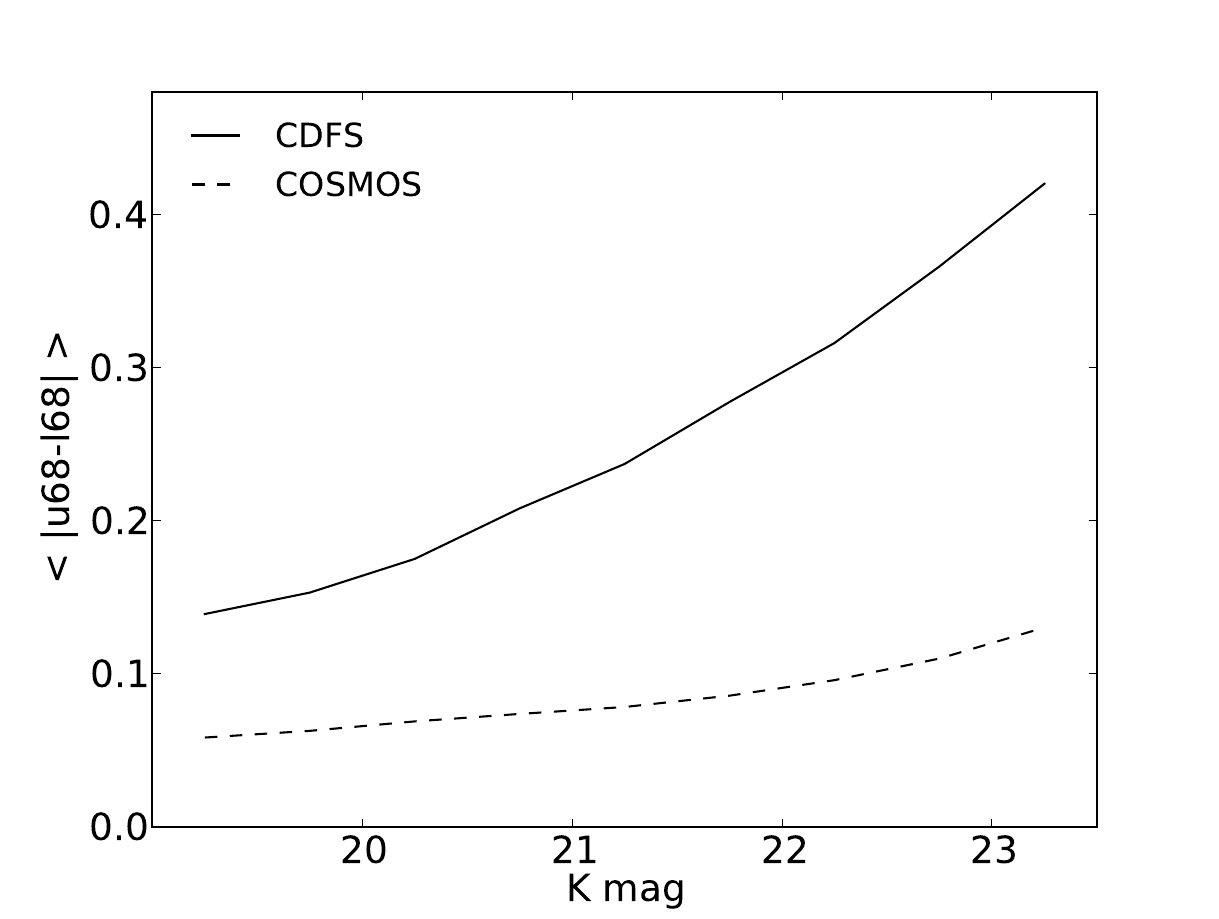} \\
\includegraphics[width=9cm]{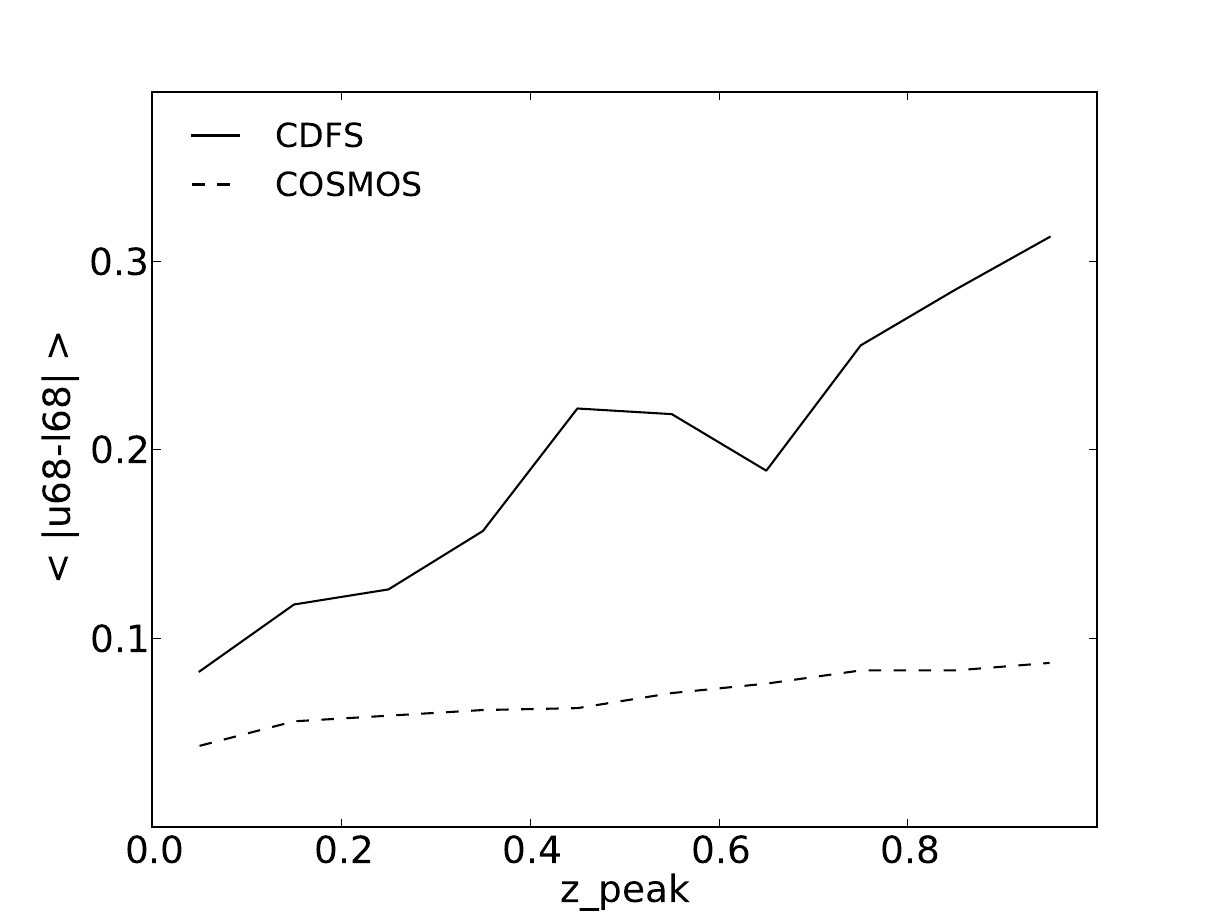}\\
\end{array}
$
\caption{The width of the 68\% confidence intervals computed from the redshift probability distribution as  a function of galaxy magnitude (top panel) and redshift (bottom panel).}
\label{l68}
\end{figure}

\subsection{Accuracy of photometric redshifts}

We explored different methods to assess the quality of photometric redshifts : $i)$ analysing the width of confidence intervals, and quality measurements  provided by {\sc EAZY}, $ii)$ comparing different redshift estimators, $iii)$ comparing the photometric redshifts with available spectroscopic redshifts, $iv)$ comparing our estimates with photometric redshifts from other groups.

\subsubsection{Internal error estimates}\label{zconf}
 
 {\sc EAZY} provides multiple estimators of the photometric redshifts among which we choose $z_{\rm peak}$  corresponding to the peak of the redshift probability distribution $P(z)$. As a measure of the uncertainty, the code provides 68, 95 and 99\% confidence intervals calculated by integrating the $P(z)$. The confidence intervals are a strong function of the galaxy apparent magnitude and redshift, as shown in Fig.\ref{l68} for the 68\% level. The narrower confidence intervals for the COSMOS field with respect to the VOICE-CDFS field are due to the better sampling of the SED for the galaxies of the former field.

{\sc EAZY} provides also a redshift quality parameter, 
$Q_z$\footnote{ $Q_z = \frac{\chi^2}{N_\mathrm{filt}-3} \, \frac{u^{99} - l^{99}}{p_{\Delta z=0.2}}$
 where  $N_\mathrm{filt}$ is the number of photometric measurements used in the fit, $u^{99}- l^{99}$ is the 99\% confidence intervals  and ${p_{\Delta z=0.2}}$ is the fractional probability that the redshift lies within $\pm 0.2$ of the nominal
value.}  that is intended as a robust estimate of the reliability of the photometric redshift \citep{Brammer2008}.
Poor fits ($Q_z>1$) may be caused by uncertainties in the photometry, poor match of the intrinsic SED from the adopted templates, or  degeneracies and nonlinear mapping in the  colour-$z$ space. We found good quality photometric redshifts ($Q_z \le 1$)  for 75\% and  for 93\% of the galaxies in CDFS and  COSMOS field, respectively.

\begin{figure}
\includegraphics[width=9cm]{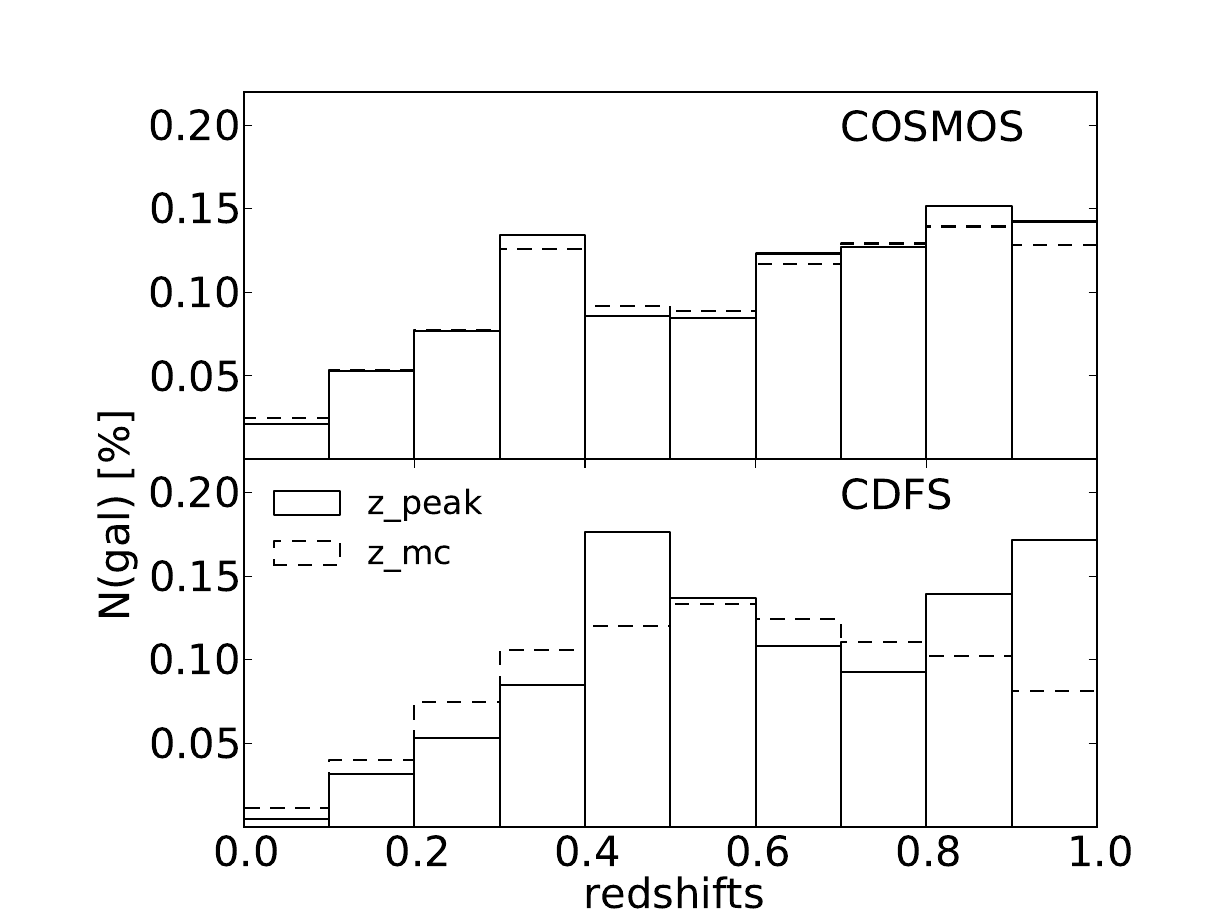} \\
\caption{The distribution of $z_{peak}$ (solid line) and $z_{MC} $ (dashed line) of the galaxies in COSMOS (top panel) and in CDFS catalogues (bottom panel).}
\label{zmc}
\end{figure}

In several cases the $P(z)$ function is multimodal, so that $z_{\rm peak}$, that corresponds to the peak of $P(z)$, does not properly reflect the probability distribution.
 \cite{Wittman:2009rt}  introduced a very simple alternative estimator that  represents the redshift probability distribution, incorporating the redshift uncertainties. This redshift estimator is drawn randomly from the $P(z)$ and denoted with $z_{MC}$ because it results from  a Monte Carlo sampling of the full $P(z)$. The difference between $z_{\rm peak}$ and $z_{\rm MC}$ can be used as an indication of the internal uncertainties of photometric redshifts. The difference in the redshift distribution obtained with different redshift estimators can be seen in Fig.~\ref{zmc}.
 
\subsubsection{Comparison with Spectroscopic Redshifts}

\begin{figure}
$
\begin{array}{c@{\hspace{.1in}}c@{\hspace{.1in}}c}
\includegraphics[width=9cm]{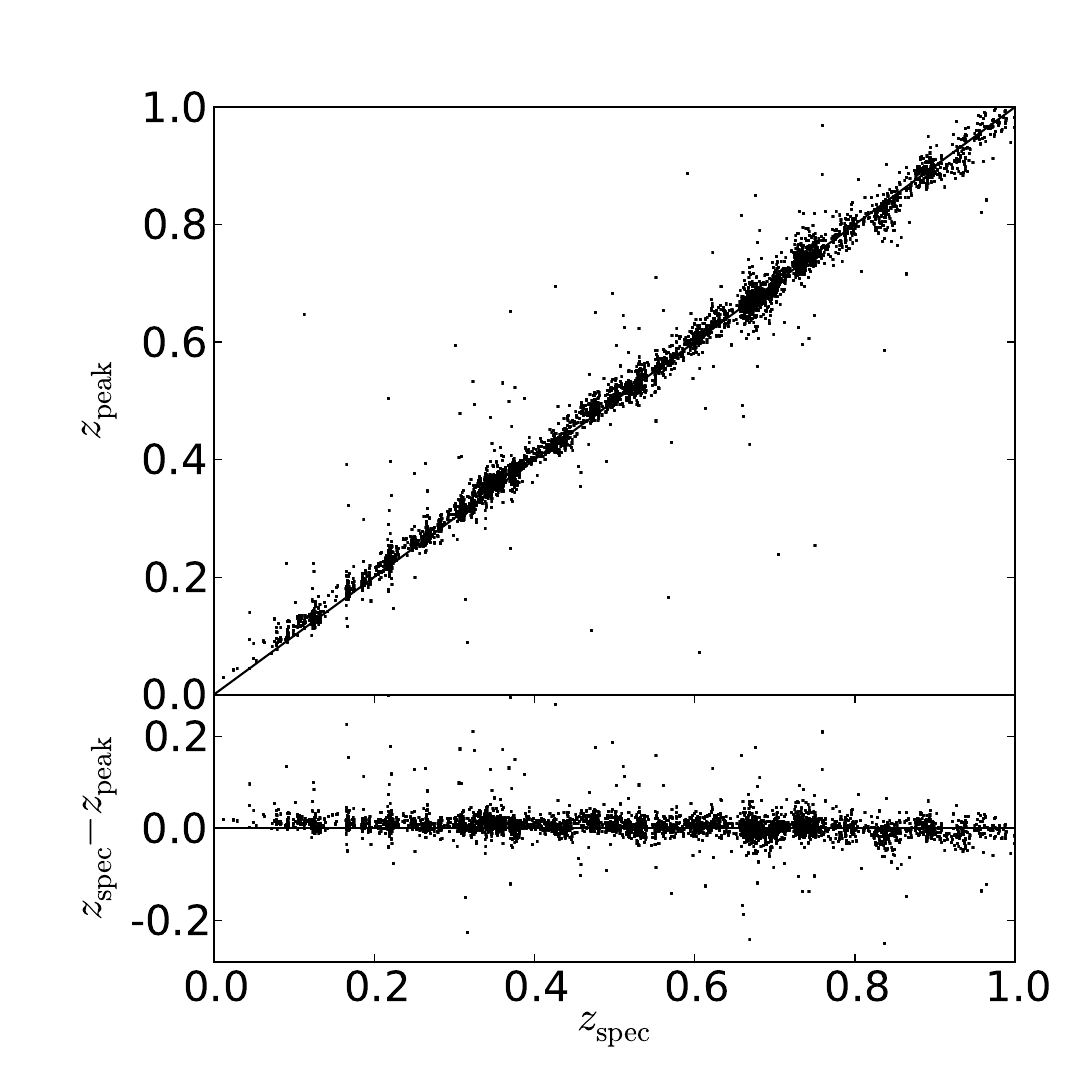}\\
\includegraphics[width=9cm]{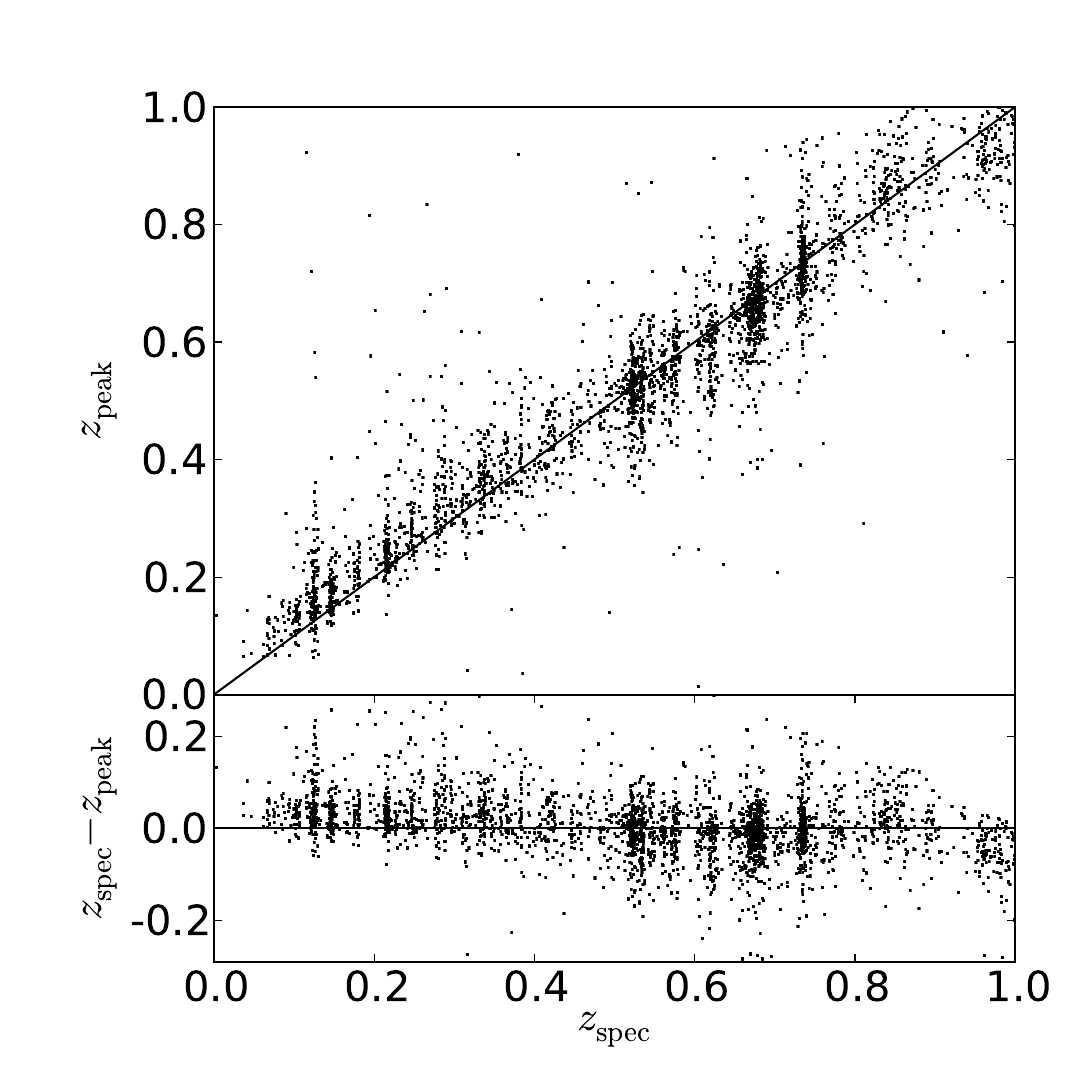} \\
\end{array}
$
\caption{Comparison of photometric vs. spectroscopic redshifts from COSMOS (top panel) and CDFS (bottom panel).}
\label{figz}
\end{figure}

Spectroscopic redshifts are available for a fairly large number of galaxies for both our fields. The spectroscopic redshifts for 4733 galaxies in the COSMOS field were taken from \cite{Muzzin2013a} while for the CDFS field the data for 3362 galaxies were collected from the literature, from different sources and with different quality flags. A comparison of photometric and spectroscopic redshifts for these subsample is shown in Fig. \ref{figz}.

We calculated the normalized, median absolute deviation\footnote{$\sigma_{\rm NMAD} = 1.48\times {\rm median \left| \frac{ \Delta{\rm z-median}(\Delta z)}{1+z_{spec}} \right|}$ as in \cite{Brammer2008} where $\Delta z=(z_{phot}-z_{spec})$. The normalization factor of 1.48 ensures that NMAD of a Gaussian distribution is equal to its standard deviation, and the subtraction of median($\Delta z$) removes possible systematic offsets}, (NMAD) which is less sensitive to outliers compared to the standard deviation \citep{Brammer2008}. For CDFS we found $\sigma_{\rm NMAD}=0.02$ which is comparable to that of other surveys with a similar number of filters, whereas for COSMOS
$\sigma_{\rm NMAD}=0.005$.

Another useful indication of the photometric redshift quality is the fraction of "catastrophic" redshifts defined as the fraction of galaxies for which $ \left| z_{\rm phot}-z_{\rm spec}\right| / (1+z_{\rm spec}) >5\sigma_{\rm NMAD}$. For the CDFS field we found a fairly large fraction of  catastrophic redshifts ($\sim 14$\%). After removing these outliers, the {\it rms} dispersion $\Delta z/(1+z)=0.02$. The same analysis for the COSMOS field \citep[see][for details]{Muzzin2013a} gives a fraction of 5$\sigma$ outliers as low as 4\% and a very small {\it rms} dispersion for the rest of the sample (0.005).

\subsubsection{Comparison with $z_{\rm phot}$ from other surveys}

The comparison between photometric and spectroscopic redshifts is biased towards brighter galaxies for which it is easier to observe the spectrum. To analyse the accuracy of our photometric redshifts in a wider luminosity range, we compare our estimates to those obtained by the Multiwavelength Survey by Yale-Chile  \citep[MUSYC,][]{cardamone:2010zr} which covers  the $\sim 30'\times 30'$ ``Extended'' Chandra Deep Field-South (that is included in CDFS1) with 18 medium-band  {\bf filter} optical imaging from the Subaru telescope, 10  broad-band optical and NIR imaging from the ESO MPG 2.2 m (Garching-Bonn Deep Survey), ESO NTT and the CTIO Blanco telescopes along with 4 MIR bands IRAC imaging from  Spitzer SIMPLE project.
The MUSYC catalog lists $BVR$-selected sources with photometric redshifts derived with the {\sc EAZY} program.  Therefore the main difference is that the MUSYC catalogue makes use of a much larger number of  filters compared with SUDARE, which significantly improves the photometric redshift accuracy.

\begin{figure}
\includegraphics[width=9cm]{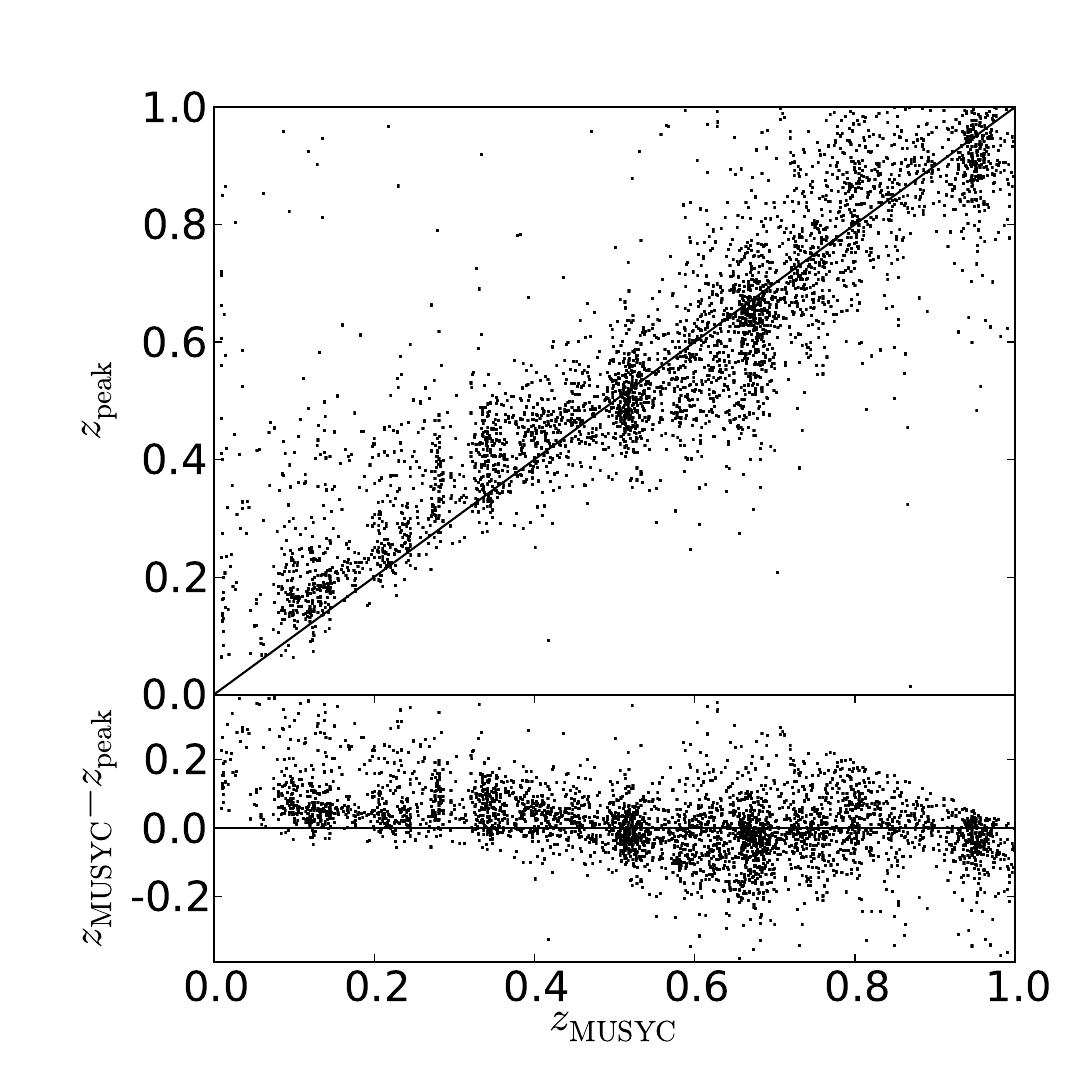} 
\caption{Comparison of our photometric redshift  for CDFS  and MUSYC photometric redshifts. }\label{musyc}
\end{figure}

Cross correlating the two catalogs with a search radius of 2$\arcsec$ we found 1830 common galaxies. In Figure \ref{musyc}, we plot the differences between the $z_{\rm phot}$ estimates as a function of the $z_{\rm phot}^{\rm MUSYC}$.  We find evidence of a some systematic difference at low redshifts  $z<0.3$, with the $z_{\rm phot}$ from SUDARE being  higher, but overall the two catalogs shows  a fair agreement with a scatter  $\Delta{\rm z/(1+z)}=0.05$ and a $5\sigma$ outlier fraction of 10\%.

\section{Computing SN rates\label{ratecalc}}

To compute the SN rate  we need to introduce the method of the control time \citep[CT,][]{zwicky:1942uq}. The CT of one observation is defined as the interval of time during which a SN occurring at a given redshift is expected to remain above the detection limit of the image. The total CT of an observing campaign is  properly computed by summing the CT of the individual observations \citep{cappellaro:1997fk}. Then, the SN rate  is computed as the number of events detected in the survey  divided by the total CT.  

The CT depends on the SN luminosity and light curve evolution and is therefore different for different SN types. We considered separately the following main SN types: Ia, Ib/c, II (including IIP and IIL), IIn and SLSN. 

\subsection{The control time}\label{controltime}

To compute the CT we select  a template light curve representative of a given SN subtype (SNi),  a redshift ($z$, in the range $0<z<1$), and an extinction value in the range $A_V=0 - 2$ mag (in the host galaxy rest frame). To take into account the diversity of the photometric evolution for SNe of different types we used a wide collection of  light curve templates (listed in  Tab.~\ref{templateslist}).
We considered four representative subtypes for thermonuclear SNe (normal, bright, faint and peculiar) ,  six subtypes for hydrogen rich SNe (IIP, IIP faint, IIL, IIb, IIn, plus peculiars) three subtypes for stripped envelope SNe  (Ib, Ic, Ic broad line) along with a template for  SLSN.   In some cases, we use a few templates for the same SN subtype to take into account the photometric variance within the class. 

Then:

\begin{itemize}
\item we define a useful range for the epochs of explosion. In order to be detectable in our search, a SN needs to explode in the interval $[ t_0 - 365d,\,t_K]$, where $t_0$ and $t_K$ are respectively the epochs of the first and last observations of the given field. In fact,  for the redshift range of interest of our survey, a SN exploded 1~yr earlier than the first observation is far too faint to be detected.

\item we compute the expected magnitude, $m_i$, at each epoch of observations, $t_i$ (with $i=1,2,.... K$, where $K$ is the number of observations), for a SN that explodes at an epoch $x_j$ included in the time interval  defined above. To derive these estimates, we use the SN template light curve, the proper K-corrections, the distance modulus for the selected redshift and the adopted extinction. 

\item the detection probability $p_i(x_j)$ of the simulated event at each observing epoch is given by the detection efficiency for the expected magnitude, $\epsilon_i(m_i)$,  estimated as described in Sect.\ref{artstar}. The detection probability for the whole observing campaign is derived as the complement of the probability of not detection at any of the epochs, that is 
 $p(x_j) = 1 - \prod_{i=0}^K{(1-p_i(x_j))}$.

\item we simulate a number $N$ of events exploring the possible epochs of explosion, in the interval $[ t_0 - 365d,\,t_K]$.   We can then compute:

 \begin{equation}
 CT _{{\rm SNi},E_{BV}}(z)= (t_K - t_0 + 365) \frac {\sum_{j=1}^N {p(x_j)}}{N} 
 \end{equation}

where $t$ is expressed in Julian Day.
The accuracy of the CT computation above depends on the sampling for the explosion epoch in the defined interval. After some experiments, we found that a sampling of 1d is more than adequate,
considering also the contribution of other error sources.
 
\item for the extinction distribution, following  \cite{neill:2006lj} we adopted an half-normal distribution with  $\sigma_{E(B-V)}=0.2$. 
We adopt the same  distribution for all SN subtypes although we may expect that different SN types, exploding in different environments, may suffer different amount of extinction. In particular, the distribution of \cite{neill:2006lj} was derived for SN~Ia and is likely to underestimate the effect for CC SNe.  
In Sect.~\ref{Syst} we verify (a posteriori) the consistency of our assumptions about the extinction distribution and estimate how its uncertainty propagates in the systematic uncertainty of SN rates.

\item
finally, the CT for each of the main SN types was computed by accounting for the subtype distribution and for the adopted extinction distribution (details below):

\begin{equation}
\label{ct}
  CT_{\rm SN}(z)=\sum_{\rm SNi}\,\sum_{\rm E_{BV}}
  f_{\rm SNi}\, \rm g_{E_{BV}}\,CT_{\rm SNi,E_{BV}}(z).
\end{equation}
where $\rm f_{SNi}$ is the SN subtype fraction and  $\rm g_{E_{BV}}$ the distribution of colour excess $E(B-V)$
 (multiple templates for a given subtype are given equal weight).

\end{itemize}

In principle, with sufficient statistics and accurate subtype classification, the fractional contribution of the different subtypes can be derived from the distribution of detected events. However, at the current stage of the project, the event statistics is not large enough and we adopted the subtype distribution from the literature.
In particular,  we adopted the fractions of different SN subtypes obtained by \cite{li:2011zr} with the exception of the fraction of faint type II SNe that is from  \cite{pastorello:2004kx}  while for bright Ic SNe  we refer to \cite{podsiadlowski:2004vn}.
Our adopted subtype distribution is:

\begin{itemize}
\item type Ia:  70\% normal, 10\% bright 1991T-like, 15\% faint 1991bg-like  and 5\% 2002cx-like ;
\item type II: 60\% IIP,  10\%  2005cs-like  , 10\%  1987A-like, 10\% IIL  and 10\% IIb,  
\item  type Ib/c:  27\%  Ib, 68\% Ic and 5\% 1998bw-like,
\item  type IIn: 45\%  1998S-like,  45\%  2010jl-like, 10\% 2005gj-like, 
\end{itemize}

We stress that these subtype distributions are obtained from a local sample and it is possible, or even expected, that they evolve with cosmic time. In Sect.~\ref{Syst} we estimate the uncertainty implied by this assumption.

\begin{figure}
\includegraphics[width=\hsize]{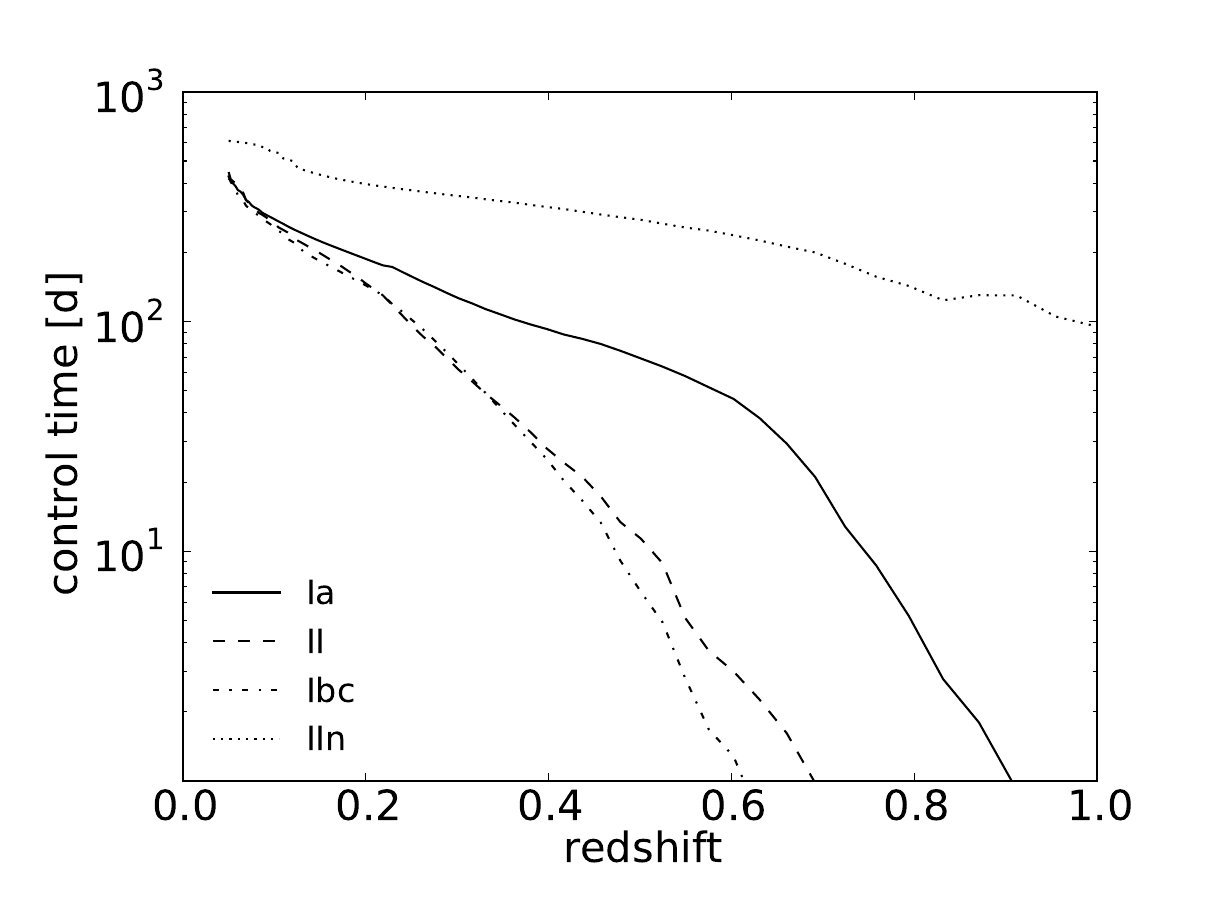}
\caption{Control time as a function of redshift for different SN types (in days per deg$^2$) averaged across the three survey fields.}
\end{figure}

\subsection{SN rate per unit volume}

The volumetric SN rates per  redshift bins in the range $0< z < 1$ is calculated as:

\begin{equation}
\label{volrate}
  r_{\mathrm{SN}}(z)=\frac{(1+z)}{V(z)}\frac{N_{\rm SN}(z)}{CT_{\rm SN}(z)}.
\end{equation}

\noindent where $\rm N_{\rm SN}(z)$ is the number of SNe of the given type in the specific redshift bin, $CT_{\rm SN}(z)$ is the control time and the factor $(1+z)$ corrects for time dilation. $V(z)$ is the comoving volume for the given redshift  bin that is computed as:

\begin{equation}
\label{eq:volume}
V(z)=\frac{4\pi}{3} \frac{\Theta}{41253}\left[
  \frac{c}{H_0}\int_{z_1}^{z_2}\frac{dz^\prime}{\sqrt{\Omega_M(1+z^\prime)^3+\Omega_\Lambda}}
  \right]^3 \mathrm{Mpc}^3
\end{equation}

\noindent where $\Theta$ is the  search area in deg$^2$ and $z$ is the mid-point of the redshift bin with extremes $z_1,\,z_2$.

\subsection{Statistical and systematic errors}\label{Syst}

We derive the 1-$\sigma$ lower and upper  confidence limits from the event statistics as in \cite{gehrels:1986fk}. Then, these values are converted into confidence limits of  SN rates through error propagation of Eq.~\ref{volrate}. 

There are many sources of systematic errors. To estimate each specific contribution we performed a number of experiments calculating  SN rates under different assumptions.\\
 
{\it  Transient Misclassification}

As described in Sect.\ref{classtool}, for a fraction of SN candidates (23\%) the SN confirmation remains uncertain. These PSNe are attributed a weight 0.5 in the rate calculation. 
To obtain an estimate of the impact of this assumption,  we compute the rate in the  extreme cases assuming for these events a weight 0 and  1 respectively. As an error estimate, we take the deviation from the reference value of the rates obtained in the two extreme cases. It turns out the error is of the order of $10-15\%$ (Tab.~\ref{systematic}). 
One concern is that we set arbitrary thresholds for $P_{\chi^2}$ and $N_{\rm pt}$ to attribute the flag of PSN.
To test the impact of this assumption, we computed the SN rate adopting different thresholds, namely $10^{-3}$ or $10^{-6}$ for $P_{\chi^2}$ and $5$ or $9$ for $N_{\rm pt}$. We found that in all cases the deviations for the reference value are $<10\%$ (typically $\sim 5\%$).
\\

{\it SN photometric typing}

For the errors of on SN typing we adopt the values discussed in Sect.~\ref{phclass} that is
10\% for type Ia, 25\% for type II and 40\% for type Ib/c and IIn, independently on redshift.
It appears that the error on SN typing has in general a moderate impact for type Ia, and  instead it is one of the dominant sources for SN~CC in general.
\\

{\it Subtype distribution}

The adopted SN subtype distribution affects the estimate of SN rates because the subtypes have different light curves hence different control times. We consider an error of 50\% for the fraction of the  subclasses and, as an estimate of the contribution to the systematic error, we take the range of values of the SN rates obtained with the extreme subtype distribution.      
This is a significant source of error, typically 10-20\% but with a peak of 40\% for type IIn SNe.
\\

{\it Detection efficiencies} 

We performed MonteCarlo simulations assuming that the value of the detection magnitude limit  of each observation has normal error distribution with $\sigma=1.0$ mag. We found that the frequent monitoring of our survey means that the large uncertainty in the detection efficiency for each single epoch 
does not have a strong impact on the overall uncertainty. The propagated error on the rates is  $ \le 10\%$. 
\\

{\it  Host galaxy extinction}\\
 In our computation we adopt a half-normal distribution of $ E(B-V)$ with  $\sigma =0.2\,{\rm mag}$  for both type Ia and CC SNe. To estimate the effect of this assumption, 
 the SN rates have been recalculated assuming a distribution with $\sigma =0.1\,{\rm  mag}$ and $\sigma =0.3\, {\rm mag}$.
We evaluate that the error on the rates is of the order of $5-10\%$.
The uncertainty is more critical for CC SNe ($11\%$) and for the highest redshift bin of type Ia SNe ($14\%$).\\

\begin{figure}
\includegraphics[width=\hsize]{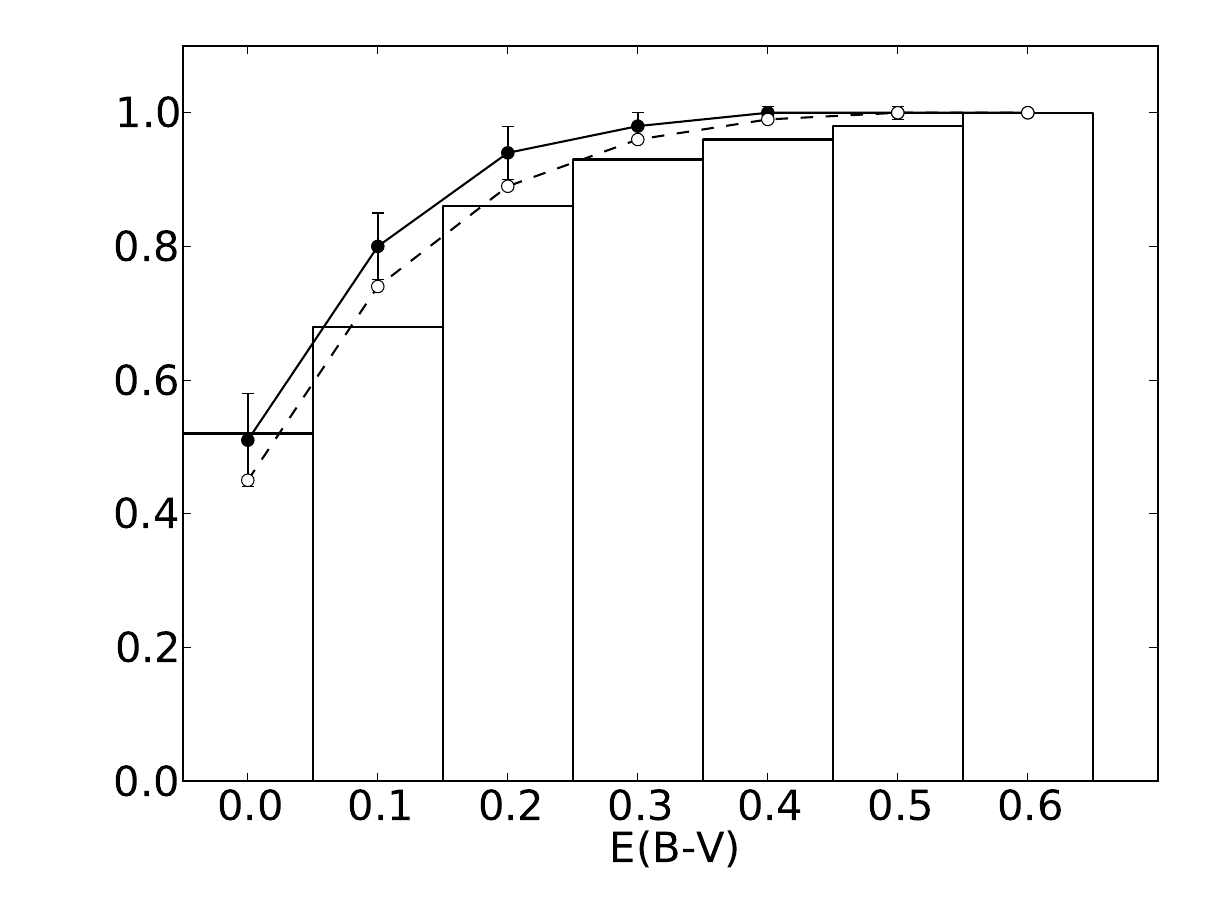}
\caption{Predicted (line) vs. observed (bars) cumulative extinction distribution.}\label{extinct}
\end{figure}

The consistency of the adopted extinction distribution was verified a-posteriori.
We computed estimates of the SN rates for a range of  $\sigma_{E(B-V)}$ values ranging from 0 to 0.5 mag. For each adopted  $\sigma_{E(B-V)}$,  we  computed the expected distribution of extinction of the detected SNe. This is different from the intrinsic distribution because of the bias against the detection of SNe with high extinction which have a shorter control time. The expected extinction distribution is compared with the observed distribution (Fig.~\ref{extinct}) and the best matching $\sigma$  is determined using a Kolmogorov-Smirnov 2-side test. We found a best match for $\sigma_{E(B-V)}=0.25$ mag using the full SN sample or $\sigma_{E(B-V)}=0.28$ mag using only the type Ia events and excluding probable SNe (PSN). 
Given the uncertainties, we consider that this values are consistent with the adopted distribution from \cite{neill:2006lj}.
We remind that the  adopted  $ E(B-V)$ distribution was used only for the CT calculation and not for the SN photometric classifier.
\\

{\it Photometric redshifts}

 We compare the results obtained using the two alternative photo-z estimatator $z_{\rm peak}$ and $z_{\rm mc}$ (c.f. Sect.\ref{zconf}).  It turns out that this is the most significant source of error, especially for type Ia SNe. A detailed analysis shows that the most important effect is for the redshift of SN host galaxies, while the effect on the control time of the galaxy population has a smaller impact.\\

{\it Cosmic variance} 

The possible under/over-density of galaxies in the field of view due to cosmic variance impacts on the SN rate measurements.
Using the cosmic variance calculator of \cite{trenti:2008fk} we found that cosmic variance can add an uncertainty of the SN rate of 5-10\% but  for the low redshift bin the variance can be as large as 15-20\%.
We note that  the cosmic variance bias  is averaged out when rate measurements from different sky fields are analyzed together as for SUDARE.
 \\

\begin{table}
\caption{Relative systematic errors}\label{systematic}
\begin{tabular}{l|cccc|cc|c}
\hline
& \multicolumn{4}{|c|}{Ia} & \multicolumn{2}{|c|}{CC}& IIn\\
$<z>$           & 0.10  & 0.25 &  0.45 & 0.65 & 0.10 & 0.25 & 0.55\\
\hline
PSN             &   -  &  0.04  &  0.11  &  0.12   &   0.26  & 0.16 & 0.23 \\
SN typing       & 0.10 &  0.10  &  0.10  &  0.10   &   0.32  & 0.31 & 0.40 \\
subtype distr.  & 0.08 &  0.10  &  0.20  &  0.24   &   0.07  & 0.12 & 0.40 \\  
detection eff.  & 0.02 &  0.05  &  0.05  &  0.07   &   0.02  & 0.05 & 0.03 \\
extinction      & 0.04 &  0.03  &  0.07  &  0.14   &   0.02  & 0.11 & 0.05  \\
z distribution  & 0.33 &  0.21  &  0.17  &  0.21   &   0.09  & 0.07 & 0.03  \\
\hline
all             & 0.36 & 0.26  &  0.31  &  0.39   &  0.43   & 0.39 & 0.61  \\
\hline  
\end{tabular}
\end{table}

In Table \ref{systematic} we report the individual systematic errors along with the overall  error obtained by their sum in quadrature. We do not include the effect of cosmic variance in the error budget, since this is not a measurement error. Rather, this is an uncertainty related to the particular galaxy sampling in our survey.

The overall systematic error is typically of the order of $30-40\%$, and is larger than the statistical error.

\section{SN rates as a function of cosmic time}

Our  SN rates per unit volume are reported in Tab~\ref{ratevol}.  
Columns 1  and 2 report the SN type and redshift bin, col.~3 gives  the number of SNe (the number of PSNe is in parenthesis), col.~4  the rate measurements,  cols. 5 and 6  the statistical and systematic errors respectively.
The redshift bins were chosen to include a significant number of SNe (a minimum number of  10 SNe) with the exception of the nearest redshift bin ($0.05<z<0.15$) , where we collected only few SNe.

\begin{table}
\caption{SN rates per unit volume [$10^{-4}\,\mathrm{yr}^{-1}\,\mathrm{Mpc}^{-3}$]. Upper limits were computed for a reference value of 3 events. In this case, from Poisson statistics, the probability of obtaining a null results is $\le 5\%$.}\label{ratevol}
\begin{tabular}{cccccc}
\hline
 SN type & $z_{bin}$  &SNe    &  rate &stat. & syst.  \\
\hline
\\
\multirow{4}{*}{Ia}      & 0.05-0.15 & 3.0    & 0.55 &$-{0.29}$ $+0.50$   & $\pm0.20$\\
                         & 0.15-0.35 & 12.7   & 0.39 &$-{0.12}$ $+0.13$   & $\pm0.10$\\ 
                         & 0.35-0.55 & 23.0   & 0.52 &$-{0.13}$ $+0.11$   & $\pm0.16$\\
                         & 0.55-0.75 & 17.4   & 0.69 &$-{0.18}$ $+0.19$   & $\pm0.27$\\
       \\
\hline
\\
\multirow{2}{*}{CC}    &  0.05-0.15  &  5.9   & 1.13 & $-0.53$ $ +0.62$   &$\pm0.49$\\
                       &  0.15-0.35  &  26.2  & 1.21 & $-0.27$ $ +0.27$   &$\pm0.47$\\
\\
II                     &  0.15-0.35  &  13.4  & 0.69 &$-0.18$ $+0.16$     & $\pm0.24$\\
Ib/c                   &  0.15-0.35  &   9.3  & 0.48 &$-0.17$ $+0.19$     & $\pm0.23$\\
\multirow{2}{*}{IIn}  &   0.15-0.35  &   3.5 &  0.043&$-0.026$ $+0.030$  & $\pm0.026$\\           
                      &   0.35-0.75  &   5.8  & 0.017&$-0.009$ $+0.009$& $\pm0.010$  \\
\\
SLSN                  &   0.35-0.75  &  $-$      & $<0.009$     \\
\hline
\end{tabular}
\end{table}

\begin{figure*}
\includegraphics[width=\textwidth]{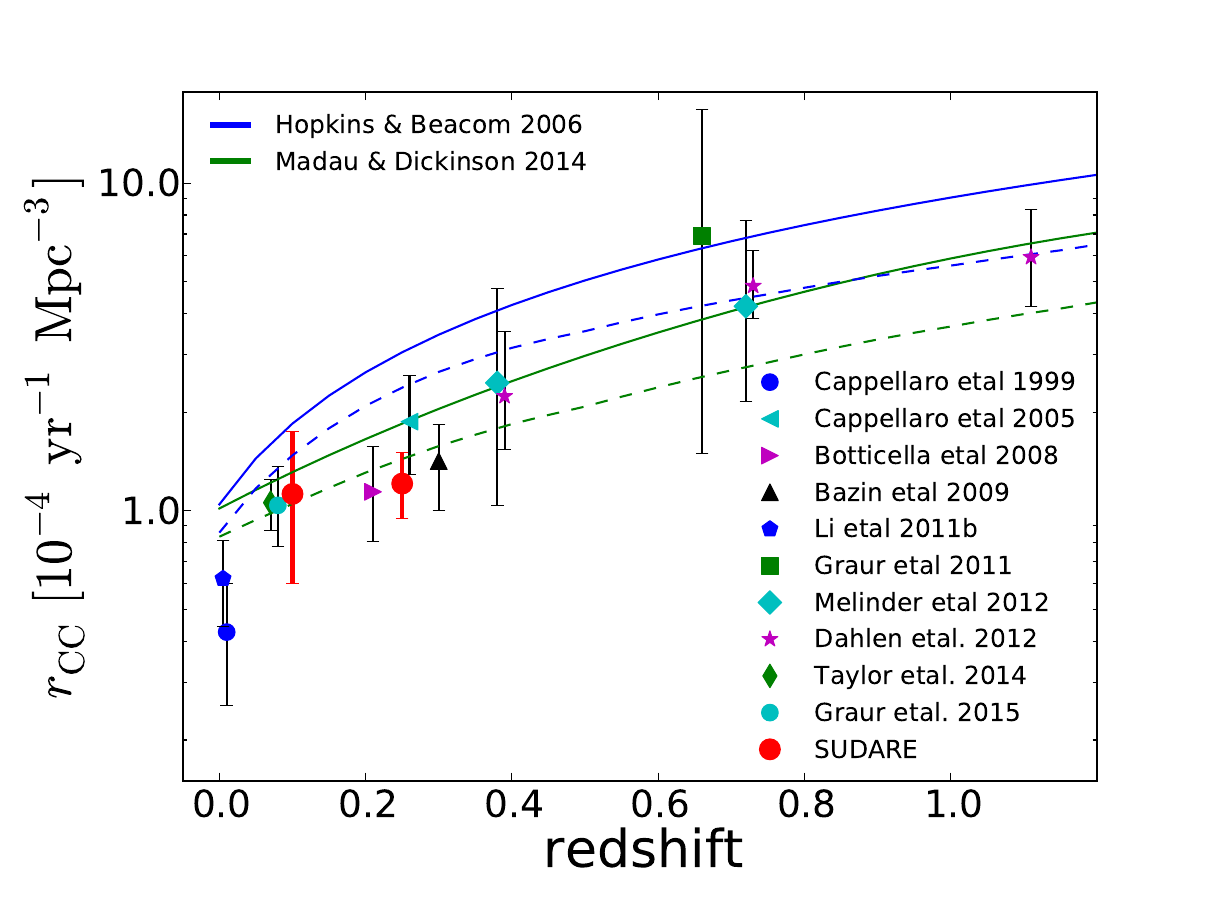}
\caption{CC SN rate per unit volume. All measurements do not account for the correction for hidden SNe. 
To obtain the predicted SN rate from the  measured SFR we adopt 8, 40 $M_\odot$ as the  lower and upper mass limits for SN CC progenitors and the proper IMF, Salpeter for  
 \cite{madau:2014uf} and SalA for \citet{hopkins:2006wj}.
The dashed lines show the 
predicted SN rate assuming the fraction of hidden SNe given in \citep{mattila:2012er}.}\label{CCfig}
\end{figure*}

\subsection{Core collapse SNe}

Fig.~\ref{CCfig} shows a comparison of our estimate of the rate of CC SNe with all measurements available in the literature. To obtain the CC SN rate we cumulated type II, Ib/c and IIn events. 

Our results are in good agreement with other measurements. Note that in Fig.~\ref{CCfig} we report the value of \cite{melinder:2012uq} and \cite{dahlen:2012vn} with no 
correction for the fraction of hidden SNe \citep{mattila:2012er}. We will return on this point later.

Given the short lifetime of their progenitors ($<30$~Myr), there is a simple, direct relation between the CC SN and the current SF rate:

\begin{equation}
r_{\rm CC}(z) = K_{\rm CC} \times \psi(z) 
\end{equation}

where $\psi(z)$ is the SFR and $K_{CC}$ is the number of stars per unit mass that produce CC SNe, that is:

\begin{equation}
K_{\rm CC} = \frac{\int_{m_{\rm L,CC}}^{m_{\rm U,CC}} \phi(m) dm}{\int_{m_L}^{m_U}m\phi(m) dm}
\label{eqkcc}
\end{equation}

where $\phi(m)$ is the initial mass function (IMF), $m_L$ and $m_U$ are the extreme limits of the stellar mass range,  $m_{\rm L,CC}$ and $m_{\rm U,CC}$ the mass range of CC SN progenitors. 

Assuming that $K_{CC}$ does not evolve significantly in the redshift range of interest, the evolution of the CC SN rates with redshift is a direct tracer of the cosmic SF history (SFH). Conversely, we can use existing estimates of the SFH to compute the expected CC SN rate, assuming a mass range for their progenitors. 
In order to do this consistently one has to use the same IMF (or $K_{\rm CC}$) adopted to derive the SFR. 
Indeed, although in Eq.\ref{eqkcc} Kcc depends on the IMF, the ratio between the cosmic SFR and CC rate does not give a real indication on the IMF, since both quantities actually trace the number of massive stars, which produce both UV photons and CC SN events. The formal dependence on the IMF of this ratio is introduced by the extrapolation factor used when deriving the SFH from luminosity measurement to convert the number of massive stars formed at the various redshifts into the total stellar mass formed.

The CC progenitor mass range is still uncertain, both for the low and upper limit.
Stellar evolution models suggest a typical range of $9-40\,{\rm M}_\sun$ \citep{heger:2003fk} for CC SNe, though the upper limit strongly depends on metallicity and other factors, e.g. rotation or binarity. In recent years, for a number of nearby CC SN it was feasible to search for the progenitor star in archival pre-explosion images\citep[][and reference therein]{smartt:2009mq,smartt:2015fk}. This allows to obtain an estimate of the masses of their progenitor stars or, if not detected, of upper limits. 
By comparing the observed mass distribution with the IMF, it was argued that the minimum initial mass is $8\pm1\,{\rm M_\sun}$. The same analysis also suggests a paucity of progenitors of SN~II with mass greater than $20\,{\rm M_\sun}$, which would indicate that these stars collapse directly to a black hole, without producing a bright optical transient \citep{smartt:2009mq}. However this result needs to be confirmed and hereafter, following the trend of the literature in the field, we adopt an upper limit of $40\,{\rm M}_\sun$.

With a mass range $8-40\, {\rm M}_{\odot}$ for the SN~CC progenitors we obtain a scale factor  $K_{\rm CC}=6.7\times 10^{-3}\, {\rm M}_\odot^{-1}$ for a standard Salpeter IMF or $K_{\rm CC}=8.8\times 10^{-3}\, {\rm M}_\odot^{-1}$ for  a modified Salpeter IMF (SalA), with a slope of -1.3 below $0.5\, {\rm M}_\odot$ \citep[similar to what adopted in][]{hopkins:2006wj}.

It has been claimed that assuming the $8-40\,{\rm M}_{\odot}$ mass range, the comparisons between the SFH from \citet{hopkins:2006wj} (hereafter HB06) and the published measurements of CC SN rates showed a discrepancy of a factor 2 at all redshifts  \citep{botticella:2008fr,bazin:2009fd}.
\citet{horiuchi:2011xv} argued that this indicated a "supernova rate problem" for which they proposed some possible explanations: either many CC SNe  are missed in the optical searches because of heavy dust-obscuration or there is a significant fraction of intrinsically very faint (or dark) SNe when, after core collapse, the whole ejecta falls back onto the black hole. 

On the other hand, \citet{botticella:2012sh} found that CC SN rate in a sample of galaxies within 11 Mpc is consistent with that expected from the SFR derived from FUV luminosities.  \cite{taylor:2014lq} based on the SDSS-II SN sample estimated that the fraction of missing events is likely of the order of 20\%.
\citet{gerke:2014yq} performed a search for failed SNe  monitoring a sample of nearby galaxies ($<10$ Mpc). After 4 yr they found only one candidate suggesting an upper limit of 40\% for the fraction of dark events among CC SNe that, unfortunately is not yet a strong constraint.

To detect the CC SNe hidden by strong extinction, a few infrared SN searches have been performed  in local starburst galaxies \citep{maiolino:2002vn,mannucci:2003wl,mattila:2001lt,miluzio:2013fq} in some cases exploiting adaptive optics \citep{cresci:2007ly,mattila:2007nb,kankare:2008ys,kankare:2012zr} to improve the spatial resolution. However, despite the efforts, it was not possible to unveil the hidden SNe.

An alternative approach to estimate the fraction of hidden SNe was made with the conservative assumption that all SNe in the nucleus of luminous and ultra-luminous infrared galaxies (LIRGs and ULIRGs) are lost by optical survey \citep{mannucci:2007tz}.  With this approach, \citet{mattila:2012er}  suggested that  the fraction of missed SNe increases  from the average local value of $\sim19\%$ to $ \sim38\%$ at  $z\sim1.2$.

The CC SN rate predicted using  two different SFH  from  HB06 and the recent results of 
\citet{madau:2014uf} (hereafter MD14) are shown in Fig.~\ref{CCfig}. The two SFH lead to different predictions  with a discrepancy of about a factor 2. Indeed, the cosmic SFR derived by MD14 is lower than the HB06's virtually at all redshifts. In addition, the MD14 SF rates assume a straight Salpeter IMF, so that the number of massive stars formed in the Universe at all epochs is further diminished, when compared to predictions obtained with HB06's SFH. Both factors concur in producing the final result shown  in Fig.\ref{CCfig}.    
The predictions based on the MD14 SFH are in good agreement with the data. We note that by applying the \citeauthor{mattila:2012er} correction for hidden SNe (dashed green lines in the figure) improves the fit at low redshift but gives a worse comparison at high redshift.

The HB06 SFH instead over-predicts the CC rate, if the progenitors come from the mass range 8 to $40\,{\rm M}_\odot$. In this case, correcting for hidden SN improves the agreement at high redshift, but still overestimates  the CC SN rates at $z < 0.4$ (dashed blue line).

On the other hand, as we mentioned above, the uncertainty in the CC progenitor mass range is still significant. Indeed, most recent data collection seems too indicate a lower mass limit as high as $9-10\,{\rm M}_\odot$ \citep{smartt:2015fk}. If we use this in combination  with the upper mass limits for CC SN progenitors of $\sim 20\,{\rm M}_\odot$  \citep[e.g.][]{smartt:2009mq,gerke:2014yq}
and also include the correction for hidden SNe, it would result that the SFR severely under-predict the CC rate (by over a factor 3 if we refer to the MD14 SFR). 

All together, it is fair to say that the both  statistic and  systematics errors  on  SN rate  and SFR measurements and the uncertainties on the progenitor mass range are too large   to invoke a "SN rate problem"  and hence to speculate on possible explanations.   

One of the goals we aim to achieve with our survey it to obtain  measurements of the evolution of specific SN subtype. While the statistics of the present sample is still small, we can however obtain some preliminary measurements.
 
\subsubsection{SN Ib/c}
 
We found that at the mean redshift $z=0.25$, type Ib/c are $40\pm 13\%$ of CC SNe. This  compares very well with the estimates of
\cite{li:2011qf} for the local Universe, measuring a fraction of Ib/c that ranges from $46\pm  17\%$ in early spiral galaxies to $20\pm 5\%$ in late spirals, with an average value 
of $33\pm 9\%$.   The physical reason of the difference in CC SN population in early and late spirals is not understood, though it is possibly related to a metallicity effect 
\citep{li:2011qf}.
 
Given the limited statistics of our sample, we can only conclude that there is no evidence for  evolution with redshift of the Ib/c fraction.

\subsection{SN IIn}

Our estimate of the rate of type IIn SNe is  uncertain for two reasons: SNe IIn are rare and the event statistics is very poor. In addition,  the variety in luminosity and light curve evolution that, in some cases, mimic those of other SNe (e.g. SN~IIL or SLSN), makes the photometric classification very uncertain (cf. Sect. ~\ref{Syst}).
However, due to the  intrinsically bright and slowly evolving luminosity  we could detect type IIn SNe in a redshfit range comparable to that of SN~Ia.

In the 0.15-0.35 redshift bin, we estimated that type IIn are  $4\pm3\%$ of all CC~SNe. This number is consistent with the $6 \pm 2\%$ value measured in the local Universe \citep{li:2011qf}.

On the other hand, the apparent decrease of the rate at higher redshift (a factor $\sim 2.5$ in the redshift bin 0.35-0.75 compared with the nearest bin) appears at odd considering that the overall CC~SN rate in the same redshift interval increases by about the same factor.  

Either there is a strong evolution of the type IIn rate with redshift or, in our search,  we are missing (or mis-classifying) over 2/3 of the distant type IIn.  Both explanations are difficult to accept: we will need to verify this result at the end of our survey with better statistics and, possibly, improved template list.

\subsection{Superluminous SNe}

Super-luminous supernovae  (SLSNe)  that radiate more than $10^{44}\,{\rm erg\, s}^{-1}$ at their peak luminosity  (about 100 times more luminous than usual Type Ia and CC SNe) have recently been discovered in faint galaxies typically at high redshifts. 
The origin of these events is still unclear: the host environment and energetics suggest massive stellar explosions  but  their  power source is  still a matter of debate.  In fact, several sub-classes have been introduced possibly related to
different explosions scenarios \citep{gal-yam:2012uq}. 

We did not detect any SLSN in our surveyed volume up to $z=0.75$.  However the null result for SLSN can provide interesting constraints on the rates of these objects. From simple Poisson statistics we find that the probability of obtaining a null result is 5\%  when the expected values is 3.0. Therefore one should expect that the rate of SLSNe is  {\bf no} higher  than 
$9 \times 10^{-7}  {\rm yr}^{-1} {\rm Mpc}^{-3}$ at a mean redshift $z\sim0.5$.
This firm upper limit is consistent with the rate estimated by   \cite{quimby:2013kx}  of 
$2.0^{+1.4}_{-0.9} \times 10^{-7}\, {\rm yr}^{-1} {\rm Mpc}^{-3}$  at a mean redshift $z = 0.16$ that is 1 SLSN for each 500 CC SNe. 

We note that between  $z=0.15$ and $z=0.5$ the CC rate measurements increases by almost a factor 3.  Our upper limit does not preclude a similar increase for the SLSN rate; clearly we need to obtain a more significative result that will become feasible  when our survey is completed.

\subsection{SNe Ia}

\begin{figure*}
\includegraphics[width=\textwidth]{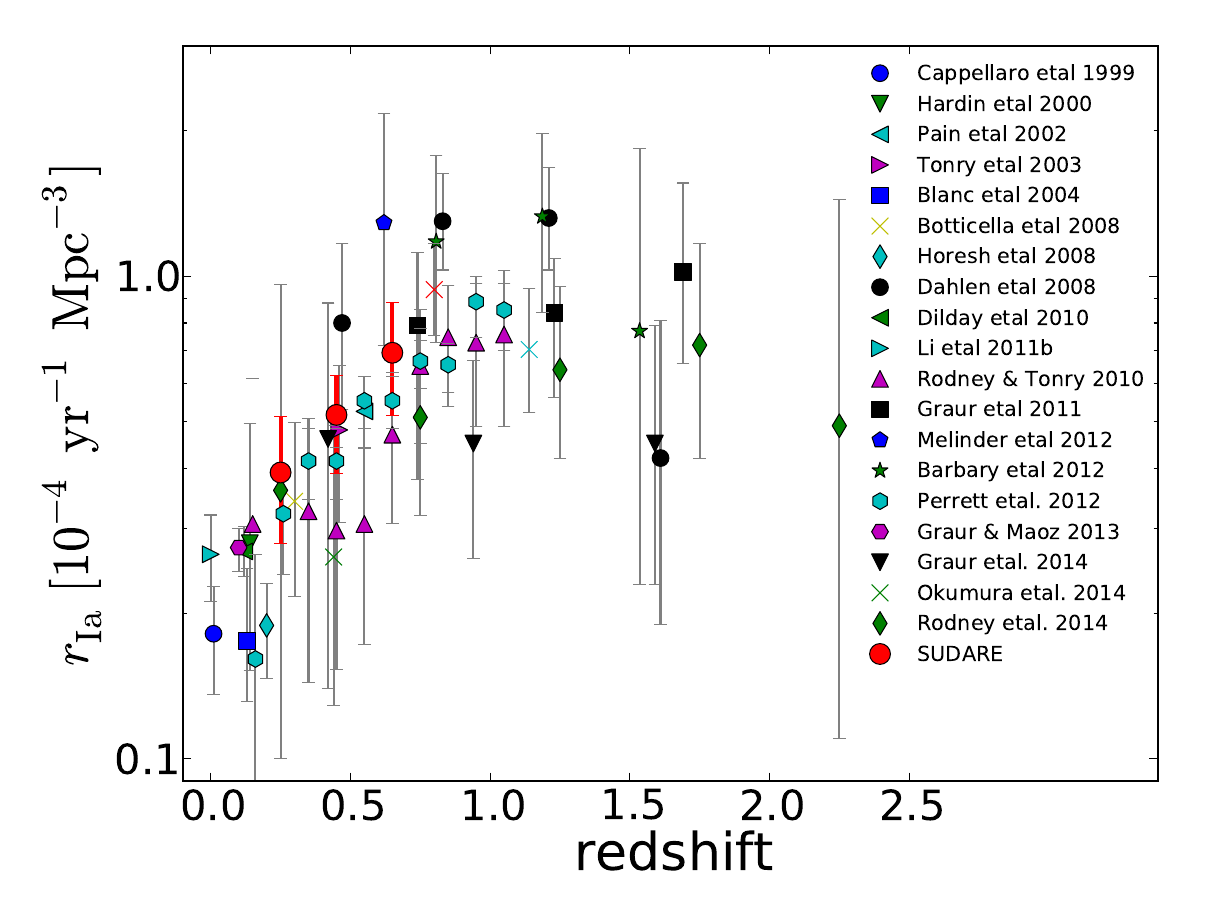}
\caption{Our estimates of the SN Ia rate at $z=0.25,0.45,0.65$ are compared with the other values from literature. The rate of \cite{cappellaro:1999dg,hardin:2000sf,Madgwick:2003vn,blanc:2004zr} were given per unit luminosity. They are converted in rate per unit volume using the following relation of the luminosity density as a function of redshift : 
$ j_B(z) = (1.03+1.76\, z) \times 10^8\, {\rm L}_\sun^B\,{\rm Mpc}^{-3}$ \citep{botticella:2008fr}. The measurements of \cite{perrett:2012uq} are scaled up by 15\% to account for the fact that they not include the faint SN~1991bg-like events.
}\label{Iafig}
\end{figure*}

Our measurements of the rate of SN~Ia  are shown in Fig.~\ref{Iafig} along with all those available  from the  literature\footnote{We do not plot original measurements that have been later revised or superseded as follow: \citet{Madgwick:2003vn} by \citet{graur:2013fd}, \citet{poznanski:2007fv}  by \citet{graur:2011ys}, \citet{dahlen:2004on,kuznetsova:2008zn}  by \citet{Dahlen:2008yq}, \citep{barris:2006vp} by \citet{rodney:2010xy}, \citet{neill:2006lj,Neill:2007hl} by \citet{perrett:2012uq} }.

Our results appear in agreement with other measurements within the statistical errors. It may be noticed, however, that our estimates seem  on the high side compared with the bulk of published measurements.

\begin{figure}
\includegraphics[width=0.5\textwidth]{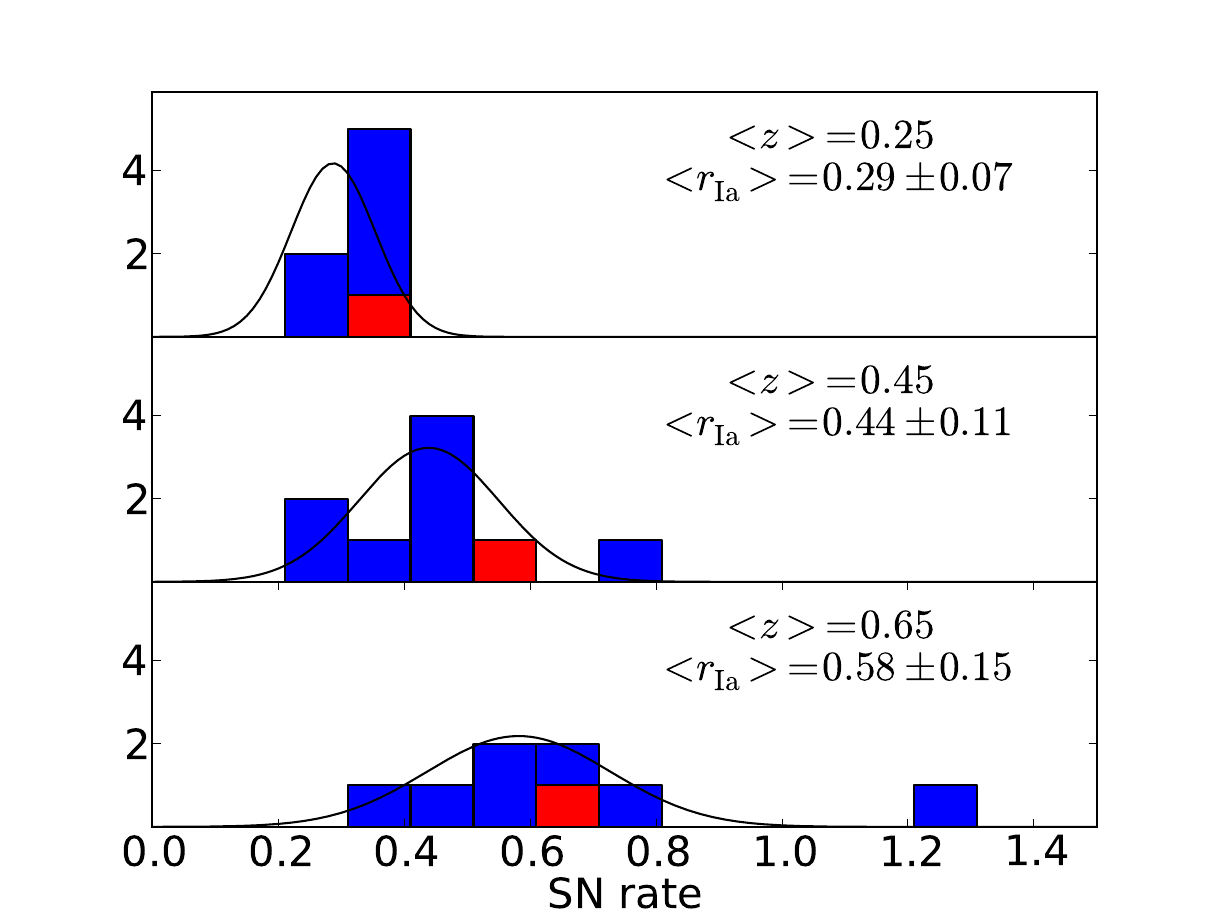}
\caption{Histogram of published estimates of the SN Ia rate in the three redshift bins of our measurements. All values are scaled to the mean redshift of the bin assuming a linear evolution of the rate with redshift (see text). In red we show our  measurements. The black lines are the Gaussian curve whose mean and variance are computed from the data and reported in each panel's legend.  The averages and dispersions were computed by weighting the individual measurements with the inverse  of their statistical errors.}\label{rate_dist} 
\end{figure}

In Fig.~\ref{rate_dist} we plot the histogram of the measurements for the three bins of redshifts corresponding to our measurements. 
The effect of the rate evolution within each of these redshift bins has been removed by scaling the measurements to the mean redshift of the bin assuming that the rates scales as
$r_{\rm Ia} \propto 0.6\times z$, that, as a first order approximation, fits the rate evolution up to redshift $\sim 1$ (the exact slope of the relation may be slightly different but it is not crucial for the comparison we are performing here).
\nocite{cappellaro:1999dg,hardin:2000sf,pain:2002kq,blanc:2004zr,neill:2006lj,horesh:2008zl,graur:2013fd,dilday:2010lr,rodney:2010xy,barbary:2012rr,okumura:2014qq,rodney:2014eu}

It appears from the Figure that for each bin the distribution of measurements (our own included) are  consistent with a normal distribution and the dispersion is well understood considering the statistical and systematic errors affecting the measurements.

Given that, in the following we will use average values as the best estimates of the SN~Ia rate for the comparison with models. The average rates per redshift bin are reported in Tab.~\ref{avrate}, where col.~1 gives the redshift bin, col.~2 the average redshift, col.~3 and 4 the average rate and dispersion and col.~5 the number of measurements per bin. Notice that for redshift $z>0.75$ in the computation of the average rate we did not correct the individual measurements for the possible rate evolution inside the bin.

\begin{table}
\caption{Average of SN~Ia rate measurements per redshift bin (units $10^{-4}\, {\rm yr^{-1}}\, {\rm Mpc}^{-3}$).}\label{avrate}
\centering
\begin{tabular}{ccccc}
\hline
$z_{\rm bin}$ & $<z>$ &  $r_{Ia}$  & $\sigma$ & N\\
\hline
0.00-0.15 & 0.05  &    0.25     &  0.05   & 6 \\
0.15-0.35 & 0.25  &    0.29     &  0.07   & 7 \\
0.35-0.55 & 0.45  &    0.44     &  0.11   & 9 \\
0.55-0.75 & 0.65  &    0.58     &  0.14   & 8 \\
0.75-1.00 & 0.84  &    0.64     &  0.20   & 11\\
1.00-1.50 & 1.16  &    0.87     &  0.22   & 7 \\
1.50-2.00 & 1.64  &    0.63     &  0.22   & 5\\
\hline
\end{tabular}
\end{table}

In Fig.~\ref{rate_sel} we compare the average SN Ia rate measurements  with  the expected evolution for different progenitor scenarios predicted by \citet{greggio:2005ph}.
  
Models in \citet{greggio:2005ph} assume that SNIa progenitors are close binary systems which attain explosion upon reaching the Chandrasekhar mass  either because of mass accretion from a companion star (single degenerate, SD) or of merging with another WD (double degenerate, DD).
The delay between the birth of the binary system and its final explosion ranges from $\sim 40$ Myr to the Hubble time, so that at each epoch the SN events in a galaxy are the result of the contributions of all past stellar generations.
Following \cite{greggio:2005ph}, the expected SN~Ia rate at the time $t$ is:

\begin{equation}\label{sniaeq}
r_{\rm Ia}(t) = K_{\rm Ia}\int_{\tau_{i}}^{min(t,\tau_{x})} f_{\rm Ia}(\tau) \psi(t- \tau)d{\tau}
\end{equation}

where $K_{\rm Ia}$ is the number of SN~Ia progenitors per unit mass of the stellar generation, $f_{\rm Ia}(\tau)$ is the distribution function of the delay times and $\psi(t-\tau)$ is the star formation rate at the epoch $t-\tau$. The integration is extended over the full range of the delay time $\tau$ in the range $\tau_{i}$ and $min(t,\tau_{x})$, with $\tau_{i}$ and $\tau_{x}$ being the minimum and maximum possible delay times for a given progenitor scenario. 
According to stellar evolution, $f_{\rm Ia}(\tau)$ is a decreasing function of 
the delay time, with a slope which depends on details of the scenario leading to the SN explosion.  One can then use Eq.~\ref{sniaeq} to constrain the progenitor's model using the trend of the SNIa rate with cosmic time, after specifying the cosmic SFH.  In the following we adopt the MD14 SFH and assume that $K_{\rm Ia}$ does not vary with cosmic time.

\begin{figure}
\includegraphics[trim=0cm 5cm 0cm 5cm,clip=true,width=0.5\textwidth,]{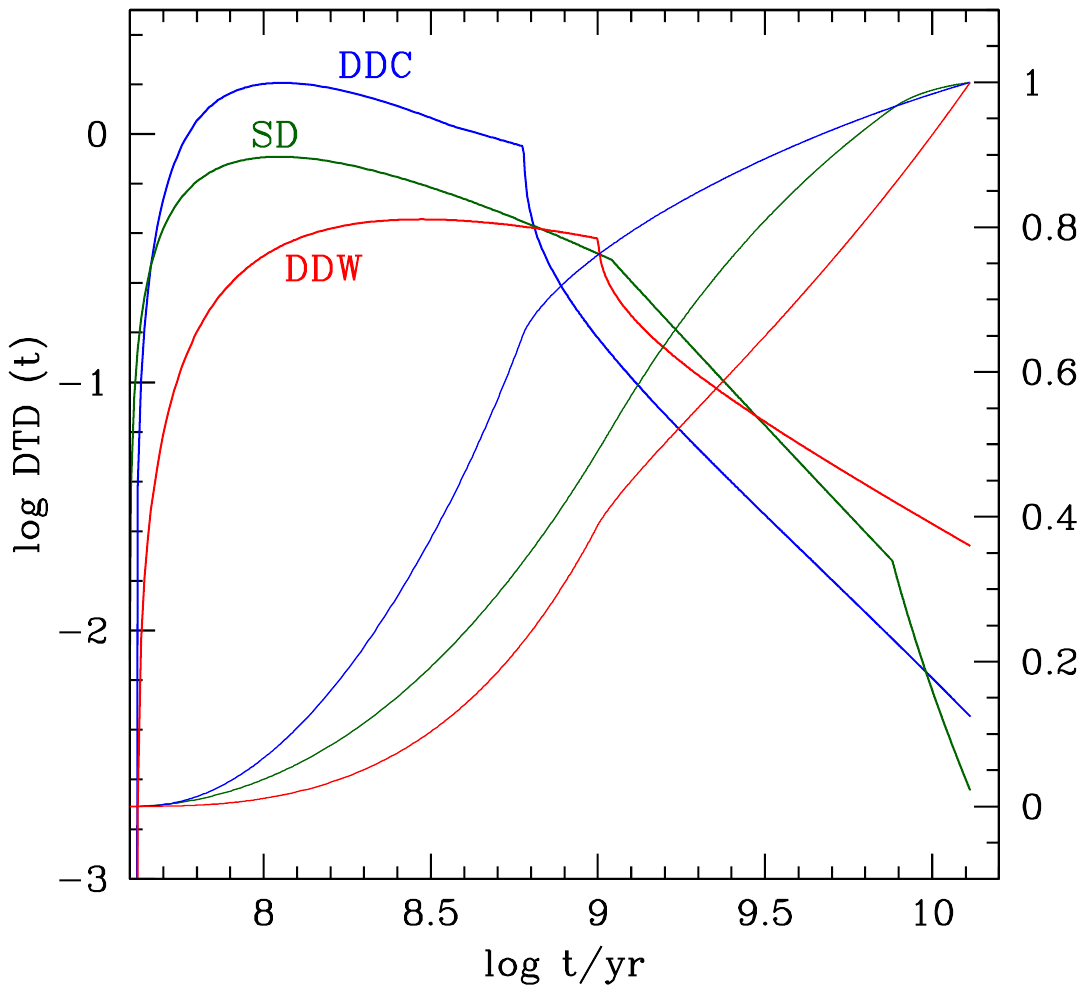}
\caption{Distribution functions of the delay times selected for our theoretical predictions for the Single Degenerate (green), and
Double Degenerate models (blue and red, see text for more details). The thin lines show the cumulative fraction of events as a function of time}\label{fias}
\end{figure}

We select three DTD models, plotted in Fig. \ref{fias}, and test their predictions for the cosmic SNIa rate. The models include a single degenerate realisation (SD), and two flavours of the double degenerates, either with a close (DDC)  or wide (DDW) binary separation predicting a steep and a mildly decreasing distribution of the delay times, respectively \citep[see][for more details]{greggio:2005ph,greggio:2010pd}. 
The selected models correspond to a very different time evolution following a burst of star formation. For the DDC, SD and DDW models, 50\% of the explosions occur within the first 0.45, 1 and 1.6 Gyr respectively, while the fraction of events within 500 Myr is 0.55, 0.3 and 0.18 of the total. The late epoch declines are also different, the rate scaling as $t^{-1.3}$ and $t^{-0.8}$ for the DDC and DDW models respectively.  
 
Fig.~\ref{rate_sel} shows the predicted rates as a function of redshift for each of the three models, having assumed the Madau and Dickinson cosmic SFH.  The best fit of the models with observations was derived by least square minimisation weighting the measurement by their $\sigma$ and gives $K_{\rm Ia}=7.5\times10^{-4}\,{\rm M}_\sun^{-1}$ for SD and DDW and  $8.5\times 10^{-4}\,{\rm M}_\sun^{-1}$ for DDC. 

 The smallest residuals in the whole redshift range is obtained for the SD model (rms$=0.0028$). The DDC model (rms$=0.0088$) gives an excellent fit up to redshift $z\sim1.2$ but predicts a relatively mild decline at higher redshift (still not in conflict with the observations). Instead, the DDW model (rms$=0.012$) shows an overall  shallow evolution with respect to the observations.
Our conclusion is that since the dispersion of the rate measurements is comparable to the scatter of theoretical tracks we cannot discriminate between SD and DD models, tough for the DD scenarios  model with close binary separation seems favoured. 

We remark that the impact of adopting different cosmic SFHs, MD14 or HB06, on the predicted rate is negligible, while it is more relevant for the estimate of constant $K_{\rm Ia}$, that is the number of stars which end up as a SNIa  per unit mass of the parent stellar population.  In particular, fitting the observations with HB06 SFH  requires $K_{\rm Ia}=5.9$, 5.7, 6.9 $\times 10^{-4}\,{\rm M}_\sun^{-1}$ for SD, DDW and DDC respectively, values which are $\sim 20\%$ smaller than those obtained when using the MD14 SFH. The number of potential progenitors per unit stellar mass depends on the IMF and, assuming from to $8\,M_\sun$ range of SN~Ia progenitors,  it is $0.021$ for a plain Salpeter IMF and $0.028$ for a SalA IMF. Stars in the selected mass range that should end up as SN~Ia t oaccount for the observed rates are 4\% and 2\% using  the MD14 and HB06 SFH, respectively. These fractions are close to the lower edge of the range reported in Maoz and Mannucci (2012).  While they remain large with respect to most theoretical predictions from binary population synthesis codes, they  still represent a minor fraction of all potential progenitors.

In any case, we  conclude that
there is no need to invoke ad-hoc delay time distributions, unrelated to the standard expectations from stellar evolution theory. 
This confirms earlier conclusions \citep[e.g.][]{forster:2006uq,blanc:2008rp,greggio:2008er}.

\begin{figure}
\includegraphics[width=9cm]{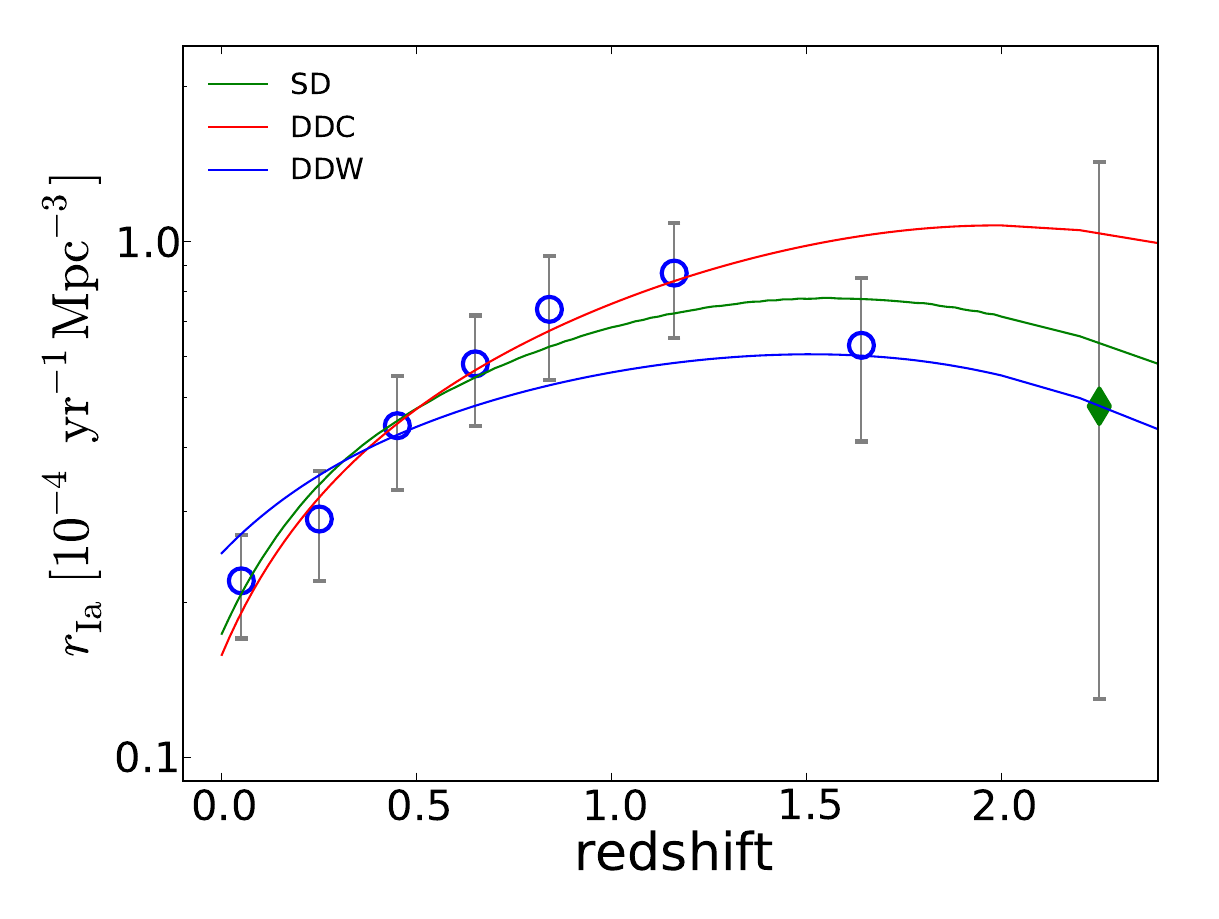}
\caption{Average values of the SN~Ia rates per unit volume as a function of redshift. The still unique measurement 
at $z>2$  of \citet{rodney:2014eu} is also plotted.
For the derivation of the SNIa rate model evolution we adopted the  SFH from \cite{madau:2014uf}.}\label{rate_sel}
\end{figure}

In a forthcoming paper (PII)  we will  
use a detailed characterization of the properties of the galaxy sample in order  to investigate the dependence of SN rates on galaxy parameters, and to obtain additional constraints on the SN progenitor scenarios. 

\section{Conclusions}

We presented the preliminary results of a new SN search, SUDARE, that was designed to measure SN rates in the redshift range $0<z<1$.  This paper describes the survey strategy, the selection and confirmation of candidates, the construction of the galaxy catalog and the rate estimates based on the first two years of the survey.

As search fields we selected two of the best studied extragalactic fields, namely  CDFS and COSMOS for which a wealth of multi-band coverage is available. Our own data, the synergy with the VOICE project, the complementary data from the VIDEO survey, along with public data from the literature allowed us to obtain a multi-band photometric catalog for the galaxies with magnitudes $K\le23.5$ that was exploited to estimate the photometric redshift using the {\sc EAZY} code \citep{Brammer2008}. 
 
We discovered 117 SNe, of which 27 were assigned a weight$=0.5$ due to a poor template match or to a low number of detections. Most of the SNe are classified of type Ia (57\%). For the core collapse, 44\% are type II,
22\% type IIn and 34\% type Ib/c.

With  this SN sample and an accurate measurement of the detection efficiency of our search, we computed the rate of SNe per unit volume.  For the CC SNe our measurements are in excellent agreement with previous results and fully consistent with the predictions from the cosmic  SFH of \cite{madau:2014uf}, assuming a standard mass range for the progenitors ($8<M<40\,{\rm M_\odot}$). Therefore, previous claims of a  significant  disagreement between SFH and CC SN rates are not confirmed. This conclusion relies on the revision of the cosmic SFH because our rate estimates are consistent withother measurements from literature.

For the SN~Ia, our measurements are consistent with literature values within the errors. We conclude that the dispersion of SN~Ia rate estimates and the marginal differences for the evolution with cosmic time of the volumetric SN rate does not allow us to discriminate between SD and DD progenitor scenarios. However, with respect to the three tested models\citep[SD, DDC and DDW from][]{greggio:2010pd}, the SD gives a better fit on the whole redshift range whereas the DDC appears to perfectly match the steep rise of the rate up to redshift 1.2. The DDW model that corresponds to a wide binary separation and a relatively flat delay time distribution appears disfavoured.

As a first attempt at searching for evolution of SN diversity, we found no evidence of evolution of the SN~Ib/c fraction. The fraction of type IIn SNe detected in the $0.15<z<.35$ redshift bin  is consistent with the measurements for the  local Universe. The rate in the higher redshift bin is formally significantly lower. Whether this is evidence for some evolution or a bias in our survey needs to be verified with more data.

\begin{acknowledgements}
We thank the referee, Steve Rodney, for useful comments and suggestions. 
We also thank G. Brammer for helping with galaxy photometric redshifts.\\
We acknowledge the support of grant ASI n.I/023/12/0  "Attivit{\'a} relative alla fase B2/C per la missione Euclid", MIUR PRIN 2010-2011 "The dark Universe and the cosmic evolution of baryons: from current surveys to Euclid" and PRIN-INAF "Galaxy Evolution With The VLT Survey  Telescope (VST)" (PI A. Grado).
G.P. acknowledge the support by Proyecto Regular FONDECYT 1140352 and by  the Ministry of Economy, Development, and Tourism's Millennium Science Initiative through grant IC12009, awarded to The Millennium Institute
of Astrophysics, MAS.
MV acknowledges support from the Square Kilometre Array South Africa project,
the South African National Research Foundation and Department of Science and
Technology (DST/CON 0134/2014), the European Commission Research Executive
Agency (FP7-SPACE-2013-1 GA 607254) and the Italian Ministry for Foreign Affairs
and International Cooperation (PGR GA ZA14GR02).

\end{acknowledgements}



\bibliographystyle{aa}
\bibliography{sudare_2014}

\end{document}